\documentclass[onecolumn,pre,floats,aps,amsmath,amssymb,nofootinbib]{revtex4-2}
\usepackage{geometry}
\usepackage{graphicx}
\usepackage{epstopdf}

\newcommand{\be}{\begin{equation}}
\newcommand{\ee}{\end{equation}}
\newcommand{\bea}{\begin{eqnarray}}
\newcommand{\eea}{\end{eqnarray}}

\begin{document}

\title{Multiple dispersive bounds. II) Sub-threshold branch-cuts}

\author{Silvano  Simula}  
\affiliation{Istituto Nazionale di Fisica Nucleare, Sezione di Roma Tre,\\ Via della Vasca Navale 84, I-00146 Rome, Italy}
\author{Ludovico  Vittorio}  
\affiliation{Physics Department and INFN Sezione di Roma, Universit\`a di Roma La Sapienza, P.le A.\,Moro 2, I-00185 Roma, Italy}

\begin{abstract}
We apply the strategy proposed in the companion paper\,\cite{Simula:2025lpc} for dealing with multiple dispersive bounds, to the case of sub-threshold branch-cuts, which is a topic addressed extensively in the literature (see, e.g., Refs.\,\cite{Caprini80, Caprini:1995wq, Boyd:1995sq, Buck:1998kp, Bhattacharya:2011ah, Epstein:2014zua, Gopal:2024mgb}). We consider the simultaneous application of a double dispersive bound as a proper way to take into account unitarity constraints within phenomenological analyses of hadronic form factors in the presence of sub-threshold branch-cuts. Accordingly, the standard $z$-expansion of hadronic form factors, commonly referred to as the Boyd-Grinstein-Lebed approach\,\cite{Boyd:1994tt, Boyd:1995cf, Boyd:1995sq, Boyd:1997kz}, is modified by including simultaneously the dispersive bounds related to the pair-production and to the sub-threshold regions. For the latter one the effects of above-threshold poles are described through a simple resonance model and the possible choices of the outer function outside the pair-production region are discussed. A detailed numerical analysis of the experimental data or lattice QCD results in the spacelike region for the charged kaon form factor is presented as a direct application of the procedure of double dispersive bound. The comparison with other methodologies present in literature and with the $z$-expansion based on the single, total dispersive bound clearly shows that the $z$-expansion including the double dispersive bound provides the most precise extrapolation at large momentum transfer as well as the most stable results with respect to the choice of the outer function outside the pair-production region.
\end{abstract}

\maketitle

\section{Introduction}
\label{sec:intro}

A proper treatment of sub-threshold branch-cuts through unitarity and analyticity is called for in the study of many physical processes, ranging from the analysis of the electromagnetic form factors to weak semileptonic decays of hadrons. Indeed, the hadronic form factors describing the transition induced by a current $J$ between two hadrons $H_1$ and $H_2$, may be characterized by a branch-cut starting below the pair-production threshold. Such \emph{sub-threshold} cut corresponds to the production of on-shell particles $\overline{H}_1^\prime H_2^\prime$ and to their subsequent scattering onto an off-shell pair $\overline{H}_1 H_2$.

The topic of sub-threshold branch-cuts has been addressed extensively in the literature (see, e.g., Refs.\,\cite{Caprini80, Caprini:1995wq, Boyd:1995sq, Buck:1998kp, Bhattacharya:2011ah, Epstein:2014zua, Gopal:2024mgb}).
In this work, we apply the strategy proposed in the companion paper\,\cite{Simula:2025lpc} for dealing with multiple dispersive bounds, to the case of of sub-threshold branch-cuts within the standard $z$-expansion of hadronic form factors, commonly referred to as the Boyd-Grinstein-Lebed (BGL) approach\,\cite{Boyd:1994tt, Boyd:1995cf, Boyd:1995sq, Boyd:1997kz}. Our goal is to implement the properties of unitarity and analyticity by imposing a double dispersive bound, which takes simultaneously into account unitarity in the pair-production region \emph{and} within the lowest sub-threshold and the pair-production branch-points. We stress the importance of considering both constraints in the BGL $z$-expansion of hadronic form factors, as already attempted in Refs.\,\cite{Caprini:1995wq, Boyd:1995sq, Caprini:2015wja, Gopal:2024mgb}, but at variance with what has been done recently in literature (see later Section\,\ref{sec:Szego}). In particular, while the first bound can be imposed through the knowledge of the so-called susceptibility, which can be estimated from the derivatives of appropriate two-point Green functions evaluated in momentum space, the second bound requires a separate estimate in a range of timelike values of the momentum transfer not accessible by experiments. 
To this end we develop a model for the effects of above-threshold resonances on the hadronic form factor. Our model successfully describes the $\rho$(770)-meson resonance in the case of the electromagnetic form factor of the pion, as well as the main poles entering the study of the electromagnetic form factors of charged and neutral kaons. 
Furthermore, we highlight that the choice of the outer function outside the pair-production arc is not unique and possible choices are discussed.

We analyze the experimental data\,\cite{Dally:1980dj, Amendolia:1986ui} or the lattice QCD (LQCD)\,\cite{Ding:2024lfj} results available for the electromagnetic form factor of the charged kaon in the spacelike-region. 
The comparison with other methodologies present in literature and with the BGL $z$-expansion based on the single, total dispersive bound clearly shows that the $z$-expansion including the double dispersive bound provides the most precise extrapolation at large momentum transfer as well as the most stable results with respect to the choice of the outer function outside the pair-production region.

This paper is organized as follows. In Section \ref{sec:subth} we review the basic properties of sub-threshold cuts, showing how double dispersive bounds can be directly imposed in the BGL $z$-expansion of hadronic form factors and, then, we compare our proposal with other expansions present in literature. In Section \ref{sec:bound} we develop a model to describe the impact of above-threshold resonances on the form factor. The model is successfully tested in the case of the $\rho$(770)-meson pole for the electromagnetic form factor of the pion. In Section \ref{sec:kaon} we focus on the electromagnetic form factors of charged and neutral kaons, decomposed in terms of isovector and isoscalar contributions. We implement our resonance model to take into account the $\rho$(770), the $\omega$(782) and the $\phi(1020)$ mesons. In Section \ref{sec:phi_>} we discuss the ambiguity in the definition of the outer function outside the pair-production area and we evaluate its impact on the unitarity bound related to the region outside the pair-production arc. Section \ref{sec:charged_kaon} contains a detailed numerical study of the charged kaon form factor based on spacelike experimental or LQCD data. Different methodologies to impose unitarity are compared and the main result is that the $z$-expansion implementing the double dispersive bound provides the most precise extrapolation at large values of the momentum transfer as well as the most stable results with respect to the choice of the outer function outside the pair-production region. In the same Section we highlight also the impact of the application of the unitarity filter  on the given set of input data (discussed in the companion paper\,\cite{Simula:2025lpc}) and we compare our model-independent findings for the charged kaon radius with the model-dependent estimates made in the last PDG review\,\cite{ParticleDataGroup:2024cfk} and in Ref.\,\cite{Ding:2024lfj}.
Finally, our conclusions are summarized in Section\,\ref{sec:conclusions}, while the Appendices contain further insights on some theoretical aspects discussed in the main text.

\section{Sub-threshold branch-cuts}
\label{sec:subth}

For sake of simplicity, let us consider the case of a single, generic form factor $f(t)$, where $t$ is the four-momentum transfer, describing the transition induced by a current $J$ between two hadrons $H_1$ and $H_2$ with masses $m_1$ and $m_2$, respectively.

As well known (see, e.g., Ref.\,\cite{Caprini:2019osi}), unitarity implies that the form factor $f(t)$ is an analytic function in the complex $t$-plane cut on the real axis from the lowest threshold $t_{th}$  to infinity. When (isolated) real poles due to bound states are present below the threshold $t_{th}$, a product of appropriate Blaschke factors can be introduced to guarantee analyticity (see later on). The discontinuity across the cut is twice the imaginary part of $f(t)$ along the cut and it is dictated by unitarity. The form factor $f(t)$ is real for real values of $t$ below $t_{th}$.
The above analytical properties imply that the form factor $f(t)$ may be represented through a dispersion relation of the form
\be
     \label{eq:DR}
     f(t) = \frac{1}{\pi} \int_{t_{th}}^\infty dt^\prime \frac{\mbox{Im}f(t^\prime)}{t^\prime - t - i \epsilon} ~ , ~
\ee
where the $i \epsilon$ defines the integral for values of $t$ on the branch-cut. Subtractions may be required to make the integral in Eq.\,(\ref{eq:DR}) convergent.

In this work, at variance with Ref.\,\cite{Simula:2025lpc}, we consider the case of a lowest branch-cut starting below the pair-production threshold
\be
     \label{eq:pair}
     t_{th} < t_+ \equiv (m_1 + m_2)^2 ~ . ~
\ee
This may occur when an intermediate on-shell state $\overline{H}_1^\prime H_2^\prime$ can be produced and scatter into an off-shell pair $\overline{H}_1 H_2$. In such a case, though the kinematical region between $t_{th}$ and $t_+$ is not experimentally accessible, the imaginary part of the form factor $f(t)$ is non-vanishing and still dictated by unitarity.
 
 Let us introduce the conformal mapping defined in terms of the variable
\be
    \label{eq:zplus}
    z_+ = z_+(t; t_0) \equiv \frac{\sqrt{t_+ - t} - \sqrt{t_+ - t_0}}{\sqrt{t_+ - t} + \sqrt{t_+ - t_0}} ~, ~ 
\ee
where $t_0 < t_+$ is an auxiliary variable, which fixes the value of $t$ at which $z_+(t_0; t_0) = 0$. For real values of $t$ up to the threshold $t_+$ the conformal variable $z_+$ is real. For $t \leq t_+$ it ranges from $z = -1$ at $t = t_+$ to $z = 1$ for $t \to - \infty$. 
For values of $t$  above the threshold $t_+$ we have to specify the principal value of the square root to distinguish the first and the second Riemann sheets. Our convention is described in Appendix\,\ref{sec:sheets}, so that the first (physical) Riemann sheet corresponds to values of the conformal variable $z_+$ always inside the unit disk (i.e., $|z_+| < 1$), while in the second Riemann sheet the values of $z_+$ lie always outside the unit disk (i.e., $|z_+| >1$). On the branch-cut one has always $|z_+| = 1$.
 
\begin{figure}[htb!]
\begin{center}
\includegraphics[scale=0.3]{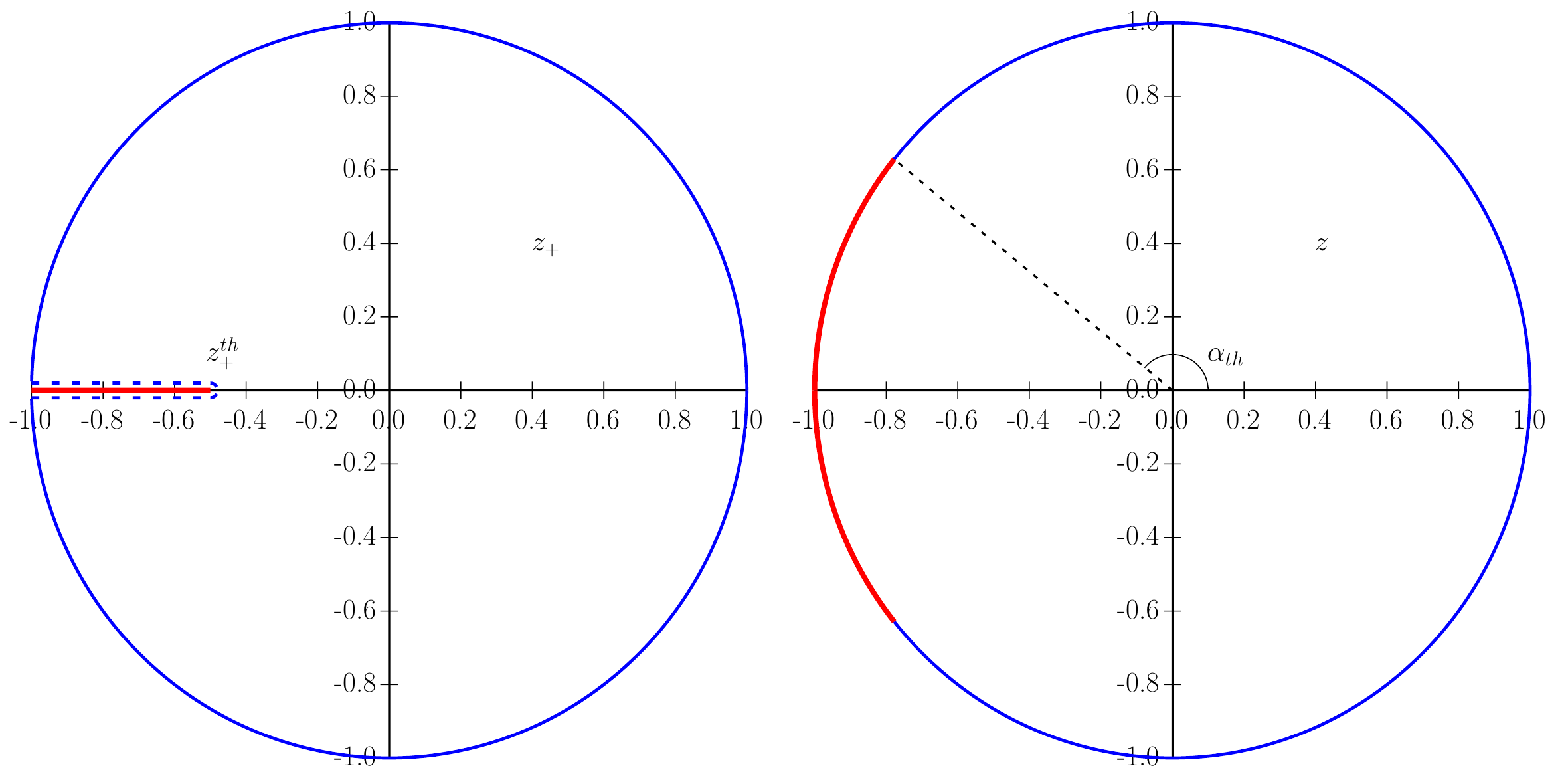}
\end{center}
\vspace{-0.75cm}
\caption{\it \small Analytical domains in terms of the conformal variables $z_+$ and $z$, given respectively by Eqs.\,(\ref{eq:zplus}) and (\ref{eq:z}) with $t_{th} < t_+ = (m_1 + m_2)^2$. The blue lines correspond to the pair-production branch-cut $t \geq t_+$, while the red ones to the extra branch-cut ranging from $t_{th}$ to $t_+$. The lowest branch-point $t_{th}$ is located at $z_+^{th} = z_+(t_{th}; t_0)$ chosen arbitrarily to be equal to $z_+^{th}  = - 0.5$.Correspondingly, the angle $\alpha_{th}$ is given by $\mbox{cos}(\alpha_{th})= 2 (1 + z_+^{th})^2 / (1 - z_+^{th})^2 - 1 \simeq - 0.78$.}
\label{fig:domains}
\end{figure}

 In terms of the conformal variable $z_+$, given by Eq.\,(\ref{eq:zplus}), the form factor $f(z_+)$ is not analytic of the real type\footnote{We remind that an analytic function is said to be of the real type when it satisfies the Schwarz reflection principle $f(z_+^*) = f^*(z_+)$, i.e.~it is real on the real axis.}  inside the unit disk, since it has an imaginary part for real values of $z_+$ between $z_+ = -1$ and $z_+ = z_+^{th} \equiv z_+(t_{th}; t_0)$ with $-1 < z_+^{th}$ (for $t_0 \leq t_{th}$ one has $-1 < z_+^{th} \leq 0$). Such an extra branch-cut is depicted in red in Fig.\,\ref{fig:domains}. 

\subsection{A model for the extra branch-cut}
\label{sec:BGL95}

In Ref.\,\cite{Boyd:1995sq} the issue of the possible presence of branch points inside the unit disc $|z_+| < 1$ was addressed with the aim of providing a reasonable model for estimating the effects of the extra branch-cuts. 
The main assumption made in Ref.\,\cite{Boyd:1995sq} is that extra branch-cuts arise from non-resonant processes with invariant masses below $\sqrt{t_+}$. Without loosing generality one can consider the case of a single extra branch-cut.

Let us indicate with $f^{th}(z_+)$ the contribution of the single extra branch-cut, starting at $t_{th} < t_+$, to the form factor $f(z_+)$.
For real values of $z_+$ above $z_+^{th}$ the function $f^{th}(z_+)$ is real, while for  $z_+$ below $z_+^{th}$ it acquires an imaginary part, which represents (half of) the discontinuity of the form factor $f(z_+)$ along the extra branch-cut.
Therefore, the difference $f(z_+) - f^{th}(z_+)$ is an analytic function of the real type inside the unit disk $|z_+| < 1$, where it has a Taylor expansion with real coefficients.
One gets\,\cite{Caprini80, Caprini:1995wq, Boyd:1995sq}
\be
     \label{eq:BGL95_fz}
     f(z_+) = f^{th}(z_+) + \frac{\sqrt{\widetilde{\chi}}}{\phi_+(z_+) B_+(z_+)} \sum_{k=0}^\infty b_k z_+^k ~ , ~
\ee
where $\phi_+(z_+)$ is a kinematical function, analytic of the real type and without zeros inside the unit disk, while the quantity $B_+(z_+)$ is a product of Blaschke factors related to (isolated) poles corresponding to bound states with masses $M_R$ below the lowest threshold $\sqrt{t_{th}}$, namely
\be
    \label{eq:Blaschke}
    B_+(z_+) = \Pi_R \frac{z_+ - z_+^R}{1 - z_+ \, z_+^R} = \Pi_R \, z_+(t; M_R^2) ~ 
\ee
with $z_+^R = z_+(M_R^2; t_0)$ being real and $|z_+^R| < |z_+^{th}| < 1$. The Blaschke product is unimodular on the unit circle (i.e., $|B_+(z_+)| = 1$ for $|z_+| = 1$).

In Eq.\,(\ref{eq:BGL95_fz}) $\widetilde{\chi}$ is a {\em unitarity bound} on the difference $f(z_+) - f^{th}(z_+)$, so that the coefficients $b_k$ should satisfy the unitarity constraint
\be
     \label{eq:BLG95_unitarity}
     \sum_{k = 0}^\infty b_k^2 \leq 1 ~ . ~
\ee
In Ref.\,\cite{Boyd:1995sq} it was argued that the quantity $\widetilde{\chi}$, appearing in Eq.\,(\ref{eq:BGL95_fz}), can be replaced by its upper limit $\widetilde{\chi}^U$, given by
\be
     \label{eq:BLG95_chi}
     \sqrt{\widetilde{\chi}} \leq \sqrt{\widetilde{\chi}^U} = \sqrt{\chi_+} + \sqrt{\chi_+^{th}} ~
\ee
with 
\bea
     \label{eq:chi_plus}
     \chi_+ &  \equiv & \frac{1}{2\pi i} \oint_{|z_+| = 1} \frac{dz_+}{z_+} |\phi_+(z_+) f(z_+)|^2 ~ , ~ \\[2mm]
     \label{eq:BLG95_chith}
     \chi_+^{th} & = & \frac{1}{2\pi i } \oint_{|z_+| = 1} \frac{dz_+}{z_+} \left| \phi_+(z_+) f^{th}(z_+) \right|^2 ~ . ~
\eea
In this work we choose the kinematical function  $\phi_+(z_+)$ to be the {\em outer function} adopted in the BGL approach of Refs.\,\cite{Boyd:1994tt, Boyd:1995cf, Boyd:1995sq, Boyd:1997kz}. 
In this way it is possible to evaluate an upper limit to the quantity $\chi_+$ from suitable two-point Green functions evaluated in momentum space (see Refs.\,\cite{Meiman63, Okubo:1971jf, Okubo:1971my, Okubo:1971wup, Boyd:1995sq, Bourrely:1980gp}), avoiding the explicit knowledge of the form factor $f(z_+)$ on the unit circle.

According to Ref.\,\cite{Boyd:1995sq} the term $f^{th}(z_+)$ can be approximated by the smooth function 
\bea
     \label{eq:fth}
     f^{th}(z_+) & = & c \left[ \frac{\sqrt{(z_+ - z_+^{th})(1 - z_+ z_+^{th})}}{(1 - z_+)(1 - z_+^{th})} - \frac{1}{2} \frac{1 + z_+}{1 - z_+} \right] \\[2mm]
                      & = & -  \frac{c}{2} \frac{1 + z_+^{th}}{1 - z_+^{th}} \frac{\sqrt{1 - z_+^{th} z_+}  - \sqrt{z_+ - z_+^{th}}}{\sqrt{1 - z_+^{th} z_+} + \sqrt{z_+ - z_+^{th}}} \nonumber \\[2mm] 
                      & = & - \frac{c}{2} \frac{1 + z_+^{th}}{1 - z_+^{th}} \frac{(1 + z_+^{th}) (1 - z_+)}{\left( \sqrt{1 - z_+^{th} z_+} + \sqrt{z_+ - z_+^{th}} \right)^2}  \nonumber ~ , ~     
\eea  
where $c$ is a coefficient incorporating the effects of the coupling of the current $J$ with the external states $\overline{H}_1 H_2$ through non-resonant on-shell intermediate states.
In the r.h.s~of Eq.\,(\ref{eq:fth}) the subtraction of the second term guarantees that $f^{th}(z_+)$ vanishes as $z_+$ goes to $1$ (i.e., when $t \to \pm \infty$). Thus, once a reasonable estimate of the coefficient $c$ has been established, one can use the model\,(\ref{eq:fth}) to evaluate both the integral\,(\ref{eq:BLG95_chith}) and the difference $f(z) - f^{th}(z)$ for a given set of known input data $\{ f_i \}$, where $f_i \equiv f[z_+(t_i; t_0)]$ with $i =1, 2, ... N$. On such differences one can apply the expansion appearing in the r.h.s.~of Eq.\,(\ref{eq:BGL95_fz}) with $\widetilde{\chi}$ replaced by $\widetilde{\chi}^U$ given in Eq.\,(\ref{eq:BLG95_chi}).

For sufficiently low values of $c$ the impact of $f^{th}(z_+)$ may be safely neglected. Such a situation may occur in specific cases, like, e.g., the one discussed in Ref.\,\cite{Boyd:1995sq} concerning the impact of intermediate $B_c^* + n \pi$ (with $n$ a positive integer) states for $B \to D$ and $B \to D^*$ semileptonic decays.
However, generally speaking, the model dependence of Eq.\,(\ref{eq:fth}) may deteriorate the accuracy of the theoretical determinations of the input data $\{ f_i \}$, coming, e.g., from lattice QCD (LQCD) simulations. Moreover, the model of Ref.\,\cite{Boyd:1995sq} is based on the assumption that the extra cut arises only from non-resonant processes.

We mention that, in the case of the $\omega - \pi$ transition form factor, the effects of sub-threshold cuts on the unitarity constraints in terms of the conformal variable\,(\ref{eq:zplus}) have been extensively studied in Refs.\,\cite{Ananthanarayan:2014pta, Caprini:2015wja}.

\subsection{The BGL expansion in terms of the conformal variable $z$}
\label{sec:z_th}

Instead of using the conformal variable $z_+$ it is convenient to introduce the conformal variable $z$ corresponding to the lowest branch-cut $t_{th}$, i.e. 
\be
     \label{eq:z}
     z \equiv z_{th}(t; t_0) = \frac{\sqrt{t_{th} - t} - \sqrt{t_{th} - t_0}}{\sqrt{t_{th} - t} + \sqrt{t_{th} - t_0}} ~ 
\ee
with $t_0 < t_{th} < t_+$.
In this way the unit circle $|z| = 1$ corresponds to $t \geq t_{th}$ and includes all the branch-cuts (see the right panel of Fig.\,\ref{fig:domains}). 

In terms of the conformal variable $z$ the BGL expansion of the form factor $f(z)$ reads as
\be
    \label{eq:BGL}
    f(z) = \frac{\sqrt{\chi^U}}{\phi(z) B(z)} \sum_{k=0}^\infty a_k z^k ~ , ~ 
\ee
where $\phi(z)$ is a kinematical function to be specified, but analytic of the real type and without zeros inside the unit disk, while $\chi^U$ is an {\em upper limit} to the quantity $\chi[f]$, associated to the form factor $f(z)$ and defined as 
\be
    \label{eq:chi_f}
    \chi[f] \equiv \frac{1}{2\pi i} \oint_{|z| = 1} \frac{dz}{z} |\phi(z) f(z)|^2 = \frac{1}{2\pi} \int_{-\pi}^\pi d\alpha |\phi(e^{i\alpha}) f(e^{i\alpha})|^2 > 0 ~ , ~
\ee
namely
\be
    \label{eq:upper}
   \chi[f] \leq  \chi^U ~ . ~ \\[2mm]
\ee
In Eq.\,(\ref{eq:BGL}) the quantity $B(z)$ is the Blaschke factor removing the dynamical poles of the bound-states with invariant masses below $\sqrt{t_{th}}$, and it is an analytic function of the real type inside the unit $z$-disc. In terms of the conformal variable $z$ one has\footnote{The Blaschke factors $B(z)$ and $B_+(z_+)$ share the same zeros (located at $t = M_R^2$), but they are different, namely $B(z) \neq B_+(z_+)$. The ratio $B_+(z_+) / B(z)$ is an analytic function of the real type inside the unit $z$-disk and it is unimodular on the pair-production arc corresponding to both $|z| =1$ and  $|z_+| =1$ (see Fig.\,\ref{fig:domains}).}
\be
    \label{eq:Blaschke_th}
    B(z) = \Pi_R \frac{z - z^R}{1 - z \, z^R} = \Pi_R \, z_{th}(t; M_R^2) ~ , ~
\ee
where $z^R = z_{th}(M_R^2; t_0)$ is real and $|z^R| < 1$. The Blaschke product is unimodular on the unit $z$-circle (i.e., $|B(z)| = 1$ for $|z| = 1$) and, therefore, it does not contribute to Eq.\,(\ref{eq:chi_f}).

Since the monomials $z^k$ are orthonormal when integrated on the unit $z$-circle, the inequality\,(\ref{eq:upper}) implies that the BGL coefficients $a_k$ must satisfy the unitarity constraint
\be
    \label{eq:unitarity}
    \sum_{k=0}^\infty a_k^2 \leq 1 ~ . ~
\ee
The important point is that in the case of a lowest branch-cut below the pair-production threshold ($t_{th} < t_+$) the bound $\chi^U$ on the form factor $f(z)$ is related to the the integral\,(\ref{eq:chi_f}) over the unit $z$-circle, i.e.\,along the full branch-cut $t \geq t_{th}$. This feature can be easily derived also within the Dispersive Matrix (DM) method of Ref.\,\cite{DiCarlo:2021dzg}, originally proposed in Refs.\,\cite{Bourrely:1980gp, Lellouch:1995yv} and based on the evaluation of a determinant of suitable inner products.

Putting $z = e^{i \alpha}$ along the unit circle, $\chi[f]$ can be split into the sum of two contributions
\be
    \label{eq:decomposition}
    \chi[f] = \chi_{\rm pair}[f] + \chi_{\rm extra}[f]
\ee
where the angle $\alpha_{th}$ corresponds to $e^{i\alpha_{th}} = z_{th}(t_+; t_0)$, namely
\be
     \label{eq:alpha_th}
     \alpha_{th} = \mbox{Arccos}\left[ 1 - 2 \, \frac{t_{th} - t_0}{t_+ - t_0} \right] ~ . ~
\ee
In Eq.\,(\ref{eq:decomposition}) the first term, $\chi_{\rm pair}[f] $, corresponds to the contribution of the pair-production arc $-\alpha_{th} \leq \alpha \leq \alpha_{th}$, viz.
\be
     \label{eq:chi_pair}
     \chi_{\rm pair}[f]  = \frac{1}{2\pi}  \int_{-\alpha_{th}}^{\alpha_{th}} d\alpha  \left| \phi(e^{i\alpha}) f(e^{i\alpha}) \right|^2 ~ , ~ 
\ee
while the second term, $\chi_{\rm extra}[f]$, arises from the integration regions $\alpha_{th} \leq |\alpha| \leq \pi$ outside the pair-production arc, namely
\be
      \label{eq:chi_extra}
      \chi_{\rm extra}[f] = \frac{1}{2\pi}  \left\{ \int_{-\pi}^{-\alpha_{th}} +  \int_{\alpha_{th}}^\pi \right\} d\alpha  \left| \phi(e^{i\alpha}) f(e^{i\alpha}) \right|^2 ~ . ~
\ee

Generally speaking, both terms, $\chi_{\rm pair}[f]$ and $\chi_{\rm extra}[f]$, depend on the form factor $f(z)$ and on the kinematical function $\phi(z)$. 
However, a convenient choice for the latter one is
\be
    \label{eq:phi_th}
    \phi(z) = \sqrt{\frac{dz_+}{dz} } \phi_+(z_+) ~ , ~
\ee
where (we remind) the kinematical function $\phi_+(z_+)$ is the {\em outer function} adopted in the BGL approach.
On the pair-production arc $-\alpha_{th} \leq \alpha \leq \alpha_{th}$, where both $z = e^{i \alpha}$ and  $z_+ = e^{i\alpha_+}$ are unimodular, one has
\be
    \frac{d\alpha_+}{d\alpha} \to \frac{z}{z_+} \frac{dz_+}{dz} = \frac{1+z}{1-z} \frac{1-z_+}{1+z_+} \to \frac{\mbox{tg}(\alpha_+ / 2)}{\mbox{tg}(\alpha / 2)} \geq 0 ~ , ~
\ee
so that $(d\alpha_+ / d\alpha)$ is real and positive. Then, it follows that
\be
     \label{eq:phi_th_arc}
     |\phi(e^{i\alpha})|^2 = \frac{d\alpha_+}{d\alpha} |\phi_+(e^{i\alpha_+})|^2 ~ 
\ee
and, consequently, one finally gets
\bea
      \label{eq:chi+}
      \chi_{\rm pair}[f] & = &  \frac{1}{2\pi} \int_{-\alpha_{th}}^{\alpha_{th}} d\alpha \left| \phi(e^{i\alpha}) f(e^{i\alpha}) \right|^2 \nonumber \\[2mm]
                                & = & \frac{1}{2\pi} \int_{-\pi}^\pi d\alpha_+ \left| \phi_+(e^{i\alpha_+}) f(e^{i\alpha_+}) \right|^2  = \chi_+[f] ~ . ~
\eea
Thus, adopting the kinematical function\,(\ref{eq:phi_th}) one has $\chi[f] = \chi_+[f] + \chi_{\rm extra}[f]$, which implies $\chi[f] > \chi_+[f]$, since in general $ \chi_{\rm extra}[f] > 0$.
Thus, it is not guaranteed that an upper limit $\chi_+^U$, known for the integral $\chi_+[f]$ related to the pair-production arc only, may act also as an upper limit to the global $\chi[f]$.
This is not surprising. Consider the bound $\chi_+^U$ obtained from the susceptibility of a suitable vacuum polarization function. The latter one takes into account only intermediate on-shell states $\overline{H}_1 H_2$, while the additional contribution $\chi_{\rm extra}[f]$ arises from intermediate off-shell states $\overline{H}_1 H_2$, generated by rescattering processes from intermediate on-shell states $\overline{H}_1^\prime H_2^\prime$ having squared mass higher than $t_{th}$ but lower than the pair production threshold $t_+$.

We point out that outside the pair-production arc, i.e.~for $\alpha_{th} < |\alpha| \leq \pi$, the choice\,(\ref{eq:phi_th}) is not unique. Indeed, one can multiply the r.h.s.~of Eq.\,(\ref{eq:phi_th}) by any function, which becomes unimodular on the pair-production arc. While such a factor does not modify the relation $\chi_{\rm pair}[f]= \chi_+[f]$, it may affect the contribution $\chi_{\rm extra}[f]$. This point will be discussed later in Section\,\ref{sec:phi_>}.

To summarize, the main result of this Section is that in the case of a lowest branch-cut lower than the pair-production threshold ($t_{th} < t_+$) the dispersive bound on the form factor $f(z)$ is related to the integral of $|\phi(z) f(z)|$ over the full unit circle $|z| = 1$, i.e.~over all the branch-cuts. The latter ones are not limited to the arc $-\alpha_{th} \leq \alpha \leq \alpha_{th}$ corresponding only to the pair-production branch-cut, but an additional contribution, which cannot be related to any susceptibility of the vacuum polarization function, should be considered (see~Section\,\ref{sec:BGL95} and, more recently, Ref.\,\cite{Gopal:2024mgb}). While the bound $\chi_+^U$ to the quantity $\chi_{\rm pair}[f] = \chi_+[f]$ can be obtained directly from the susceptibility of a suitable vacuum polarization function, the estimate of an upper limit $\chi_{\rm extra}^U$ to the quantity $\chi_{\rm extra}[f]$ will be discussed later in Section\,\ref{sec:bound}.
Thus, the bound $\chi^U$ to the global quantity $\chi[f]$ can be written as
\be
     \label{eq:bound_total}
     \chi^U = \chi_+^U + \chi_{\rm extra}^U ~ . ~
\ee

\subsection{Semi-inclusive dispersive bounds}
\label{sec:double_bound}

The decomposition\,(\ref{eq:decomposition}) of the global quantity $\chi[f]$ contains more information than the global quantity itself, which by definition is an inclusive quantity. 
The two contributions $\chi_{\rm pair}[f]$ and $\chi_{\rm extra}[f]$ can be viewed as {\em semi-inclusive} quantities, related to different partial regions of the unit circle in the conformal variable $z$.
Instead of imposing only the single, global filter $\chi[f] \leq \chi^U =  \chi_+^U + \chi_{\rm extra}^U$, we can apply simultaneously the double filter
\bea
     \label{eq:filter_pair}
     \chi_{\rm pair}[f] & \leq & \chi_+^U ~ , ~ \\[2mm]
     \label{eq:filter_extra}
     \chi_{\rm extra}[f] & \leq & \chi_{\rm extra}^U ~ . ~
\eea
This is a case of the multiple dispersive bounds discussed in Ref.\,\cite{Simula:2025lpc}. 
We expect that the simultaneous application of a double dispersive bound\footnote{In what follows we limit ourselves to the case of a double dispersive bound, since we deal with the simplest case of two branch-cuts. The generalization to the case of multiple dispersive bounds related to multiple branch-cuts is straightforward.} can be more constraining than the one related to the single, total dispersive bound $\chi^U$. If the two dispersive bounds\,(\ref{eq:filter_pair})-(\ref{eq:filter_extra}) are simultaneously fulfilled, then the global filter $\chi[f] \leq \chi^U$ is automatically satisfied.

As shown in Ref.\,\cite{Simula:2025lpc}, it is straightforward to apply multiple filters to the BGL expansion\,(\ref{eq:BGL}).
In the case of the double filter\,(\ref{eq:filter_pair})-(\ref{eq:filter_extra}) the coefficients $a_k$ of Eq.\,(\ref{eq:BGL}) must satisfy simultaneously two constraints
\be
     \label{eq:unitarity_pair}
     \sum_{k, k^\prime = 0}^\infty a_{k^\prime} U_{k^\prime k}(\alpha_{th}) a_k \leq \frac{\chi_+^U}{\chi^U} ~
\ee
and
\be
     \label{eq:unitarity_extra}
     \sum_{k, k^\prime = 0}^\infty a_{k^\prime} \left[ \delta_{k^\prime k} - U_{k^\prime k}(\alpha_{th}) \right] a_k \leq \frac{\chi_{\rm extra}^U}{\chi^U} ~ , ~
\ee
where the matrix $U$ is given by
\be
     \label{eq:U_Szego}
     U_{k^\prime k}(\alpha_{th}) \equiv \frac{1}{2\pi} \int_{-\alpha_{th}}^{\alpha_{th}} d\alpha \, e^{i (k - k^\prime) \alpha} = 
         \frac{1}{\pi} \frac{\mbox{sin}(k - k^\prime) \alpha_{th}}{k - k^\prime} ~ . ~
\ee
The two unitarity constraints\,(\ref{eq:unitarity_pair})-(\ref{eq:unitarity_extra}) automatically imply that the coefficients $a_k$ satisfy the total unitarity constraint\,(\ref{eq:unitarity}).

When a set of input data $\{ f_i \}$, where $f_i \equiv f[z_i = z(t_i; t_0)]$ with $i =1, 2, ... N$, is known, the usual procedure is to truncate the BGL expansion\,(\ref{eq:BGL}) at some order $M$, namely
\be
    \label{eq:BGL_truncated}
    f^{(M)}(z) = \frac{\sqrt{\chi^U}}{\phi(z) B(z)} \sum_{k=0}^M a_k^{(M)} z^k ~ 
\ee
where the coefficients $a_k^{(M)}$ satisfy the truncated double bound
\bea
     \label{eq:unitarity_pair_truncated}
     \sum_{k, k^\prime = 0}^M a_{k^\prime} ^{(M)}U_{k^\prime k}(\alpha_{th}) a_k^{(M)} & \leq & \frac{\chi_+^U}{\chi^U} ~ , ~ \\[2mm]
     \label{eq:unitarity_extra_truncated}
     \sum_{k, k^\prime = 0}^M a_{k^\prime}^{(M)} \left[ \delta_{k^\prime k} - U_{k^\prime k}(\alpha_{th}) \right] a_k^{(M)} & \leq & \frac{\chi_{\rm extra}^U}{\chi^U} ~ , ~
\eea
which imply the total bound
\be
    \label{eq:unitarity_truncated}
    \sum_{k=0}^M \left[ a_k^{(M)} \right]^2 \leq 1 ~ . ~
\ee
The coefficients $a_k^{(M)} $ are typically obtained from a $\chi^2$-minimization procedure applied to the set of input data $\{ f_i \}$.
The two semi-inclusive dispersive bounds\,(\ref{eq:unitarity_pair_truncated})-(\ref{eq:unitarity_extra_truncated}) can be easily imposed by adding appropriate penalty terms to the $\chi^2$-variable.

The use of a finite order $M$ in the truncated expansion\,(\ref{eq:BGL_truncated}) raises the issue of truncation errors, which were firstly discussed in Refs.\,\cite{Boyd:1995sq, Boyd:1997kz}.
There, an upper limit to the truncation error was obtained, assuming that the coefficients $a_k^{(M)}$ coincide with the first $(M+1)$ coefficients $a_k$ of the untruncated BGL expansion\,(\ref{eq:BGL}). See for details Appendix\,\ref{sec:truncation}.

Numerical applications of the double filter\,(\ref{eq:filter_pair})-(\ref{eq:filter_extra}) will be presented in the case of the charged kaon electromagnetic form factor in Section\,\ref{sec:charged_kaon}.

\subsection{Differences with other expansions}
\label{sec:Szego}

Equations\,(\ref{eq:BGL})-(\ref{eq:unitarity}), properly derived in the previous subsection, are however at variance with those proposed in Refs.\,\cite{Gubernari:2020eft, Gubernari:2022hxn, Blake:2022vfl} and adopted also in Refs.\,\cite{Flynn:2023qmi, Flynn:2023nhi, Harrison:2025yan}.
Indeed, in Ref.\,\cite{Blake:2022vfl} the BGL $z$-expansion of the form factor $f(z)$ is written in the form
\be
     \label{eq:BGL_Szego}
     f(z) = \frac{\sqrt{\chi_+^U}}{\phi(z) B(z)} \sum_{k = 0}^\infty c_k \, p_k(z; \alpha_{th}) ~ , ~
\ee
where $\chi_+^U$ is the dispersive bound to the quantity $\chi_{\rm pair}[f] = \chi_+[f]$, $p_k(z; \alpha_{th})$ is an orthonormal polynomial of degree $k$ on the arc $-\alpha_{th} \leq \alpha \leq \alpha_{th}$ (i.e.,~a (normalized) Szeg\H{o} polynomial\,\cite{simon04}), and the (real) coefficients $c_k$ satisfy the unitarity constraint
\be
     \label{eq:unitarity_Szego}
     \sum_{k = 0}^\infty c_k^2 \leq 1 ~ . ~
\ee
As already shown in Appendix B of Ref.\,\cite{Flynn:2023qmi} in terms of the monomials $z^k$ the $z$-expansion\,(\ref{eq:BGL_Szego}) can be written as (see also Appendix\,\ref{sec:orthonormal})
\be
     \label{eq:BGL_VD}
     f(z) = \frac{\sqrt{\chi_+^U}}{\phi(z) B(z)} \sum_{k = 0}^\infty b_k z^k ~ , ~
\ee
where the (real) coefficients $b_k$ satisfy an off-diagonal unitarity constraint limited only to the pair-production arc, namely 
\be
     \label{eq:unitarity_VD}
     \sum_{k, k^\prime = 0}^\infty b_{k^\prime} U_{k^\prime k}(\alpha_{th}) b_k \leq 1 ~ . ~
\ee

In Section\,\ref{sec:z_th}  we have shown that the proper global dispersive bound is given by Eq.\,(\ref{eq:upper}), which is related to the full branch-cut starting from $t_{th}$. The form factor $f(t)$ has an imaginary part not only for $t \geq t_+$, but also for $t_{th} \leq t \leq t_+$, and its structure for both branch-cuts is dictated by unitarity (see later Section\,\ref{sec:kaon}). Consequently, the global dispersive bound $\chi^U$ should include the effects of both branch-cuts, and it is not legitimate to split Eq.\,(\ref{eq:bound_total}) into two contributions and consider only one of them (in particular, the contribution of the pair-production arc). 
Furthermore, in Appendix\,\ref{sec:eigenvalues} it is shown that the matrix $U$ acts basically as a projector. In the limit of large values of its dimension it has only two possible eigenvalues equal to either $0$ or $1$. Thus, the expansion\,(\ref{eq:BGL_VD}) can be split into the sum of two contributions related respectively to the (orthogonal) subspaces of the eigenvectors of $U$ corresponding to eigenvalues equal to $0$ or $1$. The constraint\,(\ref{eq:unitarity_VD}) bounds only the contribution living in the subspace of the eigenvectors of $U$ corresponding to eigenvalues equal to $1$, while the other contribution is left unbounded.

\section{The contribution $\chi_{\rm extra}[f]$ and its dispersive bound $\chi_{\rm extra}^U$}
\label{sec:bound}

The main result of the previous Section is that the BGL expansion\,(\ref{eq:BGL}) can be straightforwardly applied to the case of a lowest threshold $t_{th}$ below the pair production one $t_+$ by using the conformal variable $z$ defined in terms of the lowest branch-point $t_{th}$, i.e.~Eq.(\ref{eq:z}). The dispersive bound is given by Eq.\,(\ref{eq:bound_total}) and it receives contributions both from the pair-production branch-cut, $\chi_{\rm pair}^U$, and from the extra branch-cut, $\chi_{\rm extra}^U$. With a suitable choice of the outer function, i.e.\,Eq.\,(\ref{eq:phi_th}), the former contribution can be expressed through the susceptibility calculable in terms of derivatives of an appropriate vacuum polarization function.

In this Section we want to address the issue of estimating the additional contribution $\chi_{\rm extra}^U$ arising from resonant processes with invariant masses above the threshold $\sqrt{t_{th}}$, but below the pair-production threshold $\sqrt{t_+}$. 

It is however instructive to start with the different case in which no extra branch-cut is present, i.e.~$t_{th} = t_+$, and to try to describe (at least qualitatively) the impact on the structure of the form factor $f(t)$ induced by the presence of a resonance $R$ with a mass $M_R$ above the pair-production threshold $\sqrt{t_+}$.

\subsection{Effects of above-threshold poles on the form factor $f(t)$}
\label{sec:poles}

It is well known\,\cite{Barton65} that a pole in the scattering amplitude or in the S-matrix in the second Riemann sheet may induce a zero of the S-matrix at the corresponding point in the first Riemann sheet (the physical sheet). This property turns out to be quite useful for searching the location of resonances in scattering processes (see for instance Ref.\,\cite{Caprini:2005zr}). 

Let us consider besides the form factor $f(t)$, describing the transition induced by a current $J$ between two hadrons $H_1$ and $H_2$ with masses $m_1$ and $m_2$, also the elastic scattering amplitude $h(t)$ (with the appropriate quantum numbers) between the two hadrons (see Refs.\,\cite{Grinstein:2015wqa, Caprini:2019osi, Caprini:2017ins}).
In the elastic region $t_+ \leq t \leq t_{in}$, where $t_{in}$ is the first inelastic threshold, unitarity implies\,\cite{Barton65} that $\mbox{Im}f(t) = \rho(t) h(t) f^*(t)$, where $\rho(t)$ is a simple phase space factor. A well-known consequence of this relation is the Fermi-Watson theorem\,\cite{Fermi:1955xrk, Watson:1954uc}, stating that in the elastic region the phases of the form factor $f(t)$ and of the scattering amplitude $h(t)$ are the same. The analytic continuation of the form factor $f(t)$ to the second Riemann sheet can be done by  matching the lower edge of the branch-cut in the first sheet with the upper edge of the branch-cut in the second sheet, namely $f^{II}(t + i \epsilon) = f^I(t - i \epsilon)$. Thus, indicating for sake of simplicity all quantities without a superscript as defined in the first Riemann sheet, unitarity implies that $f^{II}(t) = f(t) / S(t)$, where $S(t) = 1 + 2 i \rho(t) h(t)$ is the S-matrix in the first sheet. Thus, assuming that $f(t)$ does not vanish at the zeros of $S(t)$, the form factor $f(t)$  has poles in the second Riemann sheet in the same location of the zeros of $S(t)$ (or, equivalently, at the same location of the poles of the scattering amplitude $h(t)$).
 
Let us now translate the above properties in terms of the conformal variable $z_+$. The conjugate poles due to the resonance $R$ are located at $t = t_R^\pm = (M_R \pm i \Gamma_R / 2)^2$, where $\Gamma_R$ is the resonance width and $M_R^2 > t_+$. Thus, according to Eqs.\,(\ref{eq:physical_sheet})-(\ref{eq:unphysical_sheet}) the above poles occur in the second Riemann sheet (i.e.~for $|z_+| > 1$) at $z_+ = (z_R^\pm)^{II} = 1 / z_R^\pm$ with 
\bea
    \label{eq:poles}
    z_R^\pm & = &  |z_R| \, e^{\pm i \alpha_R} ~ , ~ \\[2mm]
    \label{eq:zR_abs}
    |z_R| & = & \sqrt{z_R^- z_R^+} = \sqrt{ \frac{r + t_+ - t_0 - \sqrt{2 (t_+ - t_0) (r + x)}}{r + t_+ - t_0 + \sqrt{2 (t_+ - t_0) (r + x)}} }  < 1 ~ , ~ \\[2mm]
    \label{eq:alphaR}
    \mbox{cos}\alpha_R & = & \frac{r - t_+ + t_0}{\sqrt{r^2 + (t_+ - t_0)^2 - 2 (t_+ - t_0) x}} ~ , ~
\eea
where
\bea
     \label{eq:r&x}
    r & = & \sqrt{x^2 + M_R^2 \Gamma_R^2} ~ , ~ \nonumber \\[2mm]
    x & = & t_+ + \frac{\Gamma_R^2}{4} - M_R^2 ~ . ~
\eea
Thus, one might ask whether (and to what extent) the impact of the resonance $R$ on the form factor $f(z)$ can be approximated by the following simple Ansatz
\bea
    \label{eq:fR}
    f_R(z_+) & = & f_R(0) \frac{(1 - z_+)^2}{(1 - z_R^- z_+)(1 - z_R^+ z_+)} \nonumber \\[2mm]
                   & = & f_R(0) \frac{(1 - z_+)^2}{1 - 2 |z_R| \mbox{cos}\alpha_R  \, z_+ + |z_R|^2 z_+^2}~ , ~
\eea
where $f_R(0) \equiv f_R(z_+ = 0)$ represents the value of the form factor at $z_+ = 0$ (i.e.~at $t = t_0$) and the term $(1 - z_+)^2$ in the numerator has been inserted in order to guarantee that $f_R(z) \propto 1/(-t)$ at large values of $|t|$. We stress that Eq.\,(\ref{eq:fR}) is similar to the ansatz discussed in Ref.\,\cite{Caprini:2017ins}: the resonance $R$ produces conjugate poles occurring only in the second Riemann sheet ($|z_+| > 1$) at $z_+ = 1/z_R^\pm$, and the function $f_R(z_+)$ is analytic of the real type inside the unit disk $|z_+| < 1$. Note that in this way only the knowledge of the mass $M_R$ and width $\Gamma_R$ of the resonance is required.

\subsection{The electromagnetic form factor of the charged pion}
\label{sec:pion}

We now check the quality of the approximation\,(\ref{eq:fR}) by considering the case of the electromagnetic form factor for charged pions, $F_\pi^{(em)}(t)$, which is known to receive a dominant contribution from the $\rho(770)$-meson resonance for $|t| \lesssim 1$ GeV$^2$. In this work we make use of the results of Ref.\,\cite{Colangelo:2018mtw}, based on a detailed analysis of a large set of both space-like ($t < 0$) and time-like experimental data up to $t = 1$ GeV$^2$, carried out within a dispersive approach satisfying unitarity and analyticity.

Since the pion form factor $F_\pi^{(em)}(t)$ is related to the matrix element of a vector current, angular momentum conservation imposes a specific constraint at the threshold $t_+ = 4 M_\pi^2$, namely, in terms of the conformal variable $z_+$ one must have $d F_\pi^{(em)}(z_+) / dz_+ \Big|_{z+ = -1} = 0$ (see Ref.\,\cite{Bourrely:2008za}). Therefore, we modify Eq.\,(\ref{eq:fR}) with $R = \rho$ by including as extra factor $(1 + b_1 z_+)$, with $b_1$ chosen to guarantee that $d f_\rho(z_+) / dz_+ \Big|_{z_+ = -1} = 0$.  Moreover, charge conservation implies that $F_\pi^{(em)}(t = 0) = 1$, which can be easily implemented by choosing the auxiliary variable $t_0$ in the definition of the conformal variable $z_+$ equal to $t_0 = 0$ and by putting $f_\rho(z_+ = 0) = 1$. One has (see also the similar ansatz adopted in Ref.\,\cite{Kirk:2024oyl})
\be
    \label{eq:frho}
    f_\rho(z_+) =  \frac{(1 - z_+)^2}{1 - 2 |z_\rho| \mbox{cos}\alpha_\rho  \, z_+ + |z_\rho|^2 z_+^2} (1 + b_1 z_+) ~ 
\ee
with\footnote{Explicitly one has $b_1 \simeq 0.0398$ (in the general case $0 < b_1 < 1$). Thus, the function\,(\ref{eq:frho}) has a zero only at $z_+ = 1$ on the unit circle.}
\be
     b_1 = \frac{1}{2} \frac{1 - |z_\rho|^2}{1 + |z_\rho| \mbox{cos}\alpha_\rho} ~ . ~
\ee

\begin{figure}[htb!]
\begin{center}
\includegraphics[scale=0.5]{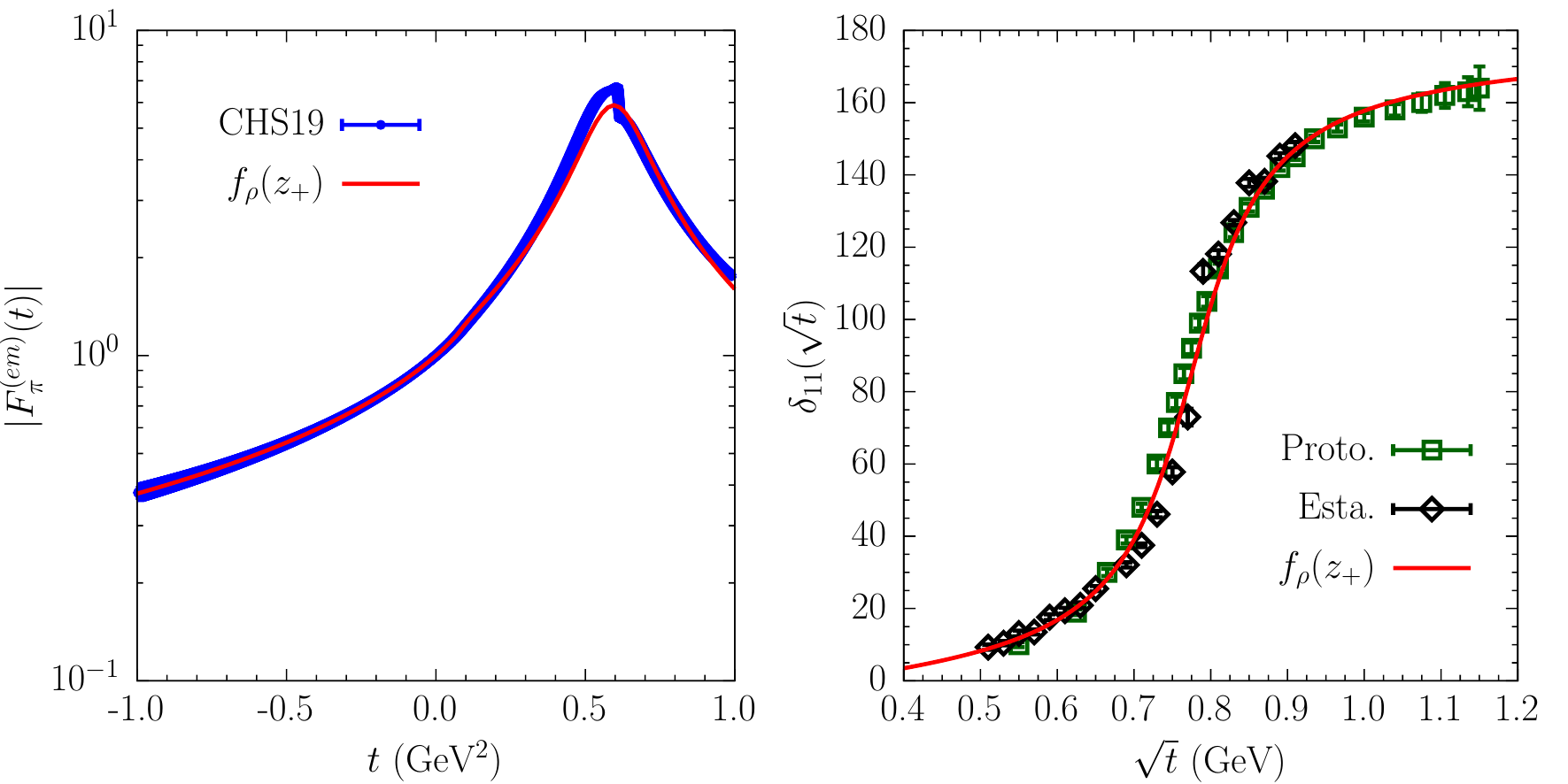}
\end{center}
\vspace{-0.75cm}
\caption{\it \small Left panel: the absolute value of the electromagnetic pion form factor $F_\pi^{(em)}(t)$, determined by the dispersive analysis of experimental data made in Ref.\,\cite{Colangelo:2018mtw}, labelled as CHS19, compared with the corresponding predictions of the approximation\,(\ref{eq:frho}), obtained using $M_\rho = 775$ MeV and $\Gamma_\rho = 147$ MeV from PDG\,\cite{ParticleDataGroup:2024cfk}. Right panel: the experimental data on $\pi-\pi$ scattering phase shifts from Ref.\,\cite{Protopopescu:1973sh} (green squares) and Ref.\,\cite{Estabrooks:1974vu} (black diamonds) compared with the phase of the form factor\,(\ref{eq:frho}).}
\label{fig:pion}
\end{figure}

In Fig.\,\ref{fig:pion} we show the predictions of Eq.\,(\ref{eq:frho}) for $t \leq 1$ GeV$^2$, obtained adopting the values of the $\rho$-meson mass and width from the latest PDG review\,\cite{ParticleDataGroup:2024cfk}, namely $M_\rho = 775$ MeV and $\Gamma_\rho = 147$ MeV. The locations of the conjugate poles\,(\ref{eq:poles}) correspond to $|z_\rho| = 0.930$ and $\alpha_\rho / \pi = 0.232$. Such predictions are compared both with the dispersive results of Ref.\,\cite{Colangelo:2018mtw} in the case of the absolute value of the pion form factor and directly with the experimental data on $\pi-\pi$ scattering phase shifts from Refs.\,\cite{Protopopescu:1973sh, Estabrooks:1974vu}.
 
It can be seen that the approximation\,(\ref{eq:frho}) works quite well almost everywhere for $|t| \leq 1$ GeV$^2$. As a matter of fact, the differences with respect to the results of Ref.\,\cite{Colangelo:2018mtw} are within the $\simeq 5 \%$ level except in the region around the $\rho$-meson peak, where they may reach the $\simeq 15 \%$ level. This is due to the fact that the analysis of Ref.\,\cite{Colangelo:2018mtw} includes the effects of the $\rho - \omega$ mixing, which is the most important isospin-breaking effect, enhanced around the peak by the small mass difference between the $\rho$ and $\omega$ resonances. 

Finally, assuming $F_\pi^{(em)}(z_+) = f_\rho(z_+)$ and adopting the kinematical function $\phi_+(z_+)$ corresponding to the pion channel, i.e.~$\phi_+(z_+) = (1536 \, \pi)^{-1/2}$ $(1 + z_+)^2$ $\sqrt{1 - z_+}$ (see, e.g., Ref.\,\cite{Simula:2023ujs}), we can evaluate directly the quantity $\chi_+[f]$ given by Eq.\,(\ref{eq:chi_plus}). One gets $\chi_+[F_\pi^{(em)} = f_\rho] \simeq 0.889 \chi_+^U$, where $\chi_+^U = 0.00574$ is the dispersive bound obtained in Ref.\,\cite{Simula:2023ujs} from lattice-based results of the light-quark vector susceptibility.
In other words, almost $90 \%$ of the pion susceptibility $\chi_+^U$ comes from the $\rho$-meson resonance.

\subsection{A modified $z$-expansion to include effects from above-threshold poles}
\label{sec:BCL_new}

The above results indicate clearly that Eq.\,(\ref{eq:fR}) is able to intercept the basic effects of an above-threshold resonance. 
In terms of the variable $t$ one has
\be
     \label{eq:fR_t}
     f_R(t) = f_R(t_0) \frac{1 + A}{1 - \frac{t - t_0}{r + t_+ - t_0} + A \sqrt{\frac{t_+ - t}{t_+ - t_0}}} ~ , ~
\ee
where the constant $A$ is given by
\be
     \label{eq:A}
     A = \frac{\sqrt{2 (t_+ - t_0)(r + x)}}{r + t_+ - t_0} 
\ee
with $r$ and $x$ given by Eq.\,(\ref{eq:r&x}).
Note that Eq.\,(\ref{eq:fR_t}) differs from the Breit-Wigner parameterization for a resonant shape\footnote{It is straightforward to check that: ~ i) for real values of $t$ Eq.\,(\ref{eq:fR_t}) has an imaginary part only for $t > t_+$, and ~ ii) taking the principal value of $\sqrt{t_+ - t}$ (see Eq.\,(\ref{eq:physical_sheet})), Eq.\,(\ref{eq:fR_t}) has no pole in the complex $t$-plane.}, while it is similar to the ansatz proposed in Ref.\,\cite{Ananthanarayan:2011uc} to describe the effects of the $D^*$-meson pole for the $D \to \pi$ transition form factors. In the zero-width limit, $\Gamma_R \to 0$, one has $(r + x) \to 0$ and $(r + t_+) \to M_R^2$, so that the function\,(\ref{eq:fR_t}) (and correspondingly also Eq.\,(\ref{eq:fR})) reduces to the simple pole Ansatz
\be
    \label{eq:pole}
    f_R(t) ~ _{\overrightarrow{\Gamma_R \to 0}} ~ f_R(t_0) \frac{M_R^2 - t_0}{M_R^2 - t} ~ . ~
\ee
Since $f_R(z_+)$ is an analytic function of the real type inside the unit disk $|z_+| < 1$, an interesting development, considered already in Ref.\,\cite{Caprini:2017ins}, is to modify the standard BGL $z$-expansion\,(\ref{eq:BGL}) or the Bourrely-Caprini-Lellouch (BCL) $z$-expansion\,\cite{Bourrely:2008za} by taking into account an above-threshold resonance through the function\,(\ref{eq:fR}). 

In the BGL case\footnote{A modified BCL-like $z$-expansion can be obtained simply by omitting the function $\sqrt{\chi_+^U} / \phi_+(z_+)$ in Eq.\,(\ref{eq:BGL_new}) and by multiplying the integrand of Eq.\,(\ref{eq:Soff}) by $|\phi_+(z_+)|^2 / \chi_+^U$.} one has
\be
     \label{eq:BGL_new}
     f(z_+) = f_R(z_+) \frac{\sqrt{\chi_+^U}}{\phi_+(z_+) B_+(z_+)} \sum_{k = 0}^\infty b_k z_+^k ~ , ~
\ee
where the (real) coefficients $b_k$ satisfy the off-diagonal unitarity constraint
\be
     \label{eq:unitarity_BGL}
     \sum_{k, k^\prime = 0}^\infty b_{k^\prime} {\cal{S}}_{k^\prime k} b_k \leq 1 ~ 
\ee
with
\be
     \label{eq:Soff}
     {\cal{S}}_{k^\prime k} \equiv \frac{1}{2\pi i} \oint_{|z_+|=1} \frac{dz_+}{z_+} |f_R(z_+)|^2 z_+^{*k^\prime} z_+^k = 
                                                   \frac{1}{2\pi} \int_{-\pi}^\pi d\alpha_+ |f_R(e^{i\alpha_+})|^2 e^{i(k - k^\prime) \alpha_+} ~ . ~
\ee
Following the simple procedure described in Appendix\,\ref{sec:orthonormal}, the $z$-expansion\,(\ref{eq:BGL_new}) can be rewritten in the form
\be
     \label{eq:BGL_pol}
     f(z_+) = f_R(z_+) \frac{\sqrt{\chi_+^U}}{B_+(z_+)} \sum_{k = 0}^\infty a_k s_k(z_+) ~ , ~
\ee
where $s_k(z_+) = \sum_{m = 0}^k L_{k m}^{-1} z^m$ is a polynomial\footnote{The polynomials $s_k(z_+)$ are an orthonormal set on the unit circle $|z_+| = 1$ with respect to a weight function given by $|f_R(z_+)|^2$, namely
\be
    \frac{1}{2\pi i} \oint_{|z_+| = 1} \frac{dz_+}{z_+} \, |f_R(z_+)|^2 \, s_k(z_+) \, s_{k^\prime}^*(z_+) = \delta_{k k^\prime} ~ . ~ \nonumber
\ee} of degree $k$, obtained from the inverse of the lower triangular matrix of the Cholensky decomposition of the matrix ${\cal{S}} = L L^T$, and the (real) coefficients $a_k$ satisfy the diagonal unitarity constraint
\be
     \label{eq:unitarity_BGL_pol}
     \sum_{k = 0}^\infty a_k^2 \leq 1 ~ . ~
\ee

Summarizing, by means of the function\,(\ref{eq:fR}) the $z$-expansion\,(\ref{eq:BGL_new}) can be constructed to include the main effects of an above-threshold resonance. This is at variance with the simple pole Ansatz\,(\ref{eq:pole}), which does not possess the correct analytical properties in the unit disk. At most, for narrow resonances ($\Gamma_R / M_R << 1$) the pole Ansatz may work as an approximation of the function\,(\ref{eq:fR}) in the kinematical region $t \lesssim t+$, but not in the timelike one $t > t_+$ (particularly, when $t \approx M_R^2$)\footnote{At first order in $\Gamma_R / M_R$ one has
\be
    f_R(t) \simeq f_R(t_0) (M_R^2 - t_0 + \widetilde{\Gamma}_R \sqrt{t_+ - t_0}) / (M_R^2 - t + \widetilde{\Gamma}_R \sqrt{t_+ - t}) ~ \nonumber
\ee
with $\widetilde{\Gamma}_R = \Gamma_R \sqrt{(M_R^2 + t_+) / 2 (M_R^2 - t_+)}$.}.
Numerical applications of the $z$-expansion\,(\ref{eq:BGL_new}), or equivalently Eq.\,(\ref{eq:BGL_pol}), to phenomenological cases of interest are beyond the aim of the present work.

We close this subsection by mentioning that another way of modifying either the BGL\,(\ref{eq:BGL}) or the BCL\,\cite{Bourrely:2008za} $z$-expansions for taking into account above-threshold resonances has been recently developed in Ref.\,\cite{Herren:2025cwv}, where a model-independent parameterization of the $B \to \pi \pi \ell \nu_\ell$ form factors is constructed using Omn\`es functions\,\cite{Omnes:1958hv} and suitable conformal variables.

\section{The electromagnetic form factor of the charged kaon}
\label{sec:kaon}

Let us now address the issue of the presence of an extra branch-cut located at $t_{th}$ generated by resonant processes with invariant masses below the pair-production threshold $t_+$. 
To this end we focus on the case of the electromagnetic form factors of charged and neutral kaon mesons. 

In the kaon case, while the pair-production threshold is located at $t_+ \equiv t_{2K} = 4m_K^2 \simeq 0.98$ GeV$^2$, an extra branch-cut starts at the pion production threshold $t_{th} \equiv t_{2\pi} = 4m_\pi^2 \simeq 0.08$ GeV$^2$. In the kinematical region between $t_{2\pi}$ and $t_{2K}$, which is not accessible directly in timelike $e^+ e^- \to \overline{K} K$ processes, the dominant reaction mechanism is expected to proceed via $\pi \pi$ (on-shell) intermediate states that scatter inelastically into the final pair of (off-shell) kaons.  Such scattering is expected to be governed mainly by the presence of $\rho$- and $\omega$-meson resonances.

We now basically refer to the dispersive analysis made in Ref.\,\cite{Stamen:2022uqh}.
The electromagnetic form factors of charged and neutral kaons, $F_{K\pm}^{(em)}(t)$ and $F_{K^0}^{(em)}(t)$, can be decomposed into isovector (V) and an isoscalar (S) contributions
\bea
     \label{eq:FFkaon_charged}
     F_{K^\pm}^{(em)}(t) & = & F_K^{(S)}(t) + F_K^{(V)}(t)  ~ , ~ \\[2mm]
     F_{K^0}^{(em)}(t) & = & F_K^{(S)}(t) - F_K^{(V)}(t) ~ , ~
\eea
where the isovector part has the lowest branch point at $t_{2\pi} = 4m_\pi^2$, while for the isoscalar part the lowest branch-cut starts at $t_{3\pi} = 9 m_\pi^2$. Charge conservation implies that $F_K^{(V)}(0) = F_K^{(S)}(0) = 1/2$. For the isovector component $F_K^{(V)}(t)$ unitarity implies\,\cite{Blatnik:1978wj}
\be
     \label{eq:unitarity_isovector}
     \mbox{Im}F_K^{(V)}(t) = \frac{t}{4 \sqrt{2}} \left( 1 - \frac{t_{2\pi}}{t} \right)^{3/2}\left[ g_1^1(t) \right]^* F_\pi^{(em)}(t) \, \Theta(t - t_{2\pi}) ~ , ~
\ee
where $F_\pi^{(em)}(t)$ is the timelike pion form factor, described in Section\,\ref{sec:pion}, and $g_1^1(t)$ represents the p-wave component of the angular decomposition of the $\pi \pi \to \overline{K} K$ scattering amplitude. Both $F_\pi^{(em)}(t)$ and $g_1^1(t)$ satisfy dispersion relations with a branch-cut starting at $t_{th} = t_{2\pi}$ (see, e.g., Ref.\,\cite{Pelaez:2020gnd}). 

Experimental data on the phase and modulus of $g_1^1(t)$ are available in the physical region $t \geq t_{2K}$ from $\pi \pi \to \overline{K} K$ scattering. Instead, in the unphysical region $t_{2\pi} \leq t \leq t_{2K}$ such data are not  experimentally accessible. However, thanks to Eq.\,(\ref{eq:unitarity_isovector}) the phase of $g_1^1(t)$ must compensate the one of the pion form factor, which in turn for $t \leq t_{2K} \sim 1$ GeV$^2$ is dominated by the phase of the elastic $\pi \pi$ scattering. Since the phase of $g_1^1(t)$ is known for $t \geq t_{2\pi}$, then its modulus can be reconstructed\,\cite{Pelaez:2020gnd} using the standard Muskhelishvili-Omn\`es method\,\cite{Muskhelishvili58, Omnes:1958hv}. Using dispersive parameterizations of $g_1^1(t)$, elaborated in Ref.\,\cite{Pelaez:2020gnd}, and of the pion form factor $F_\pi^{(em)}(t)$ (similar to the one developed in Ref.\,\cite{Colangelo:2018mtw}), the isovector part of the kaon form factor $F_K^{(V)}(t)$ was determined in Ref.\,\cite{Stamen:2022uqh} for $t \geq t_{2\pi}$ and tested successfully against experimental data on $\tau^- \to K^- K_S \nu_\tau$ decays\,\cite{BaBar:2018qry} for $t \geq t_{2K}$. 
It turned out that $F_K^{(V)}(t)$ receives an important contribution from the $\rho(770)$-meson resonance and to a lesser extent from the $\rho^\prime(1450)$-meson resonance.

As for the isoscalar part, the authors of Ref.\,\cite{Stamen:2022uqh} developed a model based on Breit-Wigner ans\"atze corresponding to the lowest-lying isoscalar vector resonances. The free parameters of the model were obtained by fitting several sets of cross-section data on the timelike reactions $e^+ e^- \to K^+ K^-$  and $e^+ e^- \to K_S K_L$ (see for details Ref.\,\cite{Stamen:2022uqh}) as well as available spacelike data for electron scattering on charged kaons from FNAL\,\cite{Dally:1980dj} and CERN\,\cite{Amendolia:1986ui}.
It turned out that $F_K^{(S)}(t)$ receives contributions from the $\omega(782)$-meson resonance in the unphysical region below $t_{2K}$ and from the $\phi(1020)$-meson resonance (and to a lesser extent from the $\omega(1420)$-meson resonance) in the physical region above $t_{2K}$.

\begin{figure}[htb!]
\begin{center}
\includegraphics[scale=0.5]{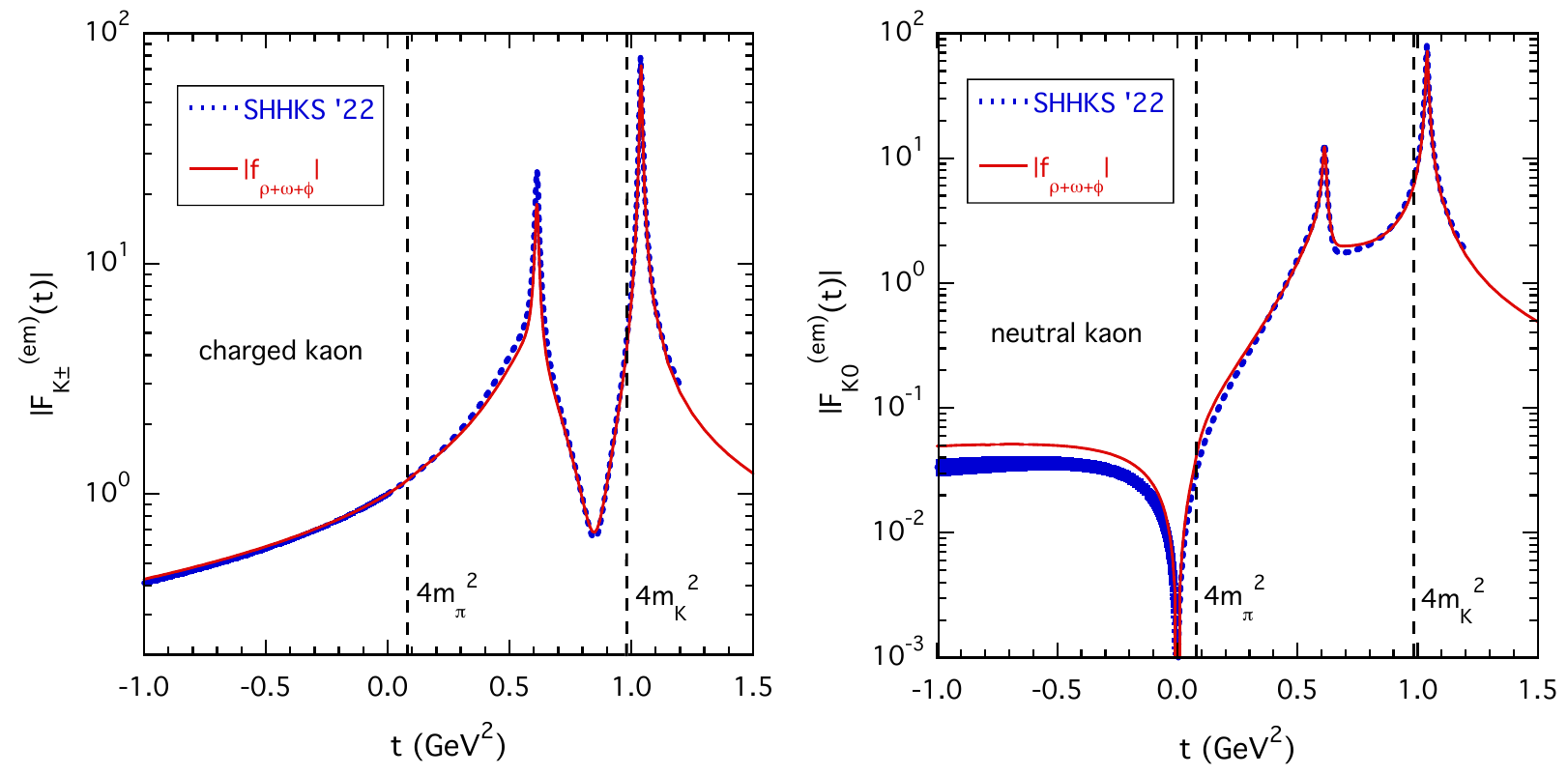}
\end{center}
\vspace{-0.75cm}
\caption{\it \small Absolute value of the electromagnetic form factors of charged (left panel) and neutral kaons (right panel) versus the squared momentum transfer $t$ both in the spacelike region ($t \leq 0$) and in the timelike one ($t > 0$). Dotted blue lines represent the results of the dispersive analysis of Ref.\,\cite{Stamen:2022uqh}, labelled as SHHKS '22, while the solid red lines correspond to the predictions of the resonance model given by Eqs.\,(\ref{eq:Kcharged_model})-(\ref{eq:Kneutral_model}) using the values $M_\rho = 775$ MeV, $\Gamma_\rho = 147$ MeV, $M_\omega = 782.7$ MeV, $\Gamma_\omega = 8.68$ MeV, $M_\phi = 1019.5$ MeV and $\Gamma_\phi = 4.25$ MeV taken from the PDG\,\cite{ParticleDataGroup:2024cfk}.}
\label{fig:kaon}
\end{figure}

The results for $F_{K^\pm}^{(em)}(t)$ and $F_{K^0}^{(em)}(t)$ obtained by the dispersive analysis of Ref.\,\cite{Stamen:2022uqh} are shown in Fig.\,\ref{fig:kaon} as the dotted blue lines for $-1 \leq t(\mbox{GeV}^2) \leq 1.2$, i.e.~both in the spacelike region ($t \leq 0$) and in the timelike one ($t > 0$).  Note that the uncertainties on both kaon form factors are available only in the spacelike region.
The results of Ref.\,\cite{Stamen:2022uqh} are compared with the predictions based on our simple model\,(\ref{eq:fR}) for above-threshold resonances. More precisely, we consider\footnote{We take into account that the electromagnetic current for $u$-, $d$- and $s$-quarks is given by the sum of an isovector $(\bar{u} \gamma^\mu u - \bar{d} \gamma^\mu d) / 2$ and an isoscalar part $(\bar{u} \gamma^\mu u + \bar{d} \gamma^\mu d - 2 \bar{s} \gamma^\mu s) / 6$. We neglect any mixing between the light and the strange sectors for the lowest-lying vector resonances and we ignore isospin-breaking effects.}
\bea
     \label{eq:Kcharged_model}
     F_{K^\pm}^{(em)}(t) & = &  \frac{1}{6} f_\omega(z_{3\pi}) + \frac{1}{3} f_\phi(z_{2K}) + \frac{1}{2} f_\rho(z_{2\pi}) ~ , ~ \\[2mm]
     \label{eq:Kneutral_model}
     F_{K^0}^{(em)}(t) & = & \frac{1}{6} f_\omega(z_{3\pi}) + \frac{1}{3} f_\phi(z_{2K}) - \frac{1}{2} f_\rho(z_{2\pi}) ~ , ~
\eea
where $f_\rho$ is given as in Eq.\,(\ref{eq:frho}) and similarly for $f_\omega$ and $f_\phi$. For each of the three resonances we adopt the conformal variable corresponding to the appropriate threshold, i.e.~$z_{2\pi}$, $z_{3\pi}$ and $z_{2K}$ corresponding respectively to the two-pion, $t_{2\pi} = 4m_\pi^2$,  three-pion, $t_{3\pi} = 9m_\pi^2$ and two-kaon, $t_{2K}=4m_K^2$, thresholds. The auxiliary variable $t_0$ is always chosen to be equal to $t_0 = 0$ in order to easily ensure charge conservation at $t =0$. Explicitly one has
\be
     \label{eq:zk}
     z_k = \frac{\sqrt{t_k - t} - \sqrt{t_k}}{\sqrt{t_k - t} + \sqrt{t_k}} ~ 
\ee
with $k = \{2\pi, 3\pi, 2K \}$.

The predictions of Eqs.\,(\ref{eq:Kcharged_model})-(\ref{eq:Kneutral_model}), using for the resonance parameters the PDG values\,\cite{ParticleDataGroup:2024cfk}, namely $M_\rho = 775$ MeV, $\Gamma_\rho = 147$ MeV, $M_\omega = 782.7$ MeV, $\Gamma_\omega = 8.68$ MeV, $M_\phi = 1019.5$ MeV and $\Gamma_\phi = 4.25$ MeV, are shown in Fig.\,\ref{fig:kaon} by the solid red lines.
The comparison with the dispersive results of Ref.\,\cite{Stamen:2022uqh} shows that the overall structure of both charged and neutral kaon form factors is reasonably reproduced for both timelike and spacelike values of $t$.  
In the pair-production region $t \geq t_{2K} \simeq 1$ GeV$^2$ the differences are within $\simeq 20 \%$, while they may reach $\simeq 30 \%$ in the unphysical region around the $\rho(770)$-$\omega(782)$ peaks for the case of the charged kaon. According to Ref.\,\cite{Stamen:2022uqh} that region is affected by SU(2)-breaking effects, which are not included in the simple resonance model of Eqs.\,(\ref{eq:Kcharged_model})-(\ref{eq:Kneutral_model}). 

Thus, the above findings suggest that a phenomenological estimate of the contribution $\chi_{\rm extra}^U$ to the dispersive bound $\chi^U$ may be obtained using for the form factor a simple resonance model, based on the $z$-dependence of Eq.\,(\ref{eq:fR}), for evaluating the integral\,(\ref{eq:chi_extra}) involving the product $|\phi(z) f(z)|^2$ over the unphysical region $\alpha_{th} \leq |\alpha| \leq \pi$ outside the pair-production arc (i.e., $t_{th} \leq t \leq t_+$). In the next Section we discuss the possible choices of the kinematical function $\phi(z)$ outside the pair-production arc.

\section{The outer function $\phi(z)$ outside the pair-production arc}
\label{sec:phi_>}

As already observed in Section\,\ref{sec:z_th}, a convenient choice for the kinematical function $\phi(z)$ is provided by Eq.\,(\ref{eq:phi_th}), namely
\be
    \label{eq:phi_00}
    \phi(z) = \sqrt{\frac{dz_+}{dz} } \phi_+(z_+) = \sqrt{\frac{1 + z}{1- z} \frac{1 - z_+}{1+ z_+} \frac{z_+}{z}} \phi_+(z_+)~ , ~
\ee
where the function $\phi_+(z_+)$ is the {\em outer function} adopted in the BGL approach\,\cite{Boyd:1994tt, Boyd:1995cf, Boyd:1995sq, Boyd:1997kz} to describe the momentum dependence of  the form factors relevant for the semileptonic weak decays of hadrons\,\cite{Boyd:1997kz, Bharucha:2010im} (see Appendix A of the companion paper\,\cite{Simula:2025lpc} for the case of semileptonic decays of pseudoscalar mesons). In this way, for obtaining an upper limit to the contribution $\chi_+^U$ inside the pair-production arc one can make use of the susceptibility $\chi_n$, related to the appropriate derivative of the vacuum polarization function and calculable nonperturbatively on the lattice (see Refs.\,\cite{Martinelli:2021frl, Melis:2024wpb, Harrison:2024iad, DiCarlo:2021dzg, Martinelli:2022tte}). We stress that this can be done without any knowledge of the $z$-dependence of the form factor on the pair-production arc. 

However, outside the pair-production arc the choice of $\phi(z)$ is not unique. Using the same auxiliary variable $t_0$ for both conformal variables $z_+$ and $z$, given respectively in Eqs.\,(\ref{eq:zplus}) and (\ref{eq:z}) with $t_0 < t_{th} < t_+$, the kinematical function $\phi(z)$ may be chosen as
\be
     \label{eq:phi_n}
     \phi(z) \equiv \sqrt{\left( \frac{z}{z_+} \right)^p \frac{dz_+}{dz} } \phi_+(z_+) ~ , ~
\ee
where $p$ is a generic constant. In this way, on the pair-production arc $-\alpha_{th} \leq \alpha \leq \alpha_{th}$, where both $z_+$ and $z$ are unimodular, Eq.\,(\ref{eq:phi_th_arc}) is still valid for any value of $p$, namely one has $|\phi(e^{i\alpha})|^2 = (d\alpha_+ / d\alpha) \, |\phi_+(e^{i\alpha_+})|^2$. However, the value of $p$ can be fixed by requiring that the integral $\chi[f]$, given by Eq.\,(\ref{eq:chi_f}), is independent of the auxiliary variable $t_0$. This can be achieved by imposing that $\phi(z) \sqrt{1 - z^2}$ is independent of $t_0$ (see the Appendix A of Ref.\,\cite{Simula:2023ujs}). Since the product $\phi_+(z_+) \sqrt{1 - z_+^2}$ does not depend on $t_0$ (by applying the independence of $\chi_+[f]$ on $t_0$), we have simply to impose that the quantity
\be
     \label{eq:t0}
     \left( \frac{z}{z_+} \right)^p \frac{dz_+}{dz} \frac{1 - z^2}{1 - z_+^2} = \left( \frac{1+z}{1+z_+} \right)^2 \left( \frac{z}{z_+} \right)^{p-1}
\ee
is independent of $t_0$. Explicitly one gets
\be
    \left( \frac{1+z}{1+z_+} \right)^2 \left( \frac{z}{z_+} \right)^{p-1} = \frac{t_{th} - t}{t_+ - t}  \left( \frac{\sqrt{t_+ - t} + \sqrt{t_+ - t_0}}{\sqrt{t_{th} - t} + \sqrt{t_{th} - t_0}} \right)^{2p} ~ , ~
\ee
so that Eq.\,(\ref{eq:t0}) is independent of $t_0$ only when $p = 0$.

Nevertheless, it is worth highlighting that also Eq.\,(\ref{eq:phi_n}) does not represent the most general form for the kinematical function $\phi(z)$. Indeed, we can still multiply Eq.\,(\ref{eq:phi_n}) by any function of $z_+$ and $z$, which is analytic of the real type inside the unit disk $|z| < 1$ and unimodular in the unit circle $|z_+| = 1$. Possible functions of this type are generic powers of specific ratios of the conformal variables corresponding to a fixed value of $t_0 < t_{th}$, namely $z_+(t; 0) / z_{th}(t; 0)$ and $z_+(t; t_-) / z_{th}(t; t_-)$, which are clearly unimodular on the pair-production arc\footnote{Any conformal variable $z_+(t; t_0)$ (or $z(t; t_0)$) becomes unimodular when  $t > t_+$ (or $t > t_{th}$) regardless the value of $t_0$.}. The conformal variables $z_+(t; 0)$ and $z_+(t; t_-)$ come into play\,\cite{Bharucha:2010im} when suitable unimodular substitutions are used to extend the kinematical function known along the unit circle inside the unit disk. 
In the case of $\phi_+(z_+)$ the following unimodular substitutions are adopted (see, e.g., Ref.\,\cite{Simula:2025lpc})
 \bea
       \label{eq:substitutions}
       \sqrt{t} & \to & \sqrt{\frac{-t}{z(t; 0)}} = \sqrt{t_+ - t} + \sqrt{t_+} ~ , ~ \\[2mm]
      \sqrt{t_- - t} & \to & \sqrt{\frac{t_- - t}{z(t; t_-)}} = \sqrt{t_+ - t} + \sqrt{t_+ - t_-} ~ , ~ \nonumber 
\eea
where $t_- \equiv (m_1 - m_2)^2$.
The above relations guarantee the analyticity and the absence of zeros of the function $\phi_+(z_+)$ inside the unit disk $|z_+| < 1$.
In the case at hand, an alternative choice is to use the conformal variables $z(t; 0)$ and $z(t; t_-)$, so that the analyticity and the absence of zeros of the kinematical function $\phi(z)$ inside the unit disk $|z| < 1$ is always guaranteed.
In other words, once $\phi_+(z_+)$ is specified, several choices of the outer function $\phi(z)$ are still possible outside the pair-production arc.
For instance, in the unit disk the outer function $\phi(z)$ can be written as:
\be
    \label{eq:phi_q1q2}
    \phi^{q_1q_2}(z) = \sqrt{\frac{dz_+}{dz}  \left[ \frac{z(t; 0)}{z_+(t; 0)} \right]^{q_1} \left[ \frac{z(t; t_-)}{z_+(t; t_-)} \right]^{q_2}} \phi_+(z_+) ~ , ~
\ee
where $q_1$ and $q_2$ are real powers. Explicitly, one has
\be
    \sqrt{\frac{z(t; 0)}{z_+(t; 0)}} = \frac{\sqrt{t_+ - t} + \sqrt{t_+}}{\sqrt{t_{th} - t} + \sqrt{t_{th}}} \quad \mbox{and} \quad
    \sqrt{\frac{z(t; t_-)}{z_+(t; t_-)}} = \frac{\sqrt{t_+ - t} + \sqrt{t_+ - t_-}}{\sqrt{t_{th} - t} + \sqrt{t_{th} - t_-}} ~ . ~ \nonumber
\ee
The kinematical function $\phi^{q_1 q_2}(z)$ is a legitimate outer function for any values of $q_1$ and $q_2$, since inside the unit disk $|z| < 1$ it fulfills the relation
\be
   \label{eq:outer_definition}
   \mbox{log}\phi^{q_1 q_2}(z) = \frac{1}{2\pi} \int_{-\pi}^\pi d\alpha  \frac{e^{i \alpha} + z}{e^{i \alpha} - z} \mbox{log}|\phi^{q_1 q_2}(e^{i \alpha})| ~ , ~
\ee
as it can be directly checked numerically.

One has $|\phi^{q_1 q_2}(z)| = |\phi^{00}(z)| = |\phi(z)|$ when $|z_+| = 1$, which implies also $|z| = 1$. Therefore, the contribution $\chi_+[f]$ on the pair-production arc is independent of the powers $q_1$ and $q_2$ and it can be always bounded by the appropriate susceptibility $\chi_+^U = \chi_n$. On the contrary, the contribution $\chi_{\rm extra}[f]$ outside the pair-production arc does depend on the powers $q_1$ and $q_2$.

\begin{figure}[htb!]
\begin{center}
\includegraphics[scale=0.5]{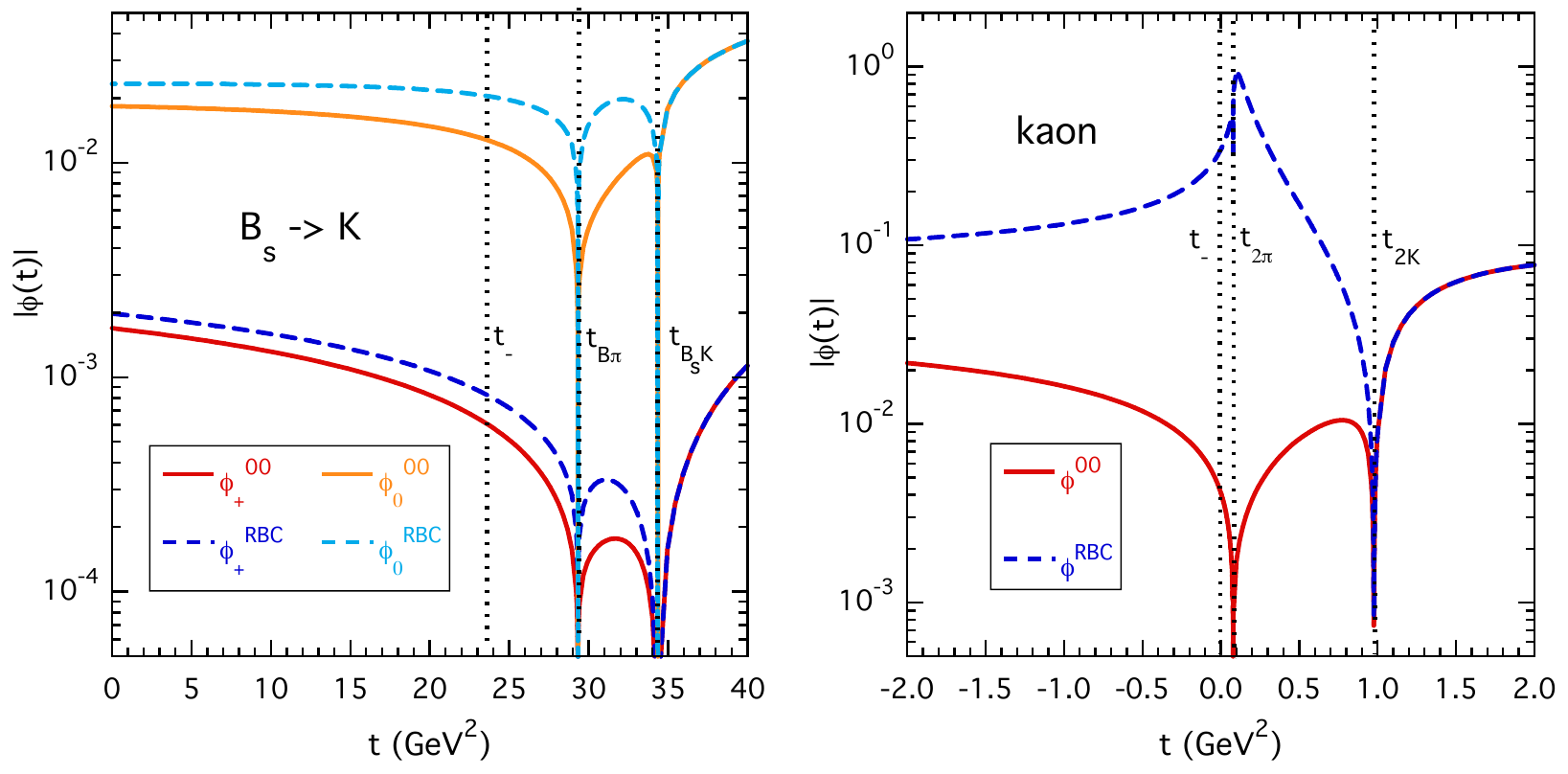}
\end{center}
\vspace{-0.75cm}
\caption{\it \small Absolute values of the kinematical function\,(\ref{eq:phi_q1q2}) versus the squared 4-momentum transfer $t$ corresponding to $q_1 = q_2 = 0$ (solid lines) and to the choice of $q_1$ and $q_2$ made in Ref.\,\cite{Flynn:2023nhi} (dashed lines) and labelled with the suffix $RBC$ (see text). The value of the auxiliary variable $t_0$ is chosen to be equal to $t_0 = t_-$. Left panel: vector and scalar kinematical functions for the $B_s \to K$ transition with $t_- = (m_{B_s} - m_K)^2 \simeq 23.7$ GeV$^2$, $t_{th} = t_{B \pi} = (m_B + m_\pi)^2 \simeq 29.3$ GeV$^2$ and $t_+ = t_{B_s K} = (m_{B_s} + m_K)^2 \simeq  34.4$ GeV$^2$. Right panel: vector kinematical function corresponding to the case of the electromagnetic form factor of the charged kaon with $t_- = 0$, $t_{th} = t_{2\pi} = 4 m_\pi^2 \simeq 0.08$ GeV$^2$ and $t_+ = t_{2K} = 4 m_K^2 \simeq  0.98$ GeV$^2$.}
\label{fig:phi}
\end{figure}

The choice\,(\ref{eq:phi_q1q2}) has been adopted in Ref.\,\cite{Blake:2022vfl} for $\Lambda_b \to \Lambda$ transition and in Ref.\,\cite{Flynn:2023nhi} for the  $B_s \to K$ form factors with specific choices of the powers $q_1$ and $q_2$.
In the case of the $B_s \to K$ form factors\,\cite{Flynn:2023nhi} the powers $q_1$ and $q_2$ are equal to $q_1 = 5$ and $q_2 = -3/2$ for the vector form factor $f_+$, while $q_1 = 4$ and $q_2 = -1/2$ for the scalar form factor $f_0$.
The corresponding $t$-dependencies of the kinematical functions\,(\ref{eq:phi_q1q2}) of the vector and scalar form factors for the $B_s \to K$ transition are shown in the left panel of Fig.\,\ref{fig:phi},   where they are compared with those related to the choice $q_1 = q_2 = 0$. The differences are limited, particularly in the semileptonic decay region $0 \leq t \leq t_- = (M_{B_s} - m_K)^2$.

In the right panel of Fig.\,\ref{fig:phi} an analogous comparison is made in the case of the electromagnetic form factor of the charged kaon.
In this case, instead, large differences, even by orders of magnitude, are visible in the unphysical region $t_{2\pi} \leq t \leq t_{2K}$, particularly close to the lowest branch-point $t_{2\pi}$, and also for spacelike values of $t$. Since in this case $t_0 = t_- = 0$ one has explicitly ($q_1 + q_2 = 7/2$)
\be
    \label{eq:phi_RBC_kaon}
   |\phi^{q_1 q_2}(z)| ~ _{\overrightarrow{q_1 = 5, \, q_2 = -3/2}} ~ \sqrt{\left| \frac{z}{z_+} \right|^{7/2}} |\phi^{00}(z)|  = 
        \left| \frac{\sqrt{t_{2K} - t} + \sqrt{t_{2K}}}{\sqrt{t_{2\pi} - t} + \sqrt{t_{2\pi} }} \right|^{7/2} |\phi^{00}(t)| ~ . ~
\ee
In the kaon case we have calculated the quantity $4 m_K^2 \chi_{\rm extra}[f]$, given by Eq.\,(\ref{eq:chi_extra}) with $\alpha_{th} / \pi = 0.1823$, adopting for the form factor either the dispersive results of Ref.\,\cite{Stamen:2022uqh} or the resonance model given by Eqs.\,(\ref{eq:Kcharged_model})-(\ref{eq:Kneutral_model}), developed in the previous Section. We remind that in the timelike region the uncertainties of the kaon form factors of Ref.\,\cite{Stamen:2022uqh} are not available (a rough estimate indicates errors of the order of $10 \%$).
The results corresponding to the kinematical function $\phi^{q_1 q_2}(z)$ (with $\phi_+(z_+) = (1536 \, \pi)^{-1/2}$ $(1 + z_+)^2$ $\sqrt{1 - z_+}$) are collected in Table\,\ref{tab:chi_>} for various values of $q \equiv q_1 + q_2$. 
\begin{table}[htb!]
\begin{center}
\begin{tabular}{|c||c|c||c|c||}
\hline \hline
$q = q_1 + q_2$ & \multicolumn{2}{c||}{charged kaon} & \multicolumn{2}{c||}{neutral kaon} \\ \cline{2-5} 
                           & $z$-model & SHHKS '22 &  $z$-model & SHHKS '22\\  \hline \hline
  - 1.0  & 0.000043 & 0.000070  & 0.000024 & 0.000025 \\ \hline
  ~ 0.0 & 0.00019   & 0.00031    & 0.000084 & 0.000089 \\ \hline 
  ~ 1.0 & 0.0010     & 0.0015      & 0.00033   & 0.00035   \\ \hline 
  ~ 2.0 & 0.0079     & 0.010        & 0.0014     & 0.0015      \\ \hline 
  ~ 3.5 & 0.63         & 0.65          & 0.015       & 0.015        \\ \hline  
\end{tabular}
\end{center}
\vspace{-0.25cm}
\caption{\it Values of the quantity $4 m_K^2 \chi_{\rm extra}[f]$, given by Eq.\,(\ref{eq:chi_extra}) with $\alpha_{th} / \pi = 0.1823$, evaluated using the kinematical function $\phi^{q_1 q_2}(z)$ with $\phi_+(z_+) = (1536 \, \pi)^{-1/2}$ $(1 + z_+)^2$ $\sqrt{1 - z_+}$ and adopting for the form factor either the dispersive results of Ref.\,\cite{Stamen:2022uqh}, labelled as SHHKS '22, or the resonance model given by Eqs.\,(\ref{eq:Kcharged_model})-(\ref{eq:Kneutral_model}), labelled as $z$-model.}
\label{tab:chi_>}
\end{table}
The simple resonance model given by Eqs.\,(\ref{eq:Kcharged_model})-(\ref{eq:Kneutral_model}) works quite well in the neutral kaon channel with respect to the results corresponding to the dispersive approach of Ref.\,\cite{Stamen:2022uqh} with differences well below the $10 \%$ level at any value of $q$. Instead, in the charged kaon channel the resonance model underestimates the contribution $\chi_{\rm extra}[f]$ up to $\simeq 30 - 40 \%$ for $q \lesssim 1$.

The values obtained for $4 m_K^2 \chi_{\rm extra}[f]$ can be compared with the contribution $4 m_K^2 \chi_+[f]$ related to the pair-production arc. Using the resonance model we get $4 m_K^2 \chi_+[f] = 0.0024$ and $0.0020$ for the charged and neutral channel, respectively. Using the dispersive form factor of Ref.\,\cite{Stamen:2022uqh} (known up to $t \simeq 1.2$ GeV$^2$ with the addition of a pQCD tail for $t > 1.2$ GeV$^2$) we obtain similar results, namely $4 m_K^2 \chi_+[f] = 0.0026$ and $0.0025$ for the charged and neutral channels\footnote{If one limits the calculation of $4 m_K^2 \chi_+[f]$ to the kinematical region of the $\phi(1020)$ resonance, i.e.~between $t_{2K} \simeq 0.98$ GeV$^2$ and $1.2$ GeV$^2$, one gets $4 m_K^2 \chi_+[f] = 0.0024$ for both charged and neutral channels.}.
As expected from the comparison made in the right panel of Fig.\,\ref{fig:phi}, the contribution $\chi_{\rm extra}[f]$ strongly depends on the value of $q$; in particular, at $q = 7/2$ it is $\simeq 2000$ times the one at $q = 0$. Moreover, while at $q = 0$ the value of $\chi_{\rm extra}[f]$ is only a fraction of the pair-production contribution $\chi_+[f]$,  at $ q = 7/2$ it becomes much larger by a factor $\simeq 240$. The impact of the above numerical values in the implementation of the double dispersive bound in the case of the electromagnetic kaon form factor will be presented in the next Section.

Before closing this Section, we stress that presently we are not aware of a strategy to calculate a dispersive bound $\chi_{\rm extra}^U$, based on first-principles only, without requiring the knowledge of the form factor outside the pair-production arc. Once this problem will be solved in terms of suitable Euclidean correlation functions calculable on the lattice, it is likely that also the choice of the kinematical function  $\phi(z)$ outside the pair-production arc will become unique.

\section{Analysis of the spacelike experimental and lattice data on the charged kaon form factor}
\label{sec:charged_kaon}

In this Section we apply the BGL $z$-expansion, described in Section\,\ref{sec:subth}, to the study of the experimental data on the electromagnetic form factor of the charged kaon, $F_{K^\pm}^{(em)}(Q^2)$, available for spacelike values of $Q^2 \equiv -t$ up to $Q^2 \simeq 0.12$ GeV$^2$ from FNAL\,\cite{Dally:1980dj} and  CERN NA7\,\cite{Amendolia:1986ui}.
Separately, we consider also the recent lattice QCD results, obtained by the HotQCD Collaboration\,\cite{Ding:2024lfj} using $N_f = 2 +1$ flavors of highly improved staggered quarks. Their results exhibit a remarkable high precision and cover a quite extended range of $Q^2$ up to $Q^2 \simeq 28$ GeV$^2$.

We want to compare the results obtained using the truncated BGL expansion\,(\ref{eq:BGL_truncated}) applying either the double bound represented by Eqs.\,(\ref{eq:unitarity_pair_truncated})-(\ref{eq:unitarity_extra_truncated}) or the single, total bound\,(\ref{eq:unitarity_truncated}). In addition we apply also the expansion\,(\ref{eq:BGL_VD}) with the unitarity constraint\,(\ref{eq:unitarity_VD}), as described in Section\,\ref{sec:Szego}. We remind that such an expansion, adopted in Refs.\,\cite{Flynn:2023qmi, Flynn:2023nhi, Harrison:2025yan}, is equivalent to the expansion of the form factor in terms of Szeg\H{o} polynomials, proposed in Refs.\,\cite{Gubernari:2020eft, Gubernari:2022hxn, Blake:2022vfl}, namely Eq.\,(\ref{eq:BGL_Szego}) with the constraint\,(\ref{eq:unitarity_Szego}). This approach implements unitarity in an incomplete way, i.e., only on the pair-production arc, and it does not guarantee unitarity on the full circle.

For the dispersive bound corresponding to the pair-production arc, $\chi_+^U$, we use the estimate 
\be
    \label{eq:chi+U}
    \chi_+^U = 0.0024 \pm 0.0005 ~, ~
\ee
obtained using the resonance model of Section\,\ref{sec:kaon}  and assuming a $20 \%$ uncertainty.
As for the contribution outside the pair-production arc, $\chi_{\rm extra}^U$, this depends upon the choice of the kinematical function $\phi^{q_1 q_2}(z)$, defined in Eq.\,(\ref{eq:phi_q1q2}).
In what follows we compare two choices: the first one is $q_1 = q_2 = 0$ and the other one is $q_1 = 5$ and $q_2 = - 3/2$, which correspond respectively to $q = q_1 + q_2$ equal to $q = 0$ and $q = 3.5$. These two choices correspond to quite different values of $\chi_{\rm extra}^U$, shown in Table\,\ref{tab:chi_>}. Here below, we adopt the values corresponding to the dispersive form factor of Ref.\,\cite{Stamen:2022uqh} (known up to $t \simeq 1.2$ GeV$^2$ with the addition of a pQCD tail for $t > 1.2$ GeV$^2$), namely
\bea
    \label{eq:chiextraU_q0}
    \chi_{\rm extra}^U & = & 0.00031 \pm 0.00006 \qquad \mbox{for } q = 0 ~ , ~ \\[2mm]
    \label{eq:chiextraU_q3.5}
    \chi_{\rm extra}^U & = & 0.65 \pm 0.13 \qquad \qquad \quad \mbox{for } q = 3.5 ~ , ~
\eea
where we have assumed again a $20 \%$ uncertainty\footnote{In this way the quantities $\chi_+^U$ and $\chi_{\rm extra}^U$ can be treated as uncorrelated.}. Indeed, in the first case $q = 0$ the contribution $\chi_{\rm extra}^U$ is a fraction of the order of $10 \%$ of the pair-production bound\,(\ref{eq:chi+U}), while in the second case $q = 3.5$, it is much larger than $\chi_+^U$ by a factor of $\approx 240$.

In the case of the experimental data on $F_{K^\pm}^{(em)}(Q^2)$ there are 10 data from FNAL\,\cite{Dally:1980dj} and 15 data from CERN\,\cite{Amendolia:1986ui}. To ensure charge conservation in the BGL $z$-expansion\,(\ref{eq:BGL_truncated}) we add a further data point $F_{K^\pm}^{(em)}(Q^2 = 0) = 1$ with a tiny error ($\simeq 10^{-12}$)\footnote{Choosing $t_0 = t_- = 0$ in Eq.\,(\ref{eq:z}), this is equivalent to fix the coefficient $a_0$ of the BGL expansion\,(\ref{eq:BGL_truncated}) to the value $a_0 = \phi(z=0) / \sqrt{\chi^U}$ without introducing the additional point at $Q^2 = 0$.}. Therefore, we have a total of $N_{data} = 26$ data points. The covariance matrices of the two experiments are known separately, so that we can construct the full covariance matrix $C_{ij}$ of the input data by considering the two experiments uncorrelated.

In the case of the LQCD results of the HotQCD Collaboration\,\cite{Ding:2024lfj} there are 28 data points available up to $Q^2 \simeq 28$ GeV$^2$. In our analysis we include only the first 7 points up to $Q^2 \simeq 0.4$ GeV$^2$ and compare our extrapolation at larger $Q^2$ with the lattice points. Adding the normalization point  $F_{K^\pm}^{(em)}(Q^2 = 0) = 1$ we have a total of $N_{data} = 8$ data points in the analysis. The covariance matrix of the data is not provided in Ref.\,\cite{Ding:2024lfj}. Therefore, we consider an uncorrelated covariance matrix $C_{ij}$.

In Section\,\ref{sec:unfiltered} we perform the BGL analyses without applying any unitary filtering to the input datasets, namely the BGL $z$-expansions are applied to samples of input data containing both non-unitary and unitary events. Instead, in Section\,\ref{sec:filtering} we filter the input datasets against the unitarity bound(s) before applying the BGL $z$-expansions.

\subsection{Unfiltered input data}
\label{sec:unfiltered}

Let us start with the case of the experimental data on the charged kaon form factor and the choice $q = 0$ for the kinematical function outside the pair-production arc. We apply the truncated BGL expansion\,(\ref{eq:BGL_truncated}) at various order $M$ of the truncation using either the double bound\,(\ref{eq:unitarity_pair_truncated})-(\ref{eq:unitarity_extra_truncated}) or the single, total bound\,(\ref{eq:unitarity_truncated}). The dispersive bounds $\chi_+^U$ and $\chi_{\rm extra}^U$ are given by Eqs.\,(\ref{eq:chi+U})-(\ref{eq:chiextraU_q0}) and, therefore, at $q = 0$ the total bound $\chi^U = \chi_+^U + \chi_{\rm extra}^U$ is largely dominated by the pair-production bound $\chi_+^U$. We observe a nice stability of the extrapolated form factor up to $Q^2 \simeq 1$ GeV$^2$ for $M \geq 6$.
We apply also the expansion\,(\ref{eq:BGL_VD}) with the unitarity constraint\,(\ref{eq:unitarity_VD}), restricted to the pair-production arc only.

\begin{figure}[htb!]
\begin{center}
\includegraphics[scale=0.4]{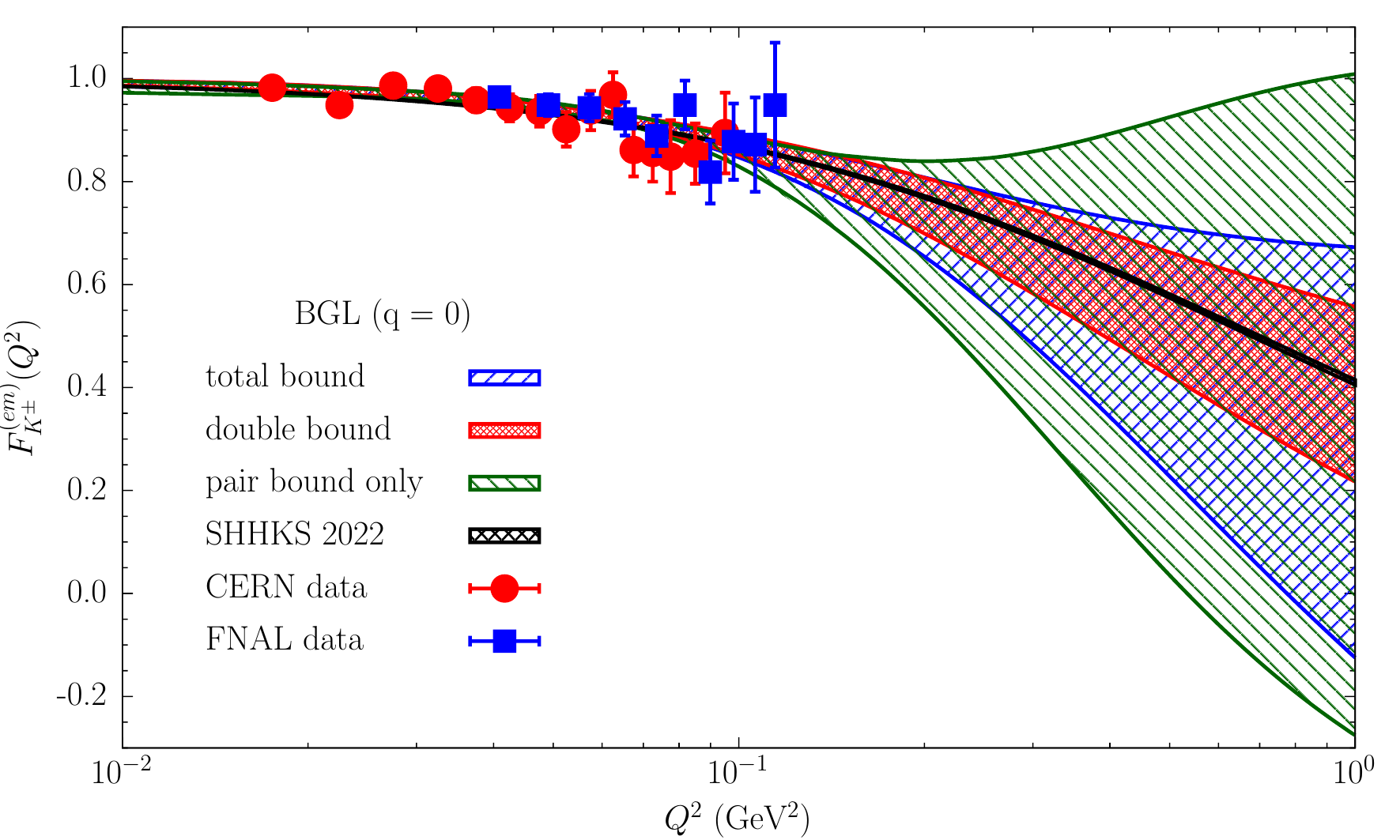} \\
\includegraphics[scale=0.4]{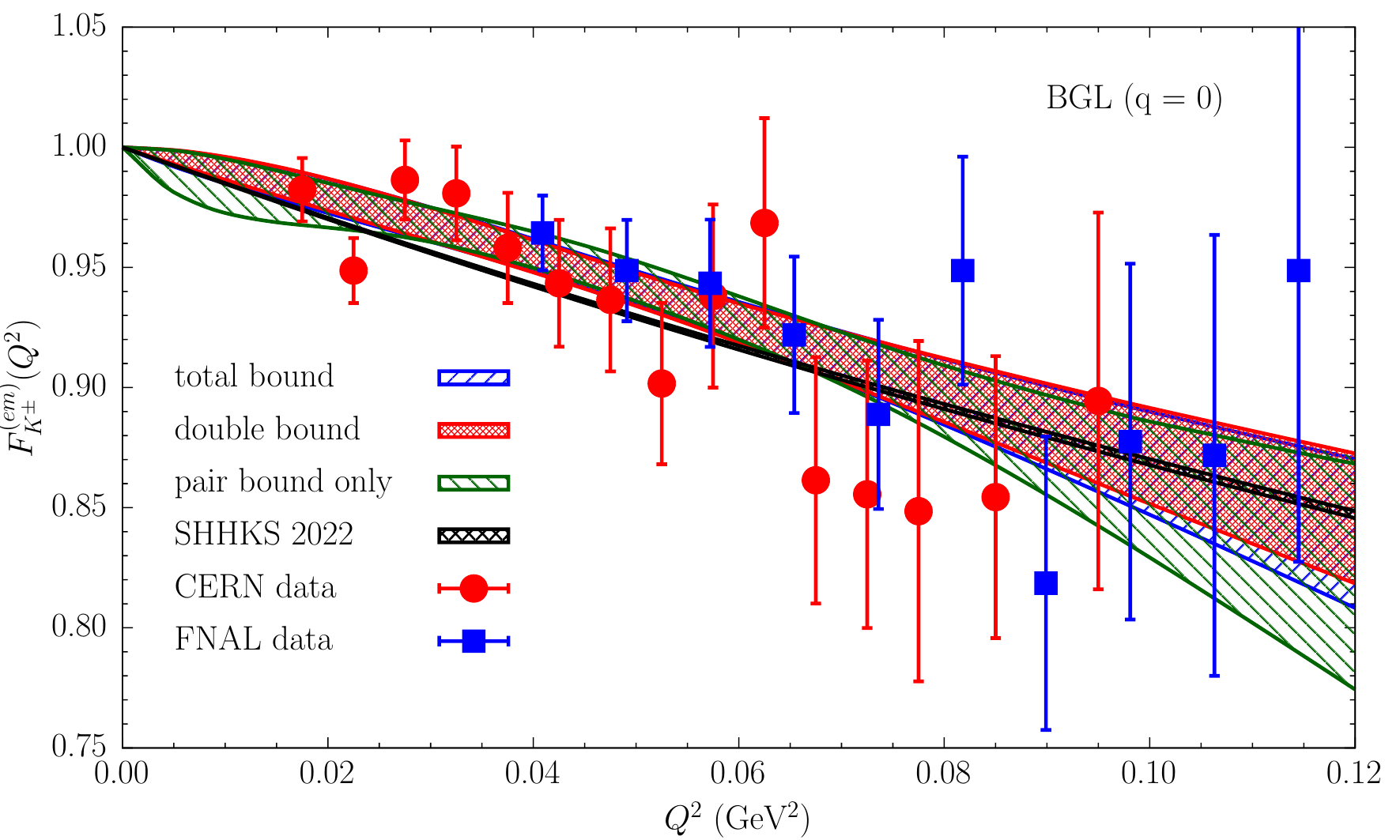}
\end{center}
\vspace{-0.75cm}
\caption{\it \small The electromagnetic form factor of the charged kaon $F_{K^\pm}^{(em)}(Q^2)$ versus the 4-momentum transfer $Q^2 = - t$. The red and blue bands, evaluated at $1\sigma$ level, correspond to the results obtained using the BGL expansion\,(\ref{eq:BGL_truncated}) at $M = 8$ and adopting respectively the double bound\,(\ref{eq:unitarity_pair_truncated})-(\ref{eq:unitarity_extra_truncated}) and the single, total bound\,(\ref{eq:unitarity_truncated}). The dispersive bounds $\chi_+^U$ and $\chi_{\rm extra}^U$ are given respectively by Eqs.\,(\ref{eq:chi+U}) and (\ref{eq:chiextraU_q0}), and the total bound $\chi^U$ is $\chi^U = \chi_+^U + \chi_{\rm extra}^U$. The green band represents the results obtained using  the expansion\,(\ref{eq:BGL_VD}) with the unitarity constraint\,(\ref{eq:unitarity_VD}) corresponding to the pair-production arc only. In all cases the choice $q = 0$ for the kinematical function $\phi^{q_1 q_2}(z)$ is considered. The black band (having tiny errors) corresponds to the results of the dispersive approach of Ref.\,\cite{Stamen:2022uqh}, while the blue squares and the red circles are respectively the experimental data from the FNAL\,\cite{Dally:1980dj} and CERN\,\cite{Amendolia:1986ui} experiments. The bottom panel is a zoom in the range of values of $Q^2$ covered by the two experiments.}
\label{fig:charged_kaon_FNAL+CERN_q0}
\end{figure}

The comparison of the predicted bands for the form factor up to $Q^2 \simeq 1$ GeV$^2$ is shown in Fig.\,\ref{fig:charged_kaon_FNAL+CERN_q0}. In what follows the width of the bands will be always provided at $1\sigma$ level. We have checked that: ~ i) the probability distributions of the calculated form factors are largely normal, and ~ ii) the calculated bands (i.e, both central values and errors) are stable against variation of the truncation order $M$. We have found that the stability starts already at $M = 4$. In Fig.\,\ref{fig:charged_kaon_FNAL+CERN_q0} all the $z$-expansions are truncated at order $M = 8$. For the three expansions the value of the reduced (correlated) $\chi^2$-variable turns out to be equal to $\chi^2 / (d.o.f.) \simeq 1.3$.
Few comments are in order.
\begin{itemize}
\item The application of the double bound\,(\ref{eq:unitarity_pair_truncated})-(\ref{eq:unitarity_extra_truncated}) on the BGL expansion\,(\ref{eq:BGL_truncated}) leads to the most precise extrapolation at large $Q^2$. In particular, it improves by a factor of $\sim 2$ the prediction at $Q^2 \sim 1$ GeV$^2$ with respect to the application of the single, total bound\,(\ref{eq:unitarity_truncated}).

\item The use of the expansion\,(\ref{eq:BGL_VD}) with the pair-production bound\,(\ref{eq:unitarity_VD}) alone, leads to a quite huge spread of values at large $Q^2$. This is related to the fact that the expansion in terms of Szeg\H{o} polynomials, proposed in Refs.\,\cite{Gubernari:2020eft, Gubernari:2022hxn, Blake:2022vfl} and adopted also in Refs.\,\cite{Flynn:2023qmi, Flynn:2023nhi, Harrison:2025yan}, represents an incomplete application of the unitarity constraints and it may produce violations of the unitarity constraint outside the pair-production arc and, consequently, of the total bound $\chi^U$ on the full arc. Indeed, in the case at hand we observe that the coefficients $b_k$ of the expansion\,(\ref{eq:BGL_VD}) have large absolute values, leading to $\sum_{k = 0}^M b_k^2 >> \chi^U / \chi_+^U \simeq 1.1$.

\item All the three $z$-expansions are consistent with the more precise dispersive results from Ref.\,\cite{Stamen:2022uqh} both at low and large values of $Q^2$. Such a high-precision level is related to the fact that the results of Ref.\,\cite{Stamen:2022uqh} are obtained from a global dispersive analysis of both spacelike and timelike data by implementing Eq.\,(\ref{eq:unitarity_isovector}) in the isovector channel, using data on the pion electromagnetic form factor and the $\pi \pi \to \overline{K} K$ partial-wave amplitude, and by considering in the isoscalar channel the contributions from $\omega$- and $\phi$-meson residues, determined from electromagnetic reactions involving kaons.
\end{itemize}

\begin{figure}[htb!]
\begin{center}
\includegraphics[scale=0.4]{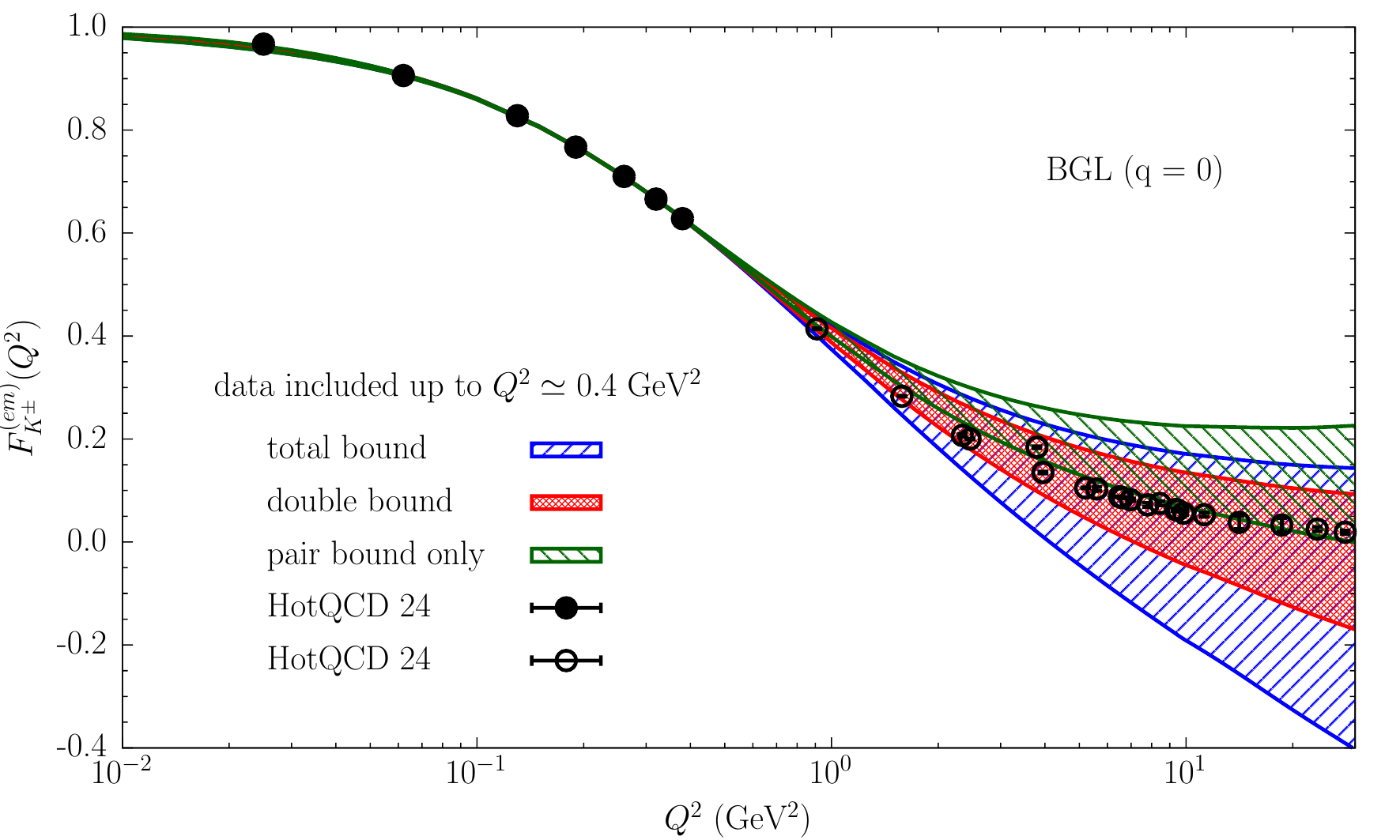}
\end{center}
\vspace{-0.75cm}
\caption{\it \small The same as in the upper panel of Fig.\,\ref{fig:charged_kaon_FNAL+CERN_q0}, but adopting the LQCD results of the HotQCD Collaboration\,\cite{Ding:2024lfj}. The data included in the analysis are those with $Q^2 \lesssim 0.4$ GeV$^2$ (full circles), while the ones not included in the analysis and extending up to $Q^2 \simeq 28$ GeV$^2$ are represented by the empty circles All $z$-expansions are truncated at order $M = 6$.}
\label{fig:charged_kaon_HotQCD24_q0}
\end{figure}

We have performed an analogous analysis by replacing the experimental data with the LQCD results of the HotQCD Collaboration\,\cite{Ding:2024lfj}. These are much more precise than the experimental determinations of $F_{K^\pm}^{(em)}(Q^2)$ by approximately one order of magnitude. As already anticipated, we include in the analysis only the lattice points up to $Q^2 \simeq 0.4$ GeV$^2$ and extrapolate our $z$-expansions up to $Q^2 \simeq 28$ GeV$^2$ to compare with the remaining LQCD data. The results are collected in Fig.\,\ref{fig:charged_kaon_HotQCD24_q0}, where the $z$-expansions are truncated at order $M = 6$.
We observe again that:  ~ i) the use of the double bound improves the extrapolation at large $Q^2$ by a factor of $\sim 2$ with respect to the use of the single, total bound;  ~ ii) the two BGL expansions implementing properly the unitarity constraints are consistent with the lattice QCD data up to $Q^2 \simeq 28$ GeV$^2$, and ~ iii) the expansion in terms of Szeg\H{o} polynomials, which takes into account only the unitarity constraint on the pair-production arc, is consistent with the lattice QCD data at large $Q^2$, i.e.\,in the region where the LQCD data points are not used in the fitting procedure; however, the coefficients $b_k$ of the expansion\,(\ref{eq:BGL_VD}) turn out to have large absolute values, leading to $\sum_{k = 0}^M b_k^2 >> \chi^U / \chi_+^U \simeq 1.1$, which implies that unitarity on the full arc is badly violated.

We now move to the other choice considered for the kinematical function $\phi^{q_1 q_2}(z)$ , namely $q = q_1 + q_2 = 3.5$. For the bound $\chi_{\rm extra}^U$ we use the estimate given by Eq.\,(\ref{eq:chiextraU_q3.5}), which is much larger than the pair-production contribution $\chi_+^U$, given by Eq.\,(\ref{eq:chi+U}), by a factor of $\approx 240$.  Thus, at $q = 3.5$ the total bound $\chi^U = \chi_+^U + \chi_{\rm extra}^U$ is largely dominated by the contribution $\chi_{\rm extra}^U$ outside the pair-production arc.
In what follows, we will leave out the results corresponding to the expansion\,(\ref{eq:BGL_Szego}), expressed in terms of Szeg\H{o} polynomials. As previously shown at $q  = 0$ and checked also at $q = 3.5$, this approach violates badly unitarity on the full arc, introducing therefore non-unitarity effects already at the level of the $z$-expansion. Our aim is to focus on the results of the two BGL $z$-expansions properly compatible with unitarity, i.e.\,those based on Eq.\,(\ref{eq:BGL_truncated}) with either the double bound\,(\ref{eq:unitarity_pair_truncated})-(\ref{eq:unitarity_extra_truncated}) or the single, total bound\,(\ref{eq:unitarity_truncated}).

\begin{figure}[htb!]
\begin{center}
\includegraphics[scale=0.4]{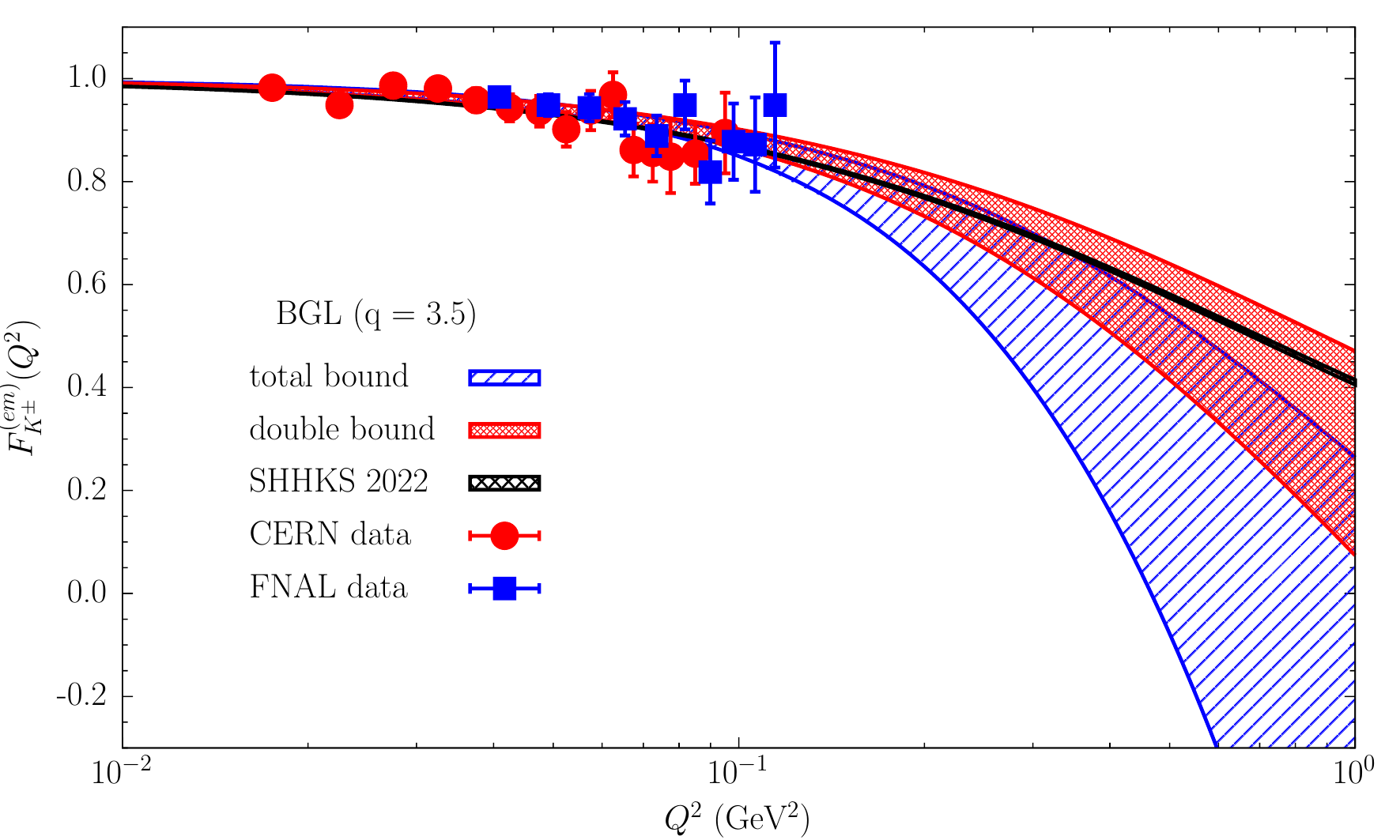} \\
\includegraphics[scale=0.4]{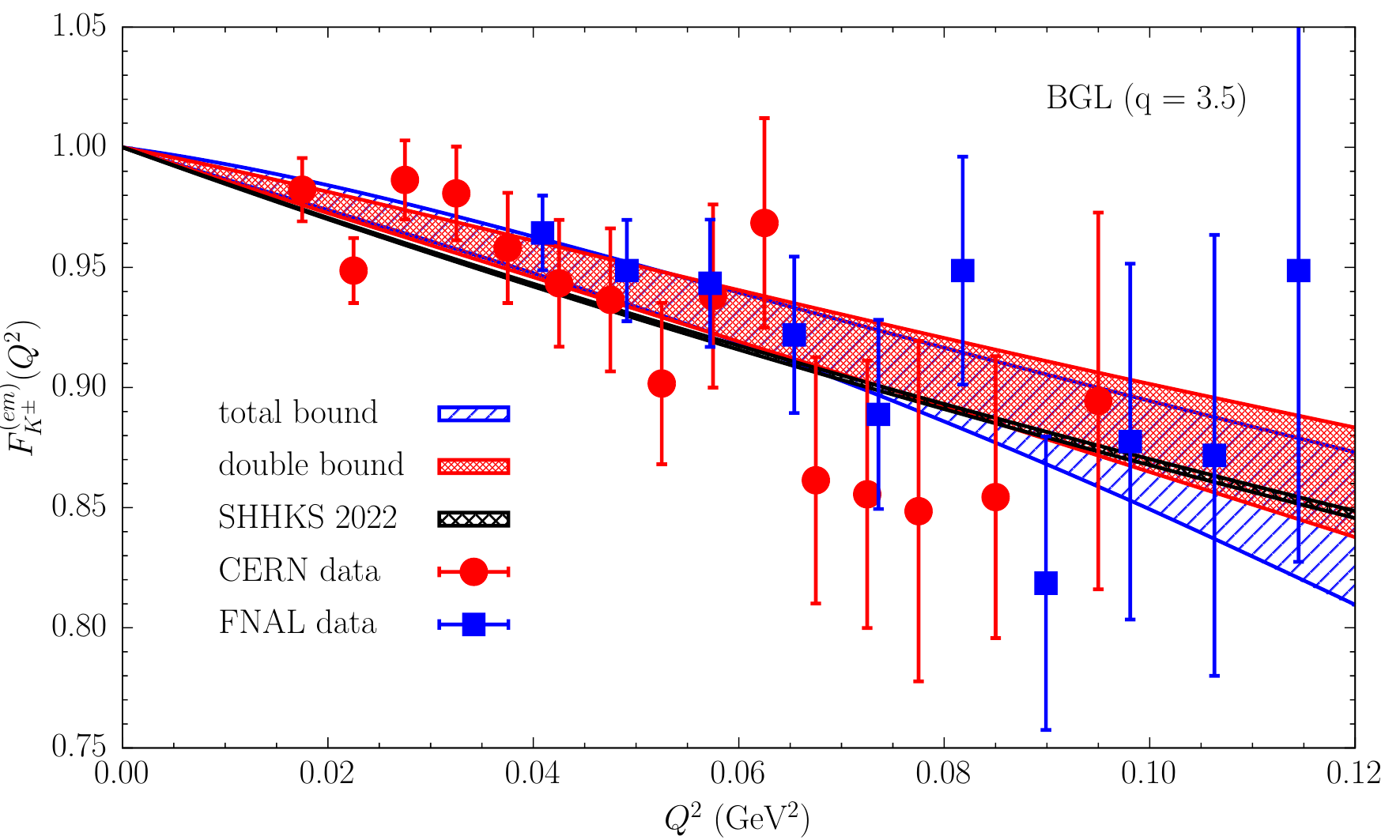}
\end{center}
\vspace{-0.75cm}
\caption{\it \small The electromagnetic form factor of the charged kaon $F_{K^\pm}^{(em)}(Q^2)$ versus the 4-momentum transfer $Q^2 = - t$. The red and blue bands, evaluated at $1\sigma$ level,  correspond to the results obtained using the BGL expansion\,(\ref{eq:BGL_truncated}) at $M = 8$ and adopting respectively the double bound\,(\ref{eq:unitarity_pair_truncated})-(\ref{eq:unitarity_extra_truncated}) and the single, total bound\,(\ref{eq:unitarity_truncated}). The dispersive bounds $\chi_+^U$ and $\chi_{\rm extra}^U$ are given respectively by Eqs.\,(\ref{eq:chi+U}) and (\ref{eq:chiextraU_q3.5}), and the total bound $\chi^U$ is $\chi^U = \chi_+^U + \chi_{\rm extra}^U$. In all cases the choice $q = 3.5$ for the kinematical function $\phi^{q_1 q_2}(z)$ is considered. The black band, blue squares and red circles are as in Fig.\,\ref{fig:charged_kaon_FNAL+CERN_q0}. The bottom panel is a zoom in the range of values of $Q^2$ covered by the two experiments.}
\label{fig:charged_kaon_FNAL+CERN_q3.5}
\end{figure}

The comparison of the predicted bands for $F_{K^\pm}^{(em)}(Q^2)$, obtained using simultaneously the two experimental datasets from FNAL\,\cite{Dally:1980dj} and CERN\,\cite{Amendolia:1986ui}, is shown in Fig.\,\ref{fig:charged_kaon_FNAL+CERN_q3.5}, where we have truncated the $z$-expansions at order $M = 8$. For the two expansions the value of the reduced (correlated) $\chi^2$-variable turns out to be equal to $\chi^2 / (d.o.f.) \simeq 1.4$.
It is confirmed that also in the case $q = 3.5$ the use of the double bound yields a much more precise band of values at large $Q^2$ with respect to the use of the single total bound. By comparing the results in Figs.\,\ref{fig:charged_kaon_FNAL+CERN_q0} and \ref{fig:charged_kaon_FNAL+CERN_q3.5} there is a clear dependence of the extrapolated band on the choice of $q$ in the case of the single total bound, while such a dependence is much more limited in the case of the double bound. 

\begin{figure}[htb!]
\begin{center}
\includegraphics[scale=0.4]{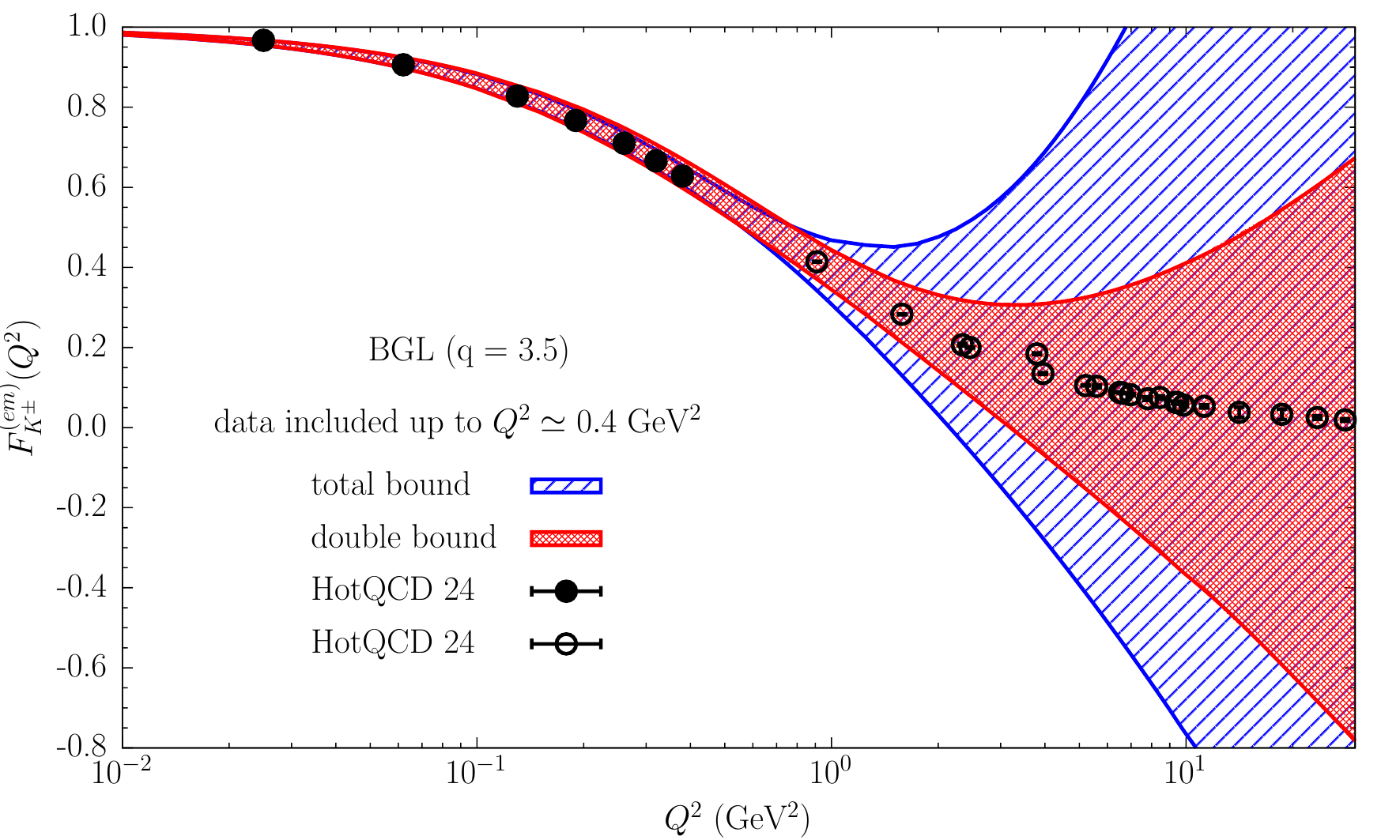}
\end{center}
\vspace{-0.75cm}
\caption{\it \small The same as in the upper panel of Fig.\,\ref{fig:charged_kaon_FNAL+CERN_q3.5}, but adopting the LQCD results of the HotQCD Collaboration\,\cite{Ding:2024lfj}. The data included in the analysis are those with $Q^2 \lesssim 0.4$ GeV$^2$ (full circles), while the ones not included in the analysis and extending up to $Q^2 \simeq 28$ GeV$^2$ are represented by the empty circles All $z$-expansions are truncated at order $M = 6$.}
\label{fig:charged_kaon_HotQCD24_q3.5}
\end{figure}

Adopting the LQCD results from the HotQCD Collaboration\,\cite{Ding:2024lfj} the results obtained with our BGL expansions, truncated at order $M = 6$, are collected in Fig.\,\ref{fig:charged_kaon_HotQCD24_q3.5}.
Both BGL expansions, including only lattice points with $Q^2 \lesssim 0.4$ GeV$^2$, predict values consistent with the lattice QCD data up to $Q^2 \simeq 28$ GeV$^2$ though within a much larger band of values with respect to the choice $q = 0$ (compare with Fig.\,\ref{fig:charged_kaon_HotQCD24_q0}). Moreover, at variance with Fig.\,\ref{fig:charged_kaon_HotQCD24_q0}, a large width of the predicted bands is visible also at low $Q^2$, i.e.\,in the $Q^2$-range of values adopted in the fitting procedure. Both unpleasant features are related to the presence of non-unitary effects in the set of input data. The removal of such effects is discussed in the next subsection.

\subsection{Unitary filtering of the input data}
\label{sec:filtering}

According to Ref.\,\cite{Simula:2025lpc}, in order to have a completely unitary BGL approach any given set of input data $\{ f_i \equiv F_{K^\pm}^{(em)}(Q_i^2) \}$ (with $i = 1, 2, ..., N_{data}$) should be filtered against the single total bound $\chi^U$; more precisely, the following constraint should be fulfilled
\be
     \label{eq:filter_data}
    \chi[\beta] \leq \chi^U = \chi_{\rm extra}^U + \chi_+^U ~ , ~
\ee
where the function $\beta(z)$ is given by
\be
    \label{eq:beta}
    \beta(z) = \frac{1}{\phi(z) B(z) d(z)} \sum_{i = 1}^{N_{data}} \phi(z_i) B(z_i) d_i f_i \frac{1 - z_i^2}{z - z_i}
\ee
and
\bea
      \label{eq:chi_beta}
      \chi[\beta]  & \equiv &  \frac{1}{2\pi i} \oint_{|z| = 1} \frac{dz}{z} |\phi(z) \beta(z)|^2 \\[2mm]
                        & = & \sum_{i, j = 1}^{N_{data}} \phi(z_i) B(z_i) d_i f_i \, \phi(z_j) B(z_j) d_j f_j \, \frac{(1 - z_i^2) (1 - z_j^2)}{1 - z_i z_j} ~ \nonumber 
\eea
with
\bea
   \label{eq:dz}
    d(z) & = & \prod_{m = 1}^{N_{data}}  \frac{1 - z z_m}{z - z_m}  ~ , ~ \nonumber \\[2mm]
    \label{eq:di}
    d_i & = & \prod_{m \neq i = 1}^{N_{data}}  \frac{1 - z_i z_m}{z_i - z_m}  ~ , ~  \nonumber
\eea

Starting from the original covariance matrix $C_{ij}$ for the input data and choosing the kinematical function $\phi(z)$ as in Eq.\,(\ref{eq:BGL}), the action of the filter\,(\ref{eq:filter_data}) is to select the unitary subset $\{ \overline{f}_i \}$ with a covariance matrix $\overline{C}_{ij}$, i.e.\,only those data that can be reproduced exactly by (an infinite number of) unitary BGL expansions, given by Eq.\,(\ref{eq:BGL}) with the constraint\,(\ref{eq:unitarity}). When the filter\,(\ref{eq:filter_data}) is not fulfilled, then the input data cannot be reproduced exactly by any unitary BGL expansion. These data contain non-unitary effects and they are removed from the analysis (see for details Ref.\,\cite{Simula:2025lpc}).
Therefore, to the subset of input data satisfying the global filter\,(\ref{eq:filter_data}) we apply the expansion\,(\ref{eq:BGL_truncated}), adopting the single total bound\,(\ref{eq:unitarity_truncated}) for the coefficients of the expansion.

In the case of the analysis employing the double bound\,(\ref{eq:unitarity_pair_truncated})-(\ref{eq:unitarity_extra_truncated}) we modify the selection of the unitary input data as follows. Since $\chi[\beta]$ is the integral over the unit circle given in Eq.\,(\ref{eq:chi_beta}), we can split such an integral into the sum of two (positive) contributions, related to the pair-production arc and outside the pair-production arc, namely
\be
    \chi[\beta] = \chi_{\rm extra}[\beta] + \chi_{\rm pair}[\beta]
\ee
where 
\bea
      \label{eq:chi_beta_extra}
      \chi_{\rm extra}[\beta] & = &  \sum_{i, j = 1}^{N_{data}} \phi(z_i) B(z_i) d_i f_i \, \phi(z_j) B(z_j) d_j f_j \, \frac{(1 - z_i^2) (1 - z_j^2)}{1 - z_i z_j} \left[ 1 - B_{ij}(\alpha_{th}) \right] ~ , ~ \\[2mm]
      \label{eq:chi_beta_pair}
      \chi_{\rm pair}[\beta] & = &  \sum_{i, j = 1}^{N_{data}} \phi(z_i) B(z_i) d_i f_i \, \phi(z_j) B(z_j) d_j f_j \, \frac{(1 - z_i^2) (1 - z_j^2)}{1 - z_i z_j} B_{ij} (\alpha_{th})~
\eea
with
\bea
    \label{eq:Bij}
    B_{ij}(\alpha_{th}) & \equiv & \frac{1 - z_i z_j}{2\pi} \int_{-\alpha_{th}}^{\alpha_{th}} d\alpha \frac{1}{1 + z_i z_j - (z_i + z_j) \mbox{cos}\alpha + i (z_j - z_i) \mbox{sin}\alpha } ~ , ~ \\[2mm]
                                 & = & 1 - \frac{1}{\pi} \left\{ \mbox{Arctg}\left[ \frac{1-z_i}{1+z_i} \mbox{cotg}\left( \frac{\alpha_{th}}{2} \right) \right] + 
                                            \mbox{Arctg}\left[ \frac{1-z_j}{1+z_j} \mbox{cotg}\left( \frac{\alpha_{th}}{2} \right) \right] \right\} ~  \nonumber
\eea
and $\alpha_{th}$ given by Eq.\,(\ref{eq:alpha_th}).
Then, we may replace the global filter\,(\ref{eq:filter_data}) with the double filter
\bea
     \label{eq:filter_extra_data}
     \chi_{\rm extra}[\beta] & \leq & \chi_{\rm extra}^U ~ , ~ \\[2mm]
     \label{eq:filter_pair_data}
     \chi_{\rm pair}[\beta] & \leq & \chi_+^U ~ . ~
\eea
Therefore, to the subset of input data satisfying the double filter\,(\ref{eq:filter_extra_data})-(\ref{eq:filter_pair_data}) we apply the expansion\,(\ref{eq:BGL_truncated}), adopting the double bound\,(\ref{eq:unitarity_pair_truncated})-(\ref{eq:unitarity_extra_truncated}) for the coefficients of the expansion. 

\begin{figure}[htb!]
\begin{center}
\includegraphics[scale=0.4]{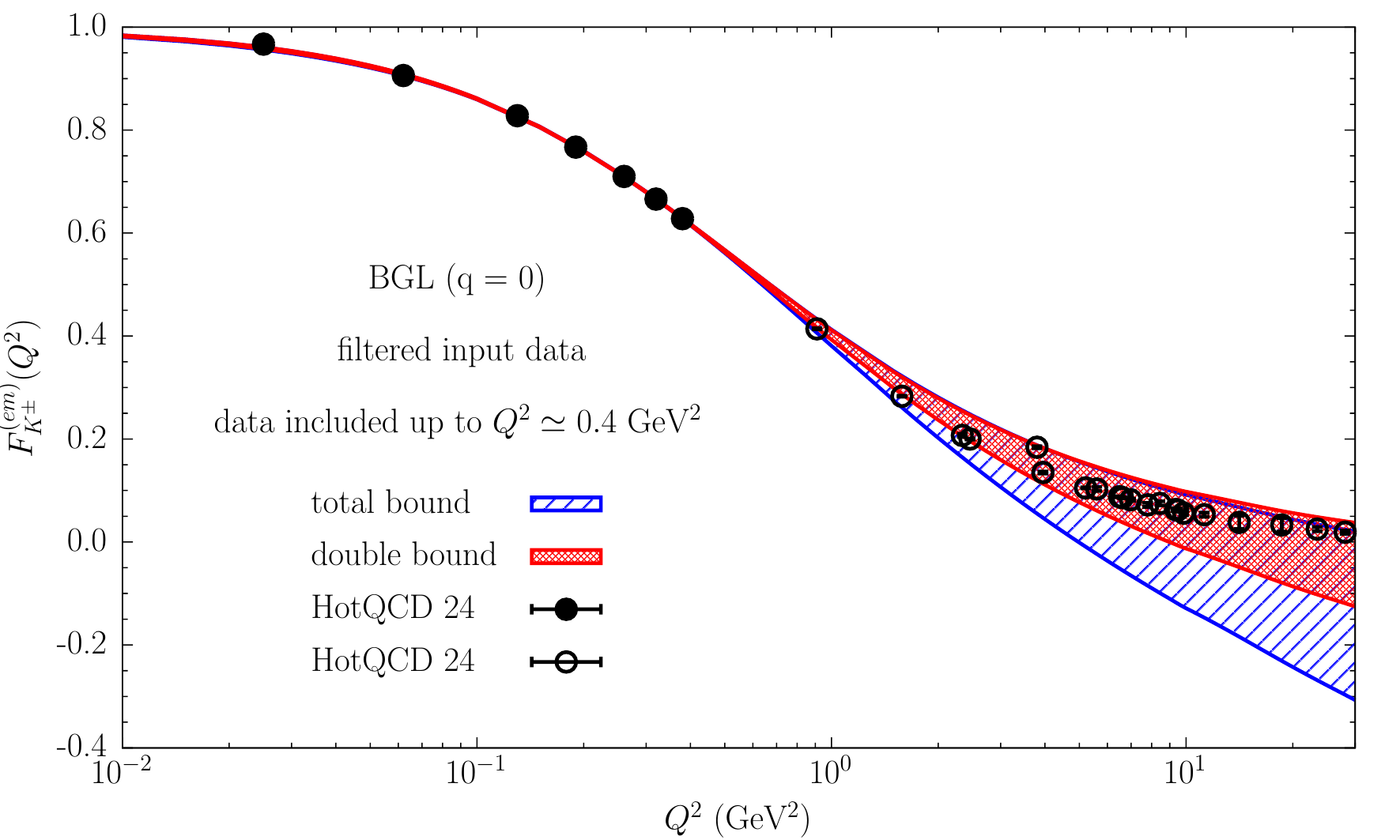} \\
\includegraphics[scale=0.4]{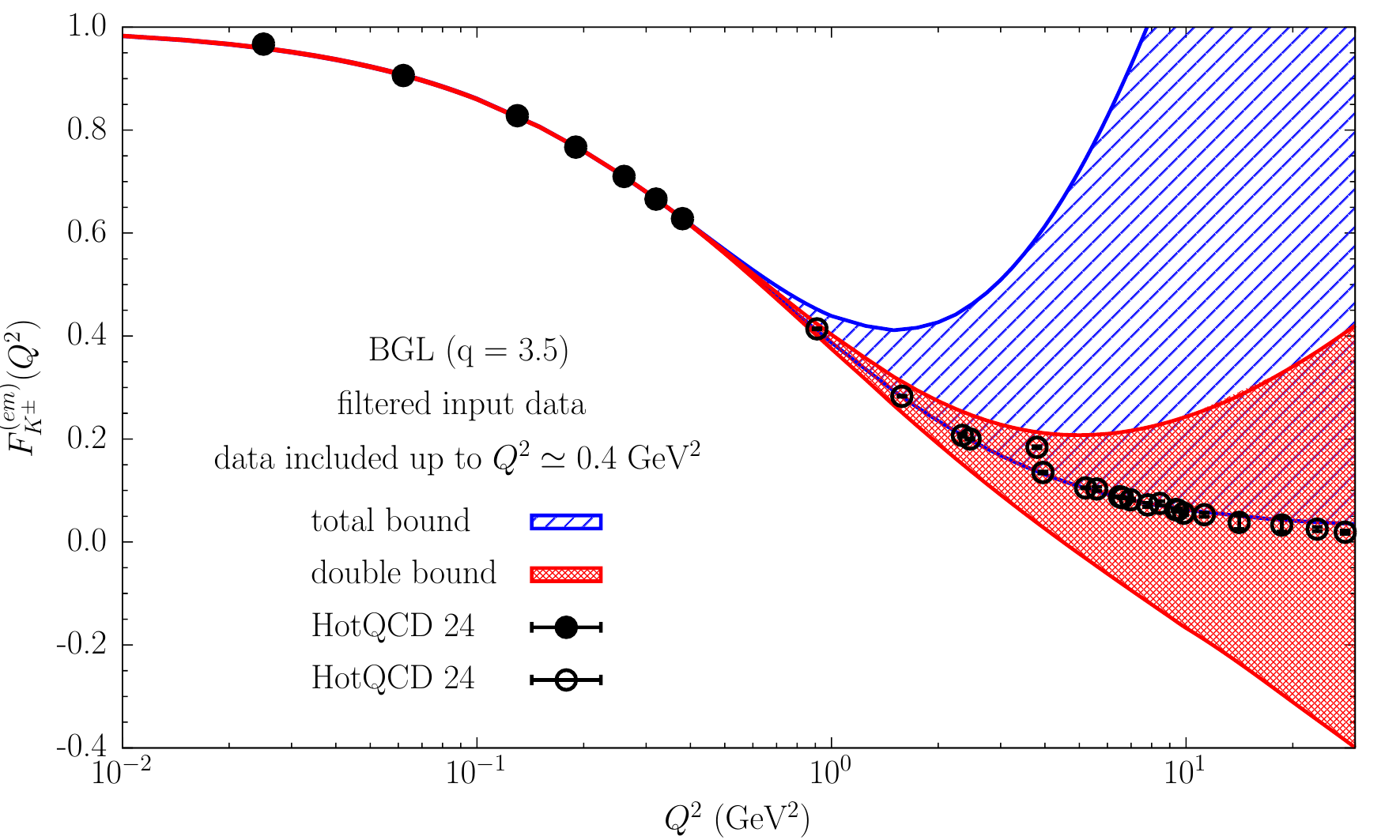}
\end{center}
\vspace{-0.75cm}
\caption{\it \small The electromagnetic form factor of the charged kaon $F_{K^\pm}^{(em)}(Q^2)$ versus the 4-momentum transfer $Q^2 = - t$, adopting the LQCD results of the HotQCD Collaboration\,\cite{Ding:2024lfj}. The data included in the analysis are those with $Q^2 \lesssim 0.4$ GeV$^2$ (full circles), while the ones not included in the analysis and extending up to $Q^2 \simeq 28$ GeV$^2$ are represented by the empty circles. The upper panel refers to the choice $q = 0$ for the kinematical function $\phi^{q_1 q_2}(z)$, while the lower panel to the choice $q = 3.5$. The red and blue bands, evaluated at $1\sigma$ level, correspond to the results obtained using the BGL expansion\,(\ref{eq:BGL_truncated}) at $M = 6$ and adopting respectively the double bound\,(\ref{eq:unitarity_pair_truncated})-(\ref{eq:unitarity_extra_truncated}) and the single, total bound\,(\ref{eq:unitarity_truncated}). The dispersive bounds $\chi_+^U$ and $\chi_{\rm extra}^U$ are described in the text (see Eqs.\,(\ref{eq:chi+U})-(\ref{eq:chiextraU_q3.5})) and the total bound $\chi^U$ is $\chi^U = \chi_+^U + \chi_{\rm extra}^U$. The input data are filtered either by the global filter\,(\ref{eq:filter_data}) or by the double filter\,(\ref{eq:filter_extra_data})-(\ref{eq:filter_pair_data}), according to the unitarity constraints applied to the coefficients of the BGL expansions.}
\label{fig:charged_kaon_HotQCD24}
\end{figure}

In the kaon case of interest in this paper there are no poles below the lowest threshold $t_{th} = 4 m_\pi^2 \simeq 0.08$ GeV$^2$ and, therefore, $B(z) = 1$.
We apply either the global filter\,(\ref{eq:filter_data}) or the double filter\,(\ref{eq:filter_extra_data})-(\ref{eq:filter_pair_data}). 
The main effect of such filters is a modification of the original covariance matrix $C_{ij}$ into a new matrix $\overline{C}_{ij}$ by introducing appropriate correlations either between the two experimental datasets or among the LQCD data. We stress that such modifications are dictated by unitarity. It turns out that in the case of the experimental data from FNAL\,\cite{Dally:1980dj} and CERN\,\cite{Amendolia:1986ui} the impact of the filtering procedure described above on the predicted bands of values for the form factor is limited, while this is not the case when the LQCD data from the HotQCD Collaboration\,\cite{Ding:2024lfj} are considered. This is likely due to the fact that the LQCD results exhibit a remarkable high precision (at the level of few permille) and also that  correlations are not provided at all in Ref.\,\cite{Ding:2024lfj}. The global filter\,(\ref{eq:filter_data}) as well as the double filter\,(\ref{eq:filter_extra_data})-(\ref{eq:filter_pair_data}), that both depend on the choice of the kinematical function, may change the mean values and errors of the form factor and may produce a new highly correlated covariance matrix $\overline{C}_{ij}$ sensitive to the value of $q$\footnote{We have checked that the subset of filtered data is always well consistent with a normal probability distribution.}. In order to give a quantitative estimate of the impact of the filtering procedure we have computed the values of the quantities $\Delta$ and $\epsilon$, defined in Eqs.\,(32)-(33) of the companion paper\,\cite{Simula:2025lpc}. For both choices of $q$ and in the case of either the single or the double dispersive bound we always find $\Delta \simeq 0.6$ and $\epsilon \simeq 0.7$, which guarantee that anomalous modifications of mean values and errors of the form factor are excluded. The corresponding predicted bands for the form factor $F_{K^\pm}^{(em)}(Q^2)$ are shown in Fig.\,\ref{fig:charged_kaon_HotQCD24} in the case of the LQCD data and the impact of the filtering procedures is clearly visible by a direct comparison with Figs.\,\ref{fig:charged_kaon_HotQCD24_q0} and \ref{fig:charged_kaon_HotQCD24_q3.5}.
In particular, though partially blurred by the size of the markers, the widths of the bands at low $Q^2$ in Fig.\,\ref{fig:charged_kaon_HotQCD24_q3.5} ($q = 3.5$) are definitely larger than the errors of the data points used in the fitting procedure, while the same does not occur in Fig.\,\ref{fig:charged_kaon_HotQCD24_q0} ($q = 0$). These findings are related to the $q$-dependent impact of non-unitary events in the (unfiltered) sample of input data. Indeed, as shown in Fig.\,\ref{fig:charged_kaon_HotQCD24}, when the fitting procedure is applied only to the unitarity-filtered subset of input data, the widths of the bands at low $Q^2$ do not exceed the errors of the data points for both values of $q$.
Moreover, at large $Q^2$ the width of the predicted bands is reduced significantly both at $q =0$ and $q = 3.5$. 

We point out that the predictions of either the total or the double bound expansion are consistent with the LQCD points above $Q^2 \simeq 0.4$ GeV$^2$ up to $Q^2 \simeq 28$ GeV$^2$ both at $q =0$ and $q =3.5$.
Moreover, only for the double bound expansion the bands of the form factor at $q = 0$ and $q = 3.5$ are consistent with each other. The most precise band is the one at $q = 0$, which therefore represents our present best estimate of the kaon form factor in the case of the LQCD inputs.

\subsection{Charged kaon radius}
\label{sec:radius}

A quantity of phenomenological interest is the charged kaon radius $r_{K^\pm}$, defined through the relation
\be
    \label{eq:slope}
    r_{K^\pm}^2 = - 6 \frac{dF_{K^\pm}^{(em)}(Q^2)}{dQ^2} \Big|_{Q^2 = 0} ~ . ~
\ee
In the experimental works from FNAL\,\cite{Dally:1980dj} and CERN\,\cite{Amendolia:1986ui} the charged kaon radius is estimated from the data assuming a simple Ansatz of the form $F_{K^\pm}^{(em)}(Q^2) = 1 / (1 +  r_{K^\pm}^2 Q^2 / 6)$, obtaining $r_{K^\pm} = 0.53 \pm 0.05$ fm\,\cite{Dally:1980dj} and $r_{K^\pm} = 0.58 \pm 0.04$ fm\,\cite{Amendolia:1986ui}. These two determinations are averaged in the last PDG review\,\cite{ParticleDataGroup:2024cfk}, leading to $r_{K^\pm} = 0.560 \pm 0.031$ fm.

We point out, however, that the monopole Ansatz used in Refs.\,\cite{Dally:1980dj, Amendolia:1986ui} represents a model-dependent assumption, which might lead to an underestimation of the uncertainty.
Using the same experimental datasets, but adopting the results of our model-independent BGL $z$-expansions with unitarity implemented through the double dispersive bounds we get
\be
     \label{eq:radius_data}
     r_{K^\pm} = 0.538 \pm 0.066 \pm 0.004_{q} ~ [0.066] ~ \qquad \mbox{[FNAL + CERN data]} ~ , ~
\ee
where the second error comes from the small dependence upon the choice of the kinematical function (estimated using half of the spread of our results obtained at $q = 0$ and $q = 3.5$). Thus, the value quoted by PDG underestimates the uncertainty by a factor of $\simeq 2$ due to the model-dependent assumption about the momentum dependence of the charged kaon form factor.

In the case of LQCD data the authors of Ref.\,\cite{Ding:2024lfj} provide the quite precise estimate $r_{K^\pm} = 0.600 \pm 0.002$ fm, based on fitting their data up to $Q^2 \simeq 0.4$ GeV$^2$ adopting a vector-meson-dominance Ansatz constructed using the three main resonances $\rho(770)$, $\omega(780)$ and $\phi(1020)$ with masses taken from the PDG\,\cite{ParticleDataGroup:2024cfk}.
Using the same LQCD dataset, but adopting the results of our {\em model-independent} BGL $z$-expansions with unitarity implemented through the double dispersive bounds we get
\be
     \label{eq:radius_LQCD}
     r_{K^\pm} = 0.641 \pm 0.022 \pm 0.001_{q} ~ [0.022] ~ \qquad \mbox{[LQCD data]} ~ , ~
\ee
which is consistent within $\simeq 2\sigma$ with the estimate made in Ref.\,\cite{Ding:2024lfj}, but with an uncertainty ten times larger. Note that, thanks to the use of the double dispersive bound, the dependence of our result\,(\ref{eq:radius_LQCD}) upon the choice of the kinematical function (i.e., the choice of $q$) is almost negligible.

\section{Conclusions}
\label{sec:conclusions}

We have discussed a specific application of the framework of multiple dispersive bounds, firstly introduced in the companion paper\,\cite{Simula:2025lpc}, to the study of sub-threshold branch-cuts. The main results are as follows:
\begin{itemize}
\item In the standard BGL $z$-expansion we have proposed a double dispersive bound as the proper way to deal with the presence of sub-threshold branch-cuts in accordance with unitarity. The key idea is that two different unitarity constraints have to be considered at the same time, namely the first one in the the pair-production arc and the second one in the extra region between the lowest sub-threshold and the pair-production branch-points. We have highlighted that the bound related to the extra region, $\chi_{\rm extra}^U$, is not related to the usual susceptibilities, i.e.\,to the quantities coming from the computation of the derivatives of appropriate two-point Green functions in momentum space. It remains an open issue to understand the proper Euclidean correlation functions that would give direct access to the bound $\chi_{\rm extra}^U$ directly from first-principles, e.g.\,from lattice QCD simulations.
\item We have developed a simple resonance model in order to take into account the effects of above-threshold poles on the description of hadronic form factors. We have successfully checked our model against the effects of the $\rho(770)$-meson resonance on the electromagnetic form factor of the pion. Then, we have extended our model to the case of the electromagnetic form factors of the charged and neutral kaons and we have successfully compared it with the dispersive results of Ref.\,\cite{Stamen:2022uqh}. 
\item We have pointed out that the choice of the outer function outside the pair-production arc is not unique. Possible choices have been discussed and used to evaluate the dispersive bound $\chi_{\rm extra}^U$.
\item We have analyzed either the experimental data or the lattice QCD results available for the electromagnetic form factor of the charged kaon in the spacelike-region. The comparison between the results corresponding to the BGL $z$-expansions based either on the single, total or on the double dispersive bound has clearly shown that the latter $z$-expansion provides the most precise extrapolation at large momentum transfer as well as the most stable results with respect to the choice of the outer function outside the pair-production region. We have considered also the predictions of the $z$-expansion based on the Szeg\H{o} polynomials, proposed in Refs.\,\cite{Gubernari:2020eft, Gubernari:2022hxn, Blake:2022vfl} and adopted also in Refs.\,\cite{Flynn:2023qmi, Flynn:2023nhi, Harrison:2025yan}, showing that the corresponding results for the kaon form factor do not fulfill unitarity on the full circle.
\item We have highlighted the impact of the application of the unitarity filter on the given set of input data (see Section\,\ref{sec:filtering} and Ref.\,\cite{Simula:2025lpc}) and we have compared our model-independent findings for the charged kaon radius with the model-dependent estimates made in the last PDG review\,\cite{ParticleDataGroup:2024cfk} and in Ref.\,\cite{Ding:2024lfj} (see Section\,\ref{sec:radius}).
\end{itemize}

We leave the investigation of multiple dispersive bounds to other physical processes of interest, such as weak semileptonic decays of hadrons, to future dedicated works.

\section*{Acknowledgements}
We warmly thank Guido Martinelli for all the insightful discussions we had together and for his continuous support.
We are deeply indebted with the authors of Ref.\,\cite{Stamen:2022uqh} for having provided us with the numerical results of their dispersive analysis for the charged kaon form factor $F_{K^\pm}^{(em)}(t)$ both in the spacelike sector down to $t \simeq -1$ GeV$^2$ and in the timelike one up to $t \simeq 1.2$ GeV$^2$.
S.S.~is supported by the Italian Ministry of University and Research (MUR) under grant PRIN 2022N4W8WR.
L.V.~is supported by the Italian Ministry of University and Research (MUR) and by the European Union’s NextGenerationEU program under the Young Researchers 2024 SoE Action, research project ‘SHYNE’, ID: SOE\_20240000025.

\appendix

\section{Riemann sheets in terms of the conformal variable}
\label{sec:sheets}

For a complex number $x + i y = r e^{i\theta}$, where $r = \sqrt{x^2 + y^2}$ and $-\pi < \theta \leq \pi$, the common choice is that in the first Riemann sheet (labelled as $I$) the real part of $\sqrt{x + i y}$ is nonnegative, namely
\be
    \label{eq:physical_sheet}
   \left[ \sqrt{x + i y} \right]^I = \sqrt{r} e^{i \theta / 2} = \sqrt{\frac{r +x}{2}} + i \, \mbox{sgn}(y) \sqrt{\frac{r-x}{2}} ~ , ~
\ee 
while in the second Riemann sheet (labelled as $II$) one has
\be
    \label{eq:unphysical_sheet}
   \left[ \sqrt{x + i y} \right]^{II} = \sqrt{r} e^{i (\theta + 2 \pi) / 2} = - \sqrt{r} e^{i \theta / 2} = - \sqrt{\frac{r +x}{2}} - i \, \mbox{sgn}(y) \sqrt{\frac{r-x}{2}} ~ . ~
\ee
This implies that $\left[ \sqrt{t_+ - t} \right]^I = - \left[ \sqrt{t_+ - t} \right]^{II}$, so that $\left[ z_+ \right]^I = 1 / \left[ z_+ \right]^{II}$. 
With the above choice of the principal value of the square root the first Riemann sheet corresponds to values of the conformal variable $z_+$ always inside the unit disk (i.e., $|z_+| < 1$), while in the second Riemann sheet the values of $z_+$ lie always outside the unit disk (i.e., $|z_+| >1$). On the branch-cut one has $\left[ |z_+| \right]^I = \left[ |z_+| \right]^{II} = 1$.

\section{Truncation errors in the BGL expansion}
\label{sec:truncation}

Following Ref.\,\cite{Boyd:1997kz}, the absolute difference between the true representation
\be
     \label{eq:true}
     \frac{\phi(z) B(z) f(z)}{\sqrt{\chi^U}} =  \sum_{k = 0}^\infty a_k z^k 
\ee
and its truncated expansion at order $M$
\be
     \label{eq:truncated}
     \frac{\phi(z) B(z) f^{(M)}(z)}{\sqrt{\chi^U}} = \sum_{k = 0}^M a_k z^k
\ee
has an upper limit for real values of $z$ given by
{\small
\be
    \frac{\left| \phi(z) B(z) \right|}{\sqrt{\chi^U}} \left| f(z) - f^{(M)}(z) \right| = \left| \sum_{k = M+1}^\infty a_k z^k \right| 
        \leq \sqrt{\sum_{k = M+1}^\infty a_k^2} \sqrt{\sum_{k = M+1}^\infty |z|^{2k}}  ~ , ~ \nonumber
\ee
}which, thanks to the unitarity constraint\,(\ref{eq:unitarity}), implies
\be
    \label{eq:truncation_error}
    \left| f(z) - f^{(M)}(z) \right| \leq \frac{\sqrt{\chi^U}}{\left| \phi(z) B(z) \right|} \, \frac{|z|^{M+1}}{\sqrt{1 - |z|^2}} 
        \sqrt{1 - \sum_{k = 0}^M a_k^2} ~ . ~
\ee
Therefore, it may be argued\,\cite{Boyd:1997kz} that, since $|z| < 1$ (and for $z$ far from the zeros of the Blaschke product), the truncation error\,(\ref{eq:truncation_error}) monotonically decreases as $M$ increases.

However, in practice the coefficients $a_k$ in Eq.\,(\ref{eq:truncated}) are typically obtained through a fitting procedure applied to a given set of input data $\{ f_i \}$. Thus, generally speaking, at fixed $k$ the coefficients $a_k$ in Eq.\,(\ref{eq:truncated}) depend on the truncation order $M$ and, therefore, they may not coincide with the corresponding coefficients of the true representation\,(\ref{eq:true}).
A further contribution, related to the differences $a_k - a_k^{(M)}$ for $k \leq M$, should be added to Eq.\,(\ref{eq:truncation_error}), obtaining the final estimate
{\small
\bea
    \label{eq:truncation_error_true}
    \left| f(z) - f^{(M)}(z) \right| & \leq & \frac{\sqrt{\chi^U}}{\left| \phi(z) B(z) \right|} \, \frac{1}{\sqrt{1 - |z|^2}}  
        \left\{ |z|^{M+1} \sqrt{1 - \sum_{k = 0}^M a_k^2} \right. ~ \nonumber \\[2mm]
        &+ &  \left. \sqrt{1 - |z|^{2(M+1)}} \sqrt{\sum_{k = 0}^M \left[a_k - a_k^{(M)} \right]^2}  \right\} ~ . ~
\eea
The size of the truncation error\,(\ref{eq:truncation_error_true}) depends also on the convergence of all the coefficients $a_k^{(M)}$ with $k \leq M$.
Thus, it is not guaranteed {\em a priori} that for a given value $|z| < 1$ the truncation error\,(\ref{eq:truncation_error_true}) decreases monotonically as $M$ increases.

\section{Orthonormal polynomials}
\label{sec:orthonormal}

Let us consider a generic truncated $z$-expansion expressed in terms of the monomials $z^k$, viz.
\be
     \label{eq:monomials}
     \eta(z) = \sum_{k = 0}^N b_k z^k ~
\ee
with $b_k$ being (real) coefficients satisfying the off-diagonal constraint
\be
     \label{eq:off_diagonal}
     \sum_{k, m = 0}^N b_k U_{k m} b_m \leq 1 ~ , ~
\ee
where the matrix $U$ is by definition real, symmetric and positive-definite.
According to the Cholensky decomposition, the matrix $U$ can be uniquely written as $U = L L^T$, where $L$ is a lower triangular matrix and $L^T$ its transpose (an upper triangular matrix).
Thus, the l.h.s.\,of Eq.\,(\ref{eq:off_diagonal}) can be rewritten in a diagonal form
\be
     \sum_{k, m = 0}^N b_k U_{k m} b_m = b^T U b = c^T c = \sum_{n = 0}^N c_n^2 ~ , ~
\ee
where $c = L^T b$. Since U is positive-definite, the matrix $L$ can be inverted and its inverse $L^{-1}$ is still a lower triangular matrix. One gets $b = (L^T)^{-1} c = (L^{-1})^T c$, so that Eq.\,(\ref{eq:monomials}) can be rewritten as
\be
    \eta(z) = \sum_{k = 0}^N c_k \, p_k(z) ~ , ~ 
\ee
where
\be
    \label{eq:orthonormal}
    p_k(z) = \sum_{m = 0}^N L^{-1}_{k m} \, z^m ~ . ~
\ee
Since $L^{-1}$ is a lower triangular matrix, i.e.~$L^{-1}_{k m} = 0$ for $ m > k$ and $L^{-1}_{k m} \neq 0$ for $m \leq k$, the polynomial $p_k(z)$ has a degree equal to $k$.
Note that the construction of the polynomial $p_k(z)$ does not depend upon the polynomials with a higher degree.

In the case of the matrix $U(\alpha_{th})$ given by Eq.\,(\ref{eq:U_Szego}) the corresponding polynomials\,(\ref{eq:orthonormal}) are known as the (normalized) Szeg\H{o} polynomials $p_k(z; \alpha_{th})$\,\cite{simon04}. They are orthonormal on the arc $-\alpha_{th} \leq \alpha \leq \alpha_{th}$, i.e.~on the unit circle $|z| =1$ with respect to a weight function given by $\Theta(\alpha_{th} - |\alpha|)$, namely
\bea
      & & \frac{1}{2\pi} \int_{-\pi}^\pi d\alpha \, \Theta(\alpha_{th} - |\alpha|) \, p_k(e^{i\alpha}; \alpha_{th}) \, p_{k^\prime}^*(e^{i\alpha}; \alpha_{th}) \nonumber \\[2mm] 
      & = & \frac{1}{2\pi} \int_{-\alpha_{th}}^{\alpha_{th}} d\alpha \, p_k(e^{i\alpha}; \alpha_{th}) \, p_{k^\prime}^*(e^{i\alpha}; \alpha_{th}) \\[2mm]  
      & = &  L_{k m}^{-1} U_{m m^\prime}(\alpha_{th}) (L^{-1})_{m^\prime k^\prime}^T = \delta_{k k^\prime} \nonumber  ~ . ~ 
\eea

\section{Eigenvalues of the matrix $U(\alpha_{th})$}
\label{sec:eigenvalues}

The eigenvalues $\lambda_k$ of the matrix $U_{k k^\prime}(\alpha_{th})$, given by Eq.\,(\ref{eq:U_Szego}), are shown in Fig.\,\ref{fig:eigenvalues} for various values of the matrix dimension $N$. 
It can be seen that all such (finite) matrices have eigenvalues in the range $(0, 1)$. This is related to the facts that: ~ i) the matrix $U(\alpha_{th})$ is positive definite (i.e.~$\vec{b}^T U(\alpha_{th}) \vec{b} > 0$ for any vector $\vec{b} \neq \vec{0}$), and ~ ii) Eq.\,(\ref{eq:U_Szego}) implies that $U_{k k}(\alpha_{th}) = \alpha_{th} / \pi < 1$, so that $\mbox{Tr}[U(\alpha_{th})] = N \alpha_{th} / \pi < N$.

The eigenvalues are approximately given by a Fermi-Dirac-like distribution\footnote{The eigenvalues $\lambda_k$ can be approximated with a good accuracy by the function $\{ 1 + e^{\frac{k / N - 1 + \alpha_{th} / \pi}{\sigma}} \}^{-1}$ with $\sigma \propto 1/N$.}. Thus, as $N$ increases, the distribution of the eigenvalues becomes closer and closer to an Heaviside step function, namely $\lambda_k \to 0$ for $k / N = 0, ..., 1 - \alpha_{th} / \pi$ and $\lambda_k \to 1$ for $k / N = 1 - \alpha_{th} / \pi, ... 1$.

\begin{figure}[htb!]
\begin{center}
\includegraphics[scale=0.45]{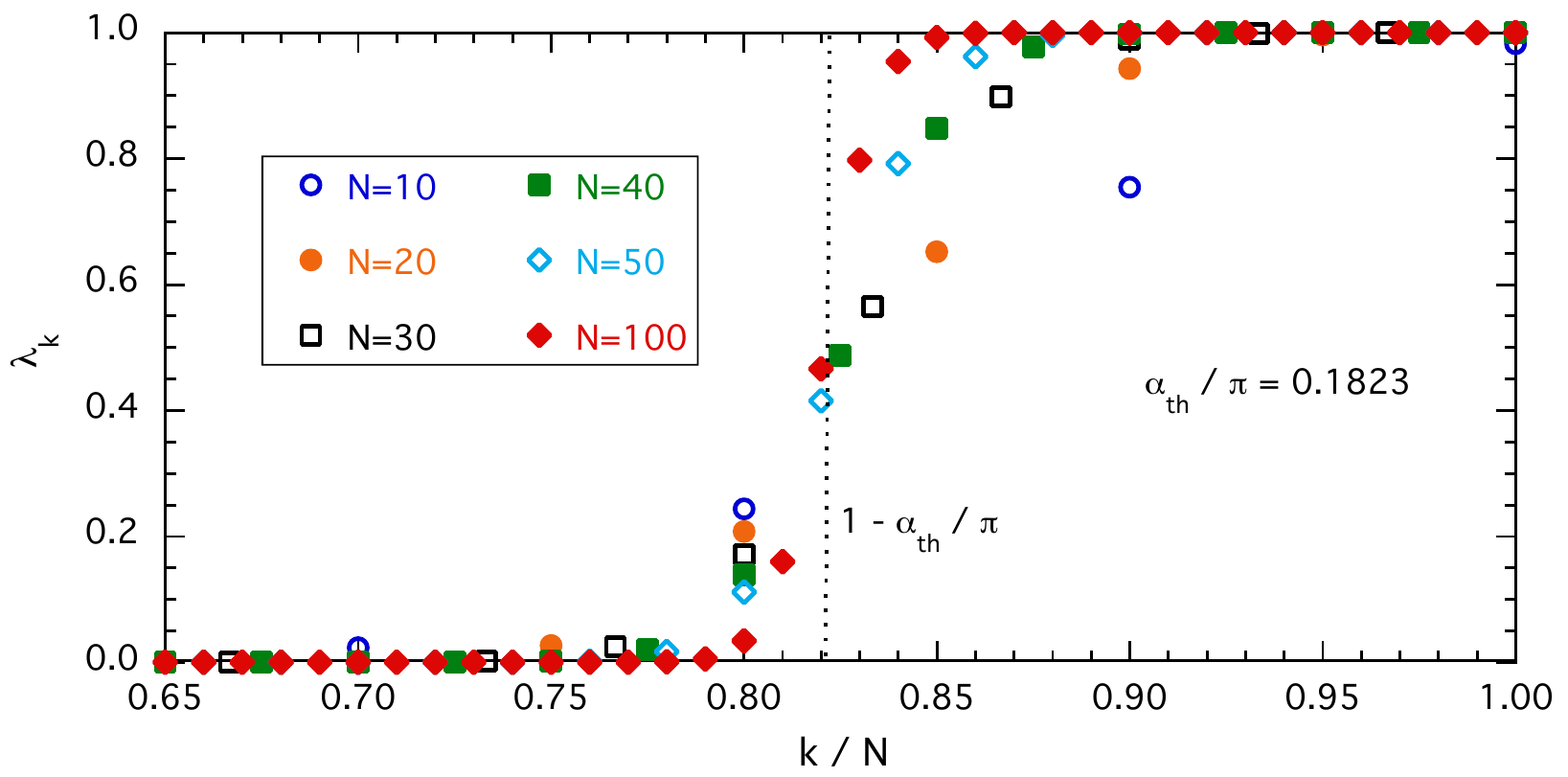}
\end{center}
\vspace{-0.75cm}
\caption{\it \small Eigenvalues $\lambda_k$ of the matrix $U_{k k^\prime}(\alpha_{th})$ with $k, k^\prime = 0, 1, ..., N$ for various values of its dimension $N$. The value of the parameter $\alpha_{th}$ is specified in the inset.}
\label{fig:eigenvalues}
\end{figure}

Few comments are in order.
\begin{itemize}

\item Since $0 < \lambda_k < 1$, one has
\be
     \sum_{k, k^\prime = 0}^N b_{k^\prime} U_{k^\prime k}(\alpha_{th}) b_k < \sum_{k = 0}^N b_k^2 ~ . ~
\ee
Thus, the diagonal constraint $\sum_{k = 0}^\infty a_k^2 \leq 1$, fulfilled by the coefficients of the expansion\,(\ref{eq:BGL}), implies that the coefficients $a_k$ satisfy also the off-diagonal constraint $\sum_{k, k^\prime = 0}^N a_{k^\prime} U_{k^\prime k}(\alpha_{th}) a_k \leq 1$ related to the pair-production arc. Instead, the opposite is not guaranteed: the off-diagonal constraint $\sum_{k, k^\prime = 0}^N b_{k^\prime} U_{k^\prime k}(\alpha_{th}) b_k \leq 1$ cannot guarantee that $\sum_{k = 0}^N b_k^2 \leq 1$. 
This in turn implies that the truncation error\,(\ref{eq:truncation_error}) of Appendix\,\ref{sec:truncation}, obtained in Refs.\,\cite{Boyd:1995sq, Boyd:1997kz}, may not be applicable to the expansion\,(\ref{eq:BGL_VD}) and, consequently, to the expansion\,(\ref{eq:BGL_Szego}).

In Fig.\,\ref{fig:Szego} the absolute values of few Szeg\H{o} polynomials $p_k(z; \alpha_{th})$ is shown for real values of $z$. It can be seen that for increasing values of $k$ the Szeg\H{o} polynomials $p_k(z; \alpha_{th})$ are not limited inside the unit disk at variance with the monomials $z^k$ (for which one has $|z^k| < 1$ for $|z| < 1$). This property confirms that the truncation error\,(\ref{eq:truncation_error}) of Appendix\,\ref{sec:truncation} may not be applicable to the expansions\,(\ref{eq:BGL_Szego}) or (\ref{eq:BGL_VD}).
\begin{figure}[htb!]
\begin{center}
\includegraphics[scale=0.40]{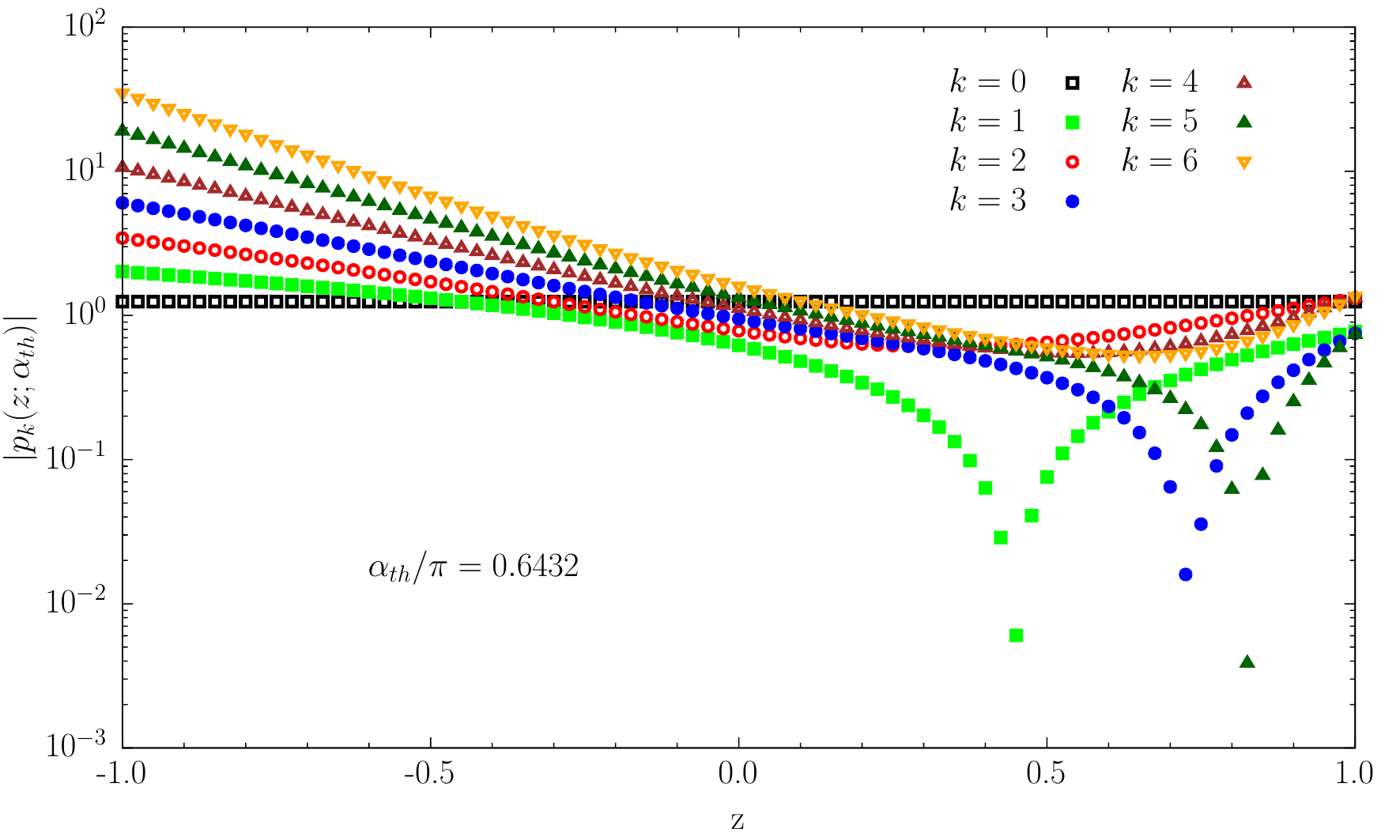}
\includegraphics[scale=0.40]{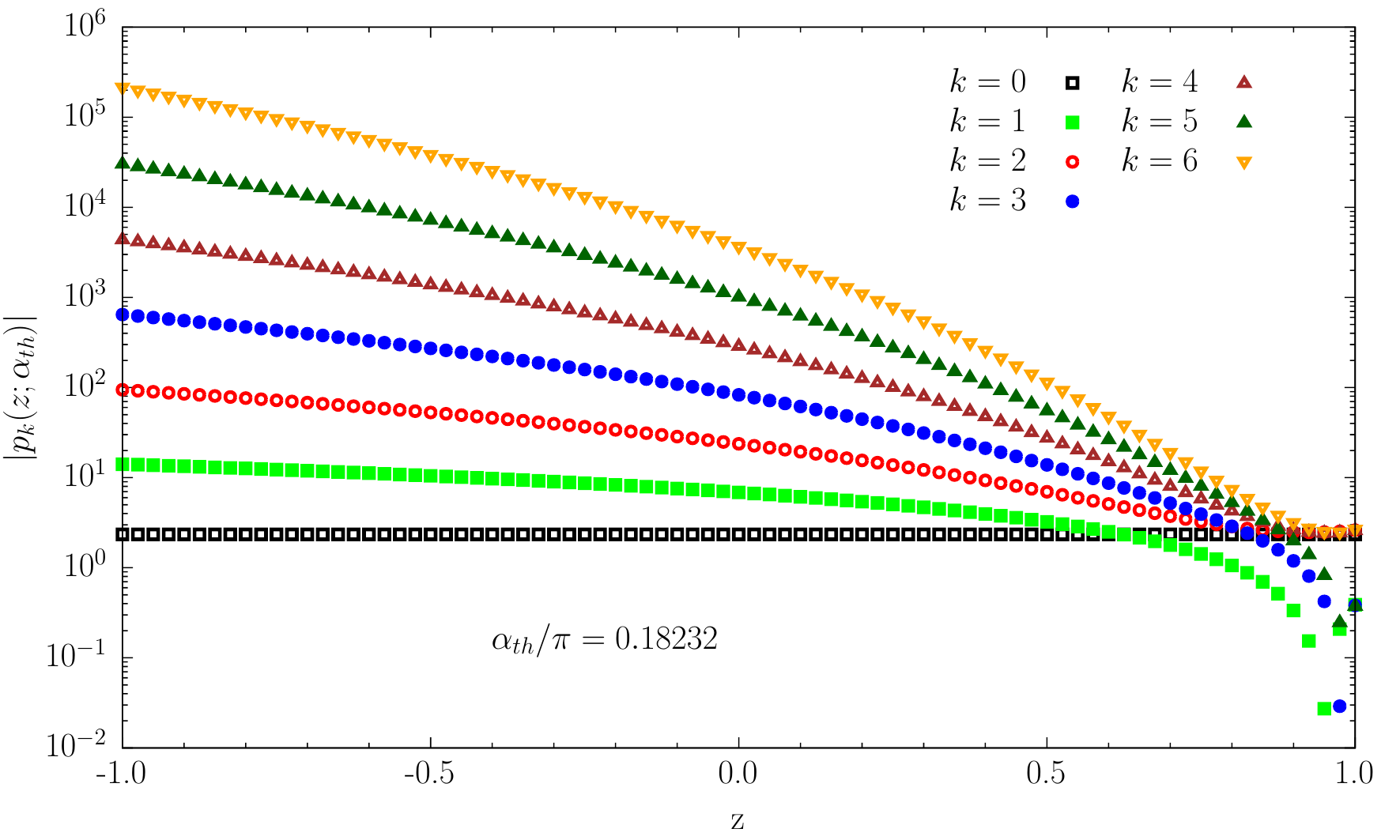}
\end{center}
\vspace{-0.75cm}
\caption{\it \small The absolute values of the (normalized) Szeg\H{o} polynomials $p_k(z; \alpha_{th})$ with $k \leq 6$ versus $z$ for a couple of values of $\alpha_{th}$ specified in the insets. The zero-order polynomial $p_0(z; \alpha_{th})$ is a constant given explicitly by $\sqrt{\pi / \alpha_{th}}$.}
\label{fig:Szego}
\end{figure}

\item In the limit $N \to \infty$ the eigenvalues of the matrix $U_{k k^\prime}(\alpha_{th})$ are either $0$ or $1$, i.e.~the matrix $U(\alpha_{th})$ acts as a projection. Thus, let us consider the coefficients $b_k$ of the expansion\,(\ref{eq:BGL_VD}) as the components of a vector $\vec{b}$ with norm given by $\vec{b}^T \vec{b} = \sum_{k = 0}^\infty b_k^2$. The vector $\vec{b}$ can be written as the sum of two vectors 
\be
    \vec{b} = \vec{b}^{(0)} + \vec{b}^{(1)} ~ , ~
\ee 
given by ($i = 0, 1$)
\be
     \vec{b}^{(i)} = \sum_{n = 0}^\infty c_n^{(i)} \vec{v}_n^{(i)} ~ , ~
\ee
where $\vec{v}_n^{(i)}$ is the $n$-th eigenvector of the matrix U corresponding to eigenvalue equal to $i$. 
The off-diagonal constraint\,(\ref{eq:unitarity_VD}) is equivalent to bound only the norm of the vector $\vec{b}^{(1)}$, namely
\be
    \sum_{k, k^\prime = 0}^\infty b_{k^\prime} U_{k^\prime k}(\alpha_{th}) b_k = \sum_{k = 0}^\infty (b_k^{(1)})^2 = \vec{b}^{(1) T} \vec{b}^{(1)} \leq 1 ~ . ~
\ee
The norm of the vector $\vec{b}^{(0)}$ as well as the norm of the full vector $\vec{b}$ are left unbounded.
Correspondingly, the$z$-expansion\,(\ref{eq:BGL_VD}) can be written as the sum of two $z$-expansions, namely
\be
    \label{eq:f0+f1}
    f(z) = f^{(0)}(z) + f^{(1)}(z) ~ , ~
\ee
where
\be
    f^{(i)}(z) =  \frac{\sqrt{\chi_+^U}}{\phi(z) B(z)} \sum_{k = 0}^\infty b_k^{(i)} z^k ~ . ~
\ee
Only the contribution $f^{(1)}(z)$ receives a true dispersive bound, i.e. $\sum_{k =0}^\infty (b_k^{(1)})^2 \leq 1$, while the quantity $f^{(0)}(z)$ is unbounded.

\end{itemize}

\bibliography{biblio}

\providecommand{\href}[2]{#2}\begingroup\raggedright\begin{thebibliography}{10}

\bibitem{Simula:2025lpc}
S.~Simula and L.~Vittorio, \emph{{Multiple dispersive bounds. I) An improved
  z-expansion}},  \href{https://arxiv.org/abs/2509.00411}{{\ttfamily
  2509.00411}}.

\bibitem{Caprini80}
I.~Caprini, \emph{{Constraints on physical amplitudes derived from a modified
  analytic interpolation problem}},
  \href{https://doi.org/10.1088/0305-4470/14/6/007}{\emph{J. Phys. A: Math.
  Gen.} {\bfseries 14} (1980) 1271}.

\bibitem{Caprini:1995wq}
I.~Caprini and M.~Neubert, \emph{{Improved bounds for the slope and curvature
  of anti-B ---{\ensuremath{>}} D(*) lepton anti-neutrino form-factors}},
  \href{https://doi.org/10.1016/0370-2693(96)00509-6}{\emph{Phys. Lett. B}
  {\bfseries 380} (1996) 376}
  [\href{https://arxiv.org/abs/hep-ph/9603414}{{\ttfamily hep-ph/9603414}}].

\bibitem{Boyd:1995sq}
C.G.~Boyd, B.~Grinstein and R.F.~Lebed, \emph{{Model independent determinations
  of anti-B ---\ensuremath{>} D (lepton), D* (lepton) anti-neutrino
  form-factors}},
  \href{https://doi.org/10.1016/0550-3213(95)00653-2}{\emph{Nucl. Phys. B}
  {\bfseries 461} (1996) 493}
  [\href{https://arxiv.org/abs/hep-ph/9508211}{{\ttfamily hep-ph/9508211}}].

\bibitem{Buck:1998kp}
W.W.~Buck and R.F.~Lebed, \emph{{New constraints on dispersive form-factor
  parameterizations from the timelike region}},
  \href{https://doi.org/10.1103/PhysRevD.58.056001}{\emph{Phys. Rev. D}
  {\bfseries 58} (1998) 056001}
  [\href{https://arxiv.org/abs/hep-ph/9802369}{{\ttfamily hep-ph/9802369}}].

\bibitem{Bhattacharya:2011ah}
B.~Bhattacharya, R.J.~Hill and G.~Paz, \emph{{Model independent determination
  of the axial mass parameter in quasielastic neutrino-nucleon scattering}},
  \href{https://doi.org/10.1103/PhysRevD.84.073006}{\emph{Phys. Rev. D}
  {\bfseries 84} (2011) 073006}
  [\href{https://arxiv.org/abs/1108.0423}{{\ttfamily 1108.0423}}].

\bibitem{Epstein:2014zua}
Z.~Epstein, G.~Paz and J.~Roy, \emph{{Model independent extraction of the
  proton magnetic radius from electron scattering}},
  \href{https://doi.org/10.1103/PhysRevD.90.074027}{\emph{Phys. Rev. D}
  {\bfseries 90} (2014) 074027}
  [\href{https://arxiv.org/abs/1407.5683}{{\ttfamily 1407.5683}}].

\bibitem{Gopal:2024mgb}
A.~Gopal and N.~Gubernari, \emph{{Unitarity bounds with subthreshold and
  anomalous cuts for b-hadron decays}},
  \href{https://doi.org/10.1103/PhysRevD.111.L031501}{\emph{Phys. Rev. D}
  {\bfseries 111} (2025) L031501}
  [\href{https://arxiv.org/abs/2412.04388}{{\ttfamily 2412.04388}}].

\bibitem{Boyd:1994tt}
C.G.~Boyd, B.~Grinstein and R.F.~Lebed, \emph{{Constraints on form-factors for
  exclusive semileptonic heavy to light meson decays}},
  \href{https://doi.org/10.1103/PhysRevLett.74.4603}{\emph{Phys. Rev. Lett.}
  {\bfseries 74} (1995) 4603}
  [\href{https://arxiv.org/abs/hep-ph/9412324}{{\ttfamily hep-ph/9412324}}].

\bibitem{Boyd:1995cf}
C.G.~Boyd, B.~Grinstein and R.F.~Lebed, \emph{{Model independent extraction of
  |V(cb)| using dispersion relations}},
  \href{https://doi.org/10.1016/0370-2693(95)00480-9}{\emph{Phys. Lett. B}
  {\bfseries 353} (1995) 306}
  [\href{https://arxiv.org/abs/hep-ph/9504235}{{\ttfamily hep-ph/9504235}}].

\bibitem{Boyd:1997kz}
C.G.~Boyd, B.~Grinstein and R.F.~Lebed, \emph{{Precision corrections to
  dispersive bounds on form-factors}},
  \href{https://doi.org/10.1103/PhysRevD.56.6895}{\emph{Phys. Rev. D}
  {\bfseries 56} (1997) 6895}
  [\href{https://arxiv.org/abs/hep-ph/9705252}{{\ttfamily hep-ph/9705252}}].

\bibitem{Caprini:2015wja}
I.~Caprini, \emph{{Testing the consistency of the $\omega\pi$ transition form
  factor with unitarity and analyticity}},
  \href{https://doi.org/10.1103/PhysRevD.92.014014}{\emph{Phys. Rev. D}
  {\bfseries 92} (2015) 014014}
  [\href{https://arxiv.org/abs/1505.05282}{{\ttfamily 1505.05282}}].

\bibitem{Dally:1980dj}
E.B.~Dally et~al., \emph{{DIRECT MEASUREMENT OF THE NEGATIVE KAON
  FORM-FACTOR}}, \href{https://doi.org/10.1103/PhysRevLett.45.232}{\emph{Phys.
  Rev. Lett.} {\bfseries 45} (1980) 232}.

\bibitem{Amendolia:1986ui}
S.R.~Amendolia et~al., \emph{{A Measurement of the Kaon Charge Radius}},
  \href{https://doi.org/10.1016/0370-2693(86)91407-3}{\emph{Phys. Lett. B}
  {\bfseries 178} (1986) 435}.

\bibitem{Ding:2024lfj}
H.-T.~Ding, X.~Gao, A.D.~Hanlon, S.~Mukherjee, P.~Petreczky, Q.~Shi et~al.,
  \emph{{QCD Predictions for Meson Electromagnetic Form Factors at High
  Momenta: Testing Factorization in Exclusive Processes}},
  \href{https://doi.org/10.1103/PhysRevLett.133.181902}{\emph{Phys. Rev. Lett.}
  {\bfseries 133} (2024) 181902}
  [\href{https://arxiv.org/abs/2404.04412}{{\ttfamily 2404.04412}}].

\bibitem{ParticleDataGroup:2024cfk}
{\scshape Particle Data Group} collaboration, \emph{{Review of particle
  physics}}, \href{https://doi.org/10.1103/PhysRevD.110.030001}{\emph{Phys.
  Rev. D} {\bfseries 110} (2024) 030001}.

\bibitem{Caprini:2019osi}
I.~Caprini, \emph{{Functional Analysis and Optimization Methods in Hadron
  Physics}}, SpringerBriefs in Physics, Springer (2019),
  \href{https://doi.org/10.1007/978-3-030-18948-8}{10.1007/978-3-030-18948-8}.

\bibitem{Meiman63}
N.N.~Meiman, \emph{{Analytic Expressions for Upper Limits of Coupling Constants
  in Quantum Field Theory}}, {\emph{Sov. Phys. JETP} {\bfseries 17} (1963)
  830}.

\bibitem{Okubo:1971jf}
S.~Okubo, \emph{{Exact bounds for k-l-3 decay parameters}},
  \href{https://doi.org/10.1103/PhysRevD.3.2807}{\emph{Phys. Rev. D} {\bfseries
  3} (1971) 2807}.

\bibitem{Okubo:1971my}
S.~Okubo, \emph{{New improved bounds for k-l-3 parameters}},
  \href{https://doi.org/10.1103/PhysRevD.4.725}{\emph{Phys. Rev. D} {\bfseries
  4} (1971) 725}.

\bibitem{Okubo:1971wup}
S.~Okubo and I.-F.~Shih, \emph{{Exact inequality and test of chiral sw(3)
  theory in k-l-3 decay problem}},
  \href{https://doi.org/10.1103/PhysRevD.4.2020}{\emph{Phys. Rev. D} {\bfseries
  4} (1971) 2020}.

\bibitem{Bourrely:1980gp}
C.~Bourrely, B.~Machet and E.~de~Rafael, \emph{{Semileptonic Decays of
  Pseudoscalar Particles ($M \to M^\prime \ell \nu_\ell$) and Short Distance
  Behavior of Quantum Chromodynamics}},
  \href{https://doi.org/10.1016/0550-3213(81)90086-9}{\emph{Nucl. Phys. B}
  {\bfseries 189} (1981) 157}.

\bibitem{Ananthanarayan:2014pta}
B.~Ananthanarayan, I.~Caprini and B.~Kubis, \emph{{Constraints on the $\mathbf
  {\omega \pi }$ form factor from analyticity and unitarity}},
  \href{https://doi.org/10.1140/epjc/s10052-014-3209-4}{\emph{Eur. Phys. J. C}
  {\bfseries 74} (2014) 3209}
  [\href{https://arxiv.org/abs/1410.6276}{{\ttfamily 1410.6276}}].

\bibitem{DiCarlo:2021dzg}
M.~Di~Carlo, G.~Martinelli, M.~Naviglio, F.~Sanfilippo, S.~Simula and
  L.~Vittorio, \emph{{Unitarity bounds for semileptonic decays in lattice
  QCD}}, \href{https://doi.org/10.1103/PhysRevD.104.054502}{\emph{Phys. Rev. D}
  {\bfseries 104} (2021) 054502}
  [\href{https://arxiv.org/abs/2105.02497}{{\ttfamily 2105.02497}}].

\bibitem{Lellouch:1995yv}
L.~Lellouch, \emph{{Lattice constrained unitarity bounds for anti-B0
  ---\ensuremath{>} pi+ lepton- anti-lepton-neutrino decays}},
  \href{https://doi.org/10.1016/0550-3213(96)00443-9}{\emph{Nucl. Phys. B}
  {\bfseries 479} (1996) 353}
  [\href{https://arxiv.org/abs/hep-ph/9509358}{{\ttfamily hep-ph/9509358}}].

\bibitem{Gubernari:2020eft}
N.~Gubernari, D.~van Dyk and J.~Virto, \emph{{Non-local matrix elements in
  $B_{(s)}\to \{K^{(*)},\phi\}\ell^+\ell^-$}},
  \href{https://doi.org/10.1007/JHEP02(2021)088}{\emph{JHEP} {\bfseries 02}
  (2021) 088} [\href{https://arxiv.org/abs/2011.09813}{{\ttfamily
  2011.09813}}].

\bibitem{Gubernari:2022hxn}
N.~Gubernari, M.~Reboud, D.~van Dyk and J.~Virto, \emph{{Improved theory
  predictions and global analysis of exclusive $b \to s\mu^+\mu^-$ processes}},
  \href{https://doi.org/10.1007/JHEP09(2022)133}{\emph{JHEP} {\bfseries 09}
  (2022) 133} [\href{https://arxiv.org/abs/2206.03797}{{\ttfamily
  2206.03797}}].

\bibitem{Blake:2022vfl}
T.~Blake, S.~Meinel, M.~Rahimi and D.~van Dyk, \emph{{Dispersive bounds for
  local form factors in
  \ensuremath{\Lambda}b\textrightarrow{}\ensuremath{\Lambda} transitions}},
  \href{https://doi.org/10.1103/PhysRevD.108.094509}{\emph{Phys. Rev. D}
  {\bfseries 108} (2023) 094509}
  [\href{https://arxiv.org/abs/2205.06041}{{\ttfamily 2205.06041}}].

\bibitem{Flynn:2023qmi}
J.M.~Flynn, A.~J\"uttner and J.T.~Tsang, \emph{{Bayesian inference for
  form-factor fits regulated by unitarity and analyticity}},
  \href{https://doi.org/10.1007/JHEP12(2023)175}{\emph{JHEP} {\bfseries 12}
  (2023) 175} [\href{https://arxiv.org/abs/2303.11285}{{\ttfamily
  2303.11285}}].

\bibitem{Flynn:2023nhi}
{\scshape RBC/UKQCD} collaboration, \emph{{Exclusive semileptonic
  Bs\textrightarrow{}K\ensuremath{\ell}\ensuremath{\nu} decays on the
  lattice}}, \href{https://doi.org/10.1103/PhysRevD.107.114512}{\emph{Phys.
  Rev. D} {\bfseries 107} (2023) 114512}
  [\href{https://arxiv.org/abs/2303.11280}{{\ttfamily 2303.11280}}].

\bibitem{Harrison:2025yan}
{\scshape HPQCD} collaboration, \emph{{Improved lattice QCD
  Bc{\textrightarrow}J/{\ensuremath{\psi}} vector, axial-vector, and tensor
  form factors}}, \href{https://doi.org/10.1103/wll6-z4cb}{\emph{Phys. Rev. D}
  {\bfseries 112} (2025) 034503}
  [\href{https://arxiv.org/abs/2503.15090}{{\ttfamily 2503.15090}}].

\bibitem{simon04}
B.~Simon, \emph{{Orthogonal polynomials on the unit circle: New results}},
  \href{https://doi.org/10.48550/arXiv.math/0405111}{\emph{International
  Mathematics Research Notices} (2004) 2837}
  [\href{https://arxiv.org/abs/math/0405111}{{\ttfamily math/0405111}}].

\bibitem{Barton65}
G.~Barton, \emph{{Introduction to Dispersion Techniques in Field Theory}},
  Lecture notes and supplements in physics, W.A. Benjamin, New York, Amstyerdam
  (1965).

\bibitem{Caprini:2005zr}
I.~Caprini, G.~Colangelo and H.~Leutwyler, \emph{{Mass and width of the lowest
  resonance in QCD}},
  \href{https://doi.org/10.1103/PhysRevLett.96.132001}{\emph{Phys. Rev. Lett.}
  {\bfseries 96} (2006) 132001}
  [\href{https://arxiv.org/abs/hep-ph/0512364}{{\ttfamily hep-ph/0512364}}].

\bibitem{Grinstein:2015wqa}
B.~Grinstein and R.F.~Lebed, \emph{{Above-Threshold Poles in Model-Independent
  Form Factor Parametrizations}},
  \href{https://doi.org/10.1103/PhysRevD.92.116001}{\emph{Phys. Rev. D}
  {\bfseries 92} (2015) 116001}
  [\href{https://arxiv.org/abs/1509.04847}{{\ttfamily 1509.04847}}].

\bibitem{Caprini:2017ins}
I.~Caprini, B.~Grinstein and R.F.~Lebed, \emph{{Model-independent constraints
  on hadronic form factors with above-threshold poles}},
  \href{https://doi.org/10.1103/PhysRevD.96.036015}{\emph{Phys. Rev. D}
  {\bfseries 96} (2017) 036015}
  [\href{https://arxiv.org/abs/1705.02368}{{\ttfamily 1705.02368}}].

\bibitem{Fermi:1955xrk}
E.~Fermi and B.T.~Feld, \emph{{Lectures on pions and nucleons}},
  \href{https://doi.org/10.1007/BF02746078}{\emph{Nuovo Cim.} {\bfseries 2}
  (1955) 17}.

\bibitem{Watson:1954uc}
K.M.~Watson, \emph{{Some general relations between the photoproduction and
  scattering of pi mesons}},
  \href{https://doi.org/10.1103/PhysRev.95.228}{\emph{Phys. Rev.} {\bfseries
  95} (1954) 228}.

\bibitem{Colangelo:2018mtw}
G.~Colangelo, M.~Hoferichter and P.~Stoffer, \emph{{Two-pion contribution to
  hadronic vacuum polarization}},
  \href{https://doi.org/10.1007/JHEP02(2019)006}{\emph{JHEP} {\bfseries 02}
  (2019) 006} [\href{https://arxiv.org/abs/1810.00007}{{\ttfamily
  1810.00007}}].

\bibitem{Bourrely:2008za}
C.~Bourrely, I.~Caprini and L.~Lellouch, \emph{{Model-independent description
  of B ---\ensuremath{>} pi l nu decays and a determination of |V(ub)|}},
  \href{https://doi.org/10.1103/PhysRevD.82.099902}{\emph{Phys. Rev. D}
  {\bfseries 79} (2009) 013008}
  [\href{https://arxiv.org/abs/0807.2722}{{\ttfamily 0807.2722}}], [Erratum:
  Phys.Rev.D 82, 099902 (2010)].

\bibitem{Kirk:2024oyl}
M.~Kirk, B.~Kubis, M.~Reboud and D.~van Dyk, \emph{{A simple parametrisation of
  the pion form factor}},
  \href{https://doi.org/10.1016/j.physletb.2025.139266}{\emph{Phys. Lett. B}
  {\bfseries 861} (2025) 139266}
  [\href{https://arxiv.org/abs/2410.13764}{{\ttfamily 2410.13764}}].

\bibitem{Protopopescu:1973sh}
S.D.~Protopopescu, M.~Alston-Garnjost, A.~Barbaro-Galtieri, S.M.~Flatte,
  J.H.~Friedman, T.A.~Lasinski et~al., \emph{{Pi pi Partial Wave Analysis from
  Reactions pi+ p ---\ensuremath{>} pi+ pi- Delta++ and pi+ p ---\ensuremath{>}
  K+ K- Delta++ at 7.1-GeV/c}},
  \href{https://doi.org/10.1103/PhysRevD.7.1279}{\emph{Phys. Rev. D} {\bfseries
  7} (1973) 1279}.

\bibitem{Estabrooks:1974vu}
P.~Estabrooks and A.D.~Martin, \emph{{pi pi Phase Shift Analysis Below the K
  anti-K Threshold}},
  \href{https://doi.org/10.1016/0550-3213(74)90488-X}{\emph{Nucl. Phys. B}
  {\bfseries 79} (1974) 301}.

\bibitem{Simula:2023ujs}
S.~Simula and L.~Vittorio, \emph{{Dispersive analysis of the experimental data
  on the electromagnetic form factor of charged pions at spacelike momenta}},
  \href{https://doi.org/10.1103/PhysRevD.108.094013}{\emph{Phys. Rev. D}
  {\bfseries 108} (2023) 094013}
  [\href{https://arxiv.org/abs/2309.02135}{{\ttfamily 2309.02135}}].

\bibitem{Ananthanarayan:2011uc}
B.~Ananthanarayan, I.~Caprini and I.~Sentitemsu~Imsong, \emph{{Implications of
  unitarity and analyticity for the D{\textbackslash}pi form factors}},
  \href{https://doi.org/10.1140/epja/i2011-11147-7}{\emph{Eur. Phys. J. A}
  {\bfseries 47} (2011) 147} [\href{https://arxiv.org/abs/1108.0284}{{\ttfamily
  1108.0284}}].

\bibitem{Herren:2025cwv}
F.~Herren, B.~Kubis and R.~van Tonder, \emph{{Model-independent parametrization
  of
  B{\textrightarrow}{\ensuremath{\pi}}{\ensuremath{\pi}}{\ensuremath{\ell}}{\ensuremath{\nu}}
  decays}}, \href{https://doi.org/10.1103/8pm6-9xzq}{\emph{Phys. Rev. D}
  {\bfseries 112} (2025) 014037}
  [\href{https://arxiv.org/abs/2502.20960}{{\ttfamily 2502.20960}}].

\bibitem{Omnes:1958hv}
R.~Omnes, \emph{{On the Solution of certain singular integral equations of
  quantum field theory}}, \href{https://doi.org/10.1007/BF02747746}{\emph{Nuovo
  Cim.} {\bfseries 8} (1958) 316}.

\bibitem{Stamen:2022uqh}
D.~Stamen, D.~Hariharan, M.~Hoferichter, B.~Kubis and P.~Stoffer, \emph{{Kaon
  electromagnetic form factors in dispersion theory}},
  \href{https://doi.org/10.1140/epjc/s10052-022-10348-3}{\emph{Eur. Phys. J. C}
  {\bfseries 82} (2022) 432}
  [\href{https://arxiv.org/abs/2202.11106}{{\ttfamily 2202.11106}}].

\bibitem{Blatnik:1978wj}
S.~Blatnik, J.~Stahov and C.B.~Lang, \emph{{The Isovector Part of the Kaon
  Form-factor and the Kaon Charge Radii}},
  \href{https://doi.org/10.1007/BF02725742}{\emph{Lett. Nuovo Cim.} {\bfseries
  24} (1979) 39}.

\bibitem{Pelaez:2020gnd}
J.R.~Pel\'aez and A.~Rodas, \emph{{Dispersive
  \ensuremath{\pi}K\textrightarrow{}\ensuremath{\pi}K and
  \ensuremath{\pi}\ensuremath{\pi}\textrightarrow{}K$\overline{K}$amplitudes
  from scattering data, threshold parameters, and the lightest strange
  resonance \ensuremath{\kappa} or K0\ensuremath{*}(700)}},
  \href{https://doi.org/10.1016/j.physrep.2022.03.004}{\emph{Phys. Rept.}
  {\bfseries 969} (2022) 1} [\href{https://arxiv.org/abs/2010.11222}{{\ttfamily
  2010.11222}}].

\bibitem{Muskhelishvili58}
N.~Muskhelishvili, \emph{{Singular Integral Equations}}, Springer Dordrecht
  (1958),
  \href{https://doi.org/https://doi.org/10.1007/978-94-009-9994-7}{https://doi.org/10.1007/978-94-009-9994-7}.

\bibitem{BaBar:2018qry}
{\scshape BaBar} collaboration, \emph{{Measurement of the spectral function for
  the $\tau^-\to K^-K_S\nu_{\tau}$ decay}},
  \href{https://doi.org/10.1103/PhysRevD.98.032010}{\emph{Phys. Rev. D}
  {\bfseries 98} (2018) 032010}
  [\href{https://arxiv.org/abs/1806.10280}{{\ttfamily 1806.10280}}].

\bibitem{Bharucha:2010im}
A.~Bharucha, T.~Feldmann and M.~Wick, \emph{{Theoretical and Phenomenological
  Constraints on Form Factors for Radiative and Semi-Leptonic B-Meson Decays}},
  \href{https://doi.org/10.1007/JHEP09(2010)090}{\emph{JHEP} {\bfseries 09}
  (2010) 090} [\href{https://arxiv.org/abs/1004.3249}{{\ttfamily 1004.3249}}].

\bibitem{Martinelli:2021frl}
G.~Martinelli, S.~Simula and L.~Vittorio, \emph{{Constraints for the
  semileptonic B\textrightarrow{}D(*) form factors from lattice QCD simulations
  of two-point correlation functions}},
  \href{https://doi.org/10.1103/PhysRevD.104.094512}{\emph{Phys. Rev. D}
  {\bfseries 104} (2021) 094512}
  [\href{https://arxiv.org/abs/2105.07851}{{\ttfamily 2105.07851}}].

\bibitem{Melis:2024wpb}
A.~Melis, F.~Sanfilippo and S.~Simula, \emph{{Hadronic susceptibilities for b
  to c transitions from two point correlation functions}},
  \href{https://doi.org/10.22323/1.453.0243}{\emph{PoS} {\bfseries LATTICE2023}
  (2024) 243} [\href{https://arxiv.org/abs/2401.03920}{{\ttfamily
  2401.03920}}].

\bibitem{Harrison:2024iad}
J.~Harrison, \emph{{b\textasciimacron{}c susceptibilities from fully
  relativistic lattice QCD}},
  \href{https://doi.org/10.1103/PhysRevD.110.054506}{\emph{Phys. Rev. D}
  {\bfseries 110} (2024) 054506}
  [\href{https://arxiv.org/abs/2405.01390}{{\ttfamily 2405.01390}}].

\bibitem{Martinelli:2022tte}
G.~Martinelli, S.~Simula and L.~Vittorio, \emph{{Exclusive semileptonic B
  \textrightarrow{} \ensuremath{\pi}\ensuremath{\ell}\ensuremath{\nu}$_\ell$
  and B$_{s}$ \textrightarrow{} K\ensuremath{\ell}\ensuremath{\nu}$_\ell$
  decays through unitarity and lattice QCD}},
  \href{https://doi.org/10.1007/JHEP08(2022)022}{\emph{JHEP} {\bfseries 08}
  (2022) 022} [\href{https://arxiv.org/abs/2202.10285}{{\ttfamily
  2202.10285}}].

\end{thebibliography}\endgroup
\bibliographystyle{JHEP}

\end{document}